\documentclass[a4paper,11pt]{article}
\usepackage{jheppub} 
\usepackage{lineno}
\usepackage{graphicx}
\usepackage{lipsum}
\usepackage{slashed}
\usepackage{mathtools}
\usepackage{bm}
\usepackage[all]{hypcap}
\usepackage{tikz}
\usepackage{tikz-feynman}
\usepackage{tikz-cd}
\usepackage{bbm}
\usepackage{subcaption}
\usepackage[font=footnotesize,labelfont=bf]{caption}
\usepackage{xcolor}
\usepackage[normalem]{ulem}
\usepackage{array}
\allowdisplaybreaks

\allowdisplaybreaks

\newtheorem{theorem}{Theorem}[section]

\newtheorem{lemma}[theorem]{Lemma}

\def\Z{\mathbb{Z}}
\def\R{\mathbb{R}}

\def\Spin{\text{Spin}}

\def\de{\partial}

\def\sS{\sigma_{\text{S}}}
\def\sH{\sigma_{\text{H}}}

\usepackage{xspace}

\newcommand{\be}{\begin{equation}}
\newcommand{\ee}{\end{equation}}
\newcommand{\bea}{\begin{eqnarray}}
\newcommand{\eea}{\end{eqnarray}}
\newcommand{\beq}{\begin{equation}}
\newcommand{\eeq}{\end{equation}}

\newenvironment{nalign}{ 
	\begin{equation}
	\begin{aligned}
}{
	\end{aligned}
	\end{equation}
	\ignorespacesafterend
}

\newcommand{\cL}{{\cal L}}

\begin{document}

\title{Jet Bundle Geometry of Scalar Field Theories}

\author[a]{Mohammad Alminawi,}
\author[a,b]{Ilaria Brivio,}
\author[a]{and Joe Davighi}

\affiliation[a]{Physik Institut, Universit\"at Z\"urich, Switzerland}
\emailAdd{mohammad.alminawi@physik.uzh.ch}
\affiliation[b]{Dipartimento di Fisica e Astronomia, Università di Bologna, Italy}
\emailAdd{ilaria.brivio@unibo.it}

\emailAdd{joe.davighi@physik.uzh.ch}

\abstract{ For scalar field theories, such as those EFTs describing the Higgs, it is well-known that the 2-derivative Lagrangian is captured by geometry. That is, the set of operators with exactly 2 derivatives can be obtained by pulling back a metric from a field space manifold $M$ to spacetime $\Sigma$. We here generalise this geometric understanding of scalar field theories to higher- (and lower-) derivative Lagrangians. We show how the entire EFT Lagrangian with up to 4-derivatives can be obtained from geometry by pulling back a metric to $\Sigma$ from the \emph{1-jet bundle} that is (roughly) associated with maps from $\Sigma$ to $M$. More precisely, our starting point is to trade the field space $M$ for a fibre bundle $\pi:E \to \Sigma$, with fibre $M$, of which the scalar field $\phi$ is a local section. We discuss symmetries and field redefinitions in this bundle formalism, before showing how everything can be `prolongated' to the 1-jet bundle $J^1 E$ which, as a manifold, is the space of sections $\phi$ that agree in their zeroth and first derivatives above each spacetime point. Equipped with a notion of (spacetime and internal) symmetry on $J^1 E$, the idea is that one can write down the most general metric on $J^1 E$ consistent with symmetries, in the spirit of the effective field theorist, and pull it back to spacetime to build an invariant Lagrangian; because $J^1 E$ has `derivative coordinates', one naturally obtains operators with more than 2-derivatives from this geometry. We apply this formalism to various examples, including a single real scalar in 4d and a quartet of real scalars with $O(4)$ symmetry that describes the Higgs EFTs. We show how an entire non-redundant basis of 0-, 2-, and 4-derivative operators is obtained from jet bundle geometry in this way. Finally, we study the connection to amplitudes and the role of geometric invariants.
}

\maketitle

\section{Introduction}

Effective Field Theories (EFTs) have long played a crucial role in guiding our understanding of fundamental physics to higher and higher energies. In the past decade, EFTs have been increasingly regarded as essential tools to explore physics beyond the Standard Model (SM), becoming prominent in both the theoretical and experimental programs of new physics searches~\cite{Brivio:2017vri,Isidori:2023pyp}.
Two EFTs are particularly pertinent in this context: the so-called Standard Model EFT (SMEFT)~\cite{Buchmuller:1985jz,Grzadkowski:2010es} and the Higgs EFT (HEFT)~\cite{Feruglio:1992wf,Buchalla:2012qq,Alonso:2012px,Buchalla:2013rka,Brivio:2016fzo,Dong:2022jru}. Both theories extend the SM Lagrangian with higher-dimensional interactions, but they differ in the choice of scalar fields and in the power-counting adopted to organize the EFT series.
While the SMEFT is formulated in terms of the $SU(2)$ Higgs doublet, that transforms linearly under the (gauge) symmetries, the HEFT builds upon the formalism of the electroweak chiral Lagrangian, whereby the three Goldstone bosons are embedded in a chiral field that transforms non-linearly~\cite{Coleman:1969sm,Appelquist:1974tg,Longhitano:1980iz,Longhitano:1980tm}, and the physical Higgs particle is treated as a singlet excitation~\cite{Grinstein:2007iv}. 
This difference is expected to capture different dynamics underlying the electroweak symmetry breaking (EWSB) mechanism, which gives great significance
to the question of which EFT should be used to go beyond the SM~\cite{Falkowski:2019tft}. 

The phenomenological characterization of SMEFT and HEFT is a long-standing open challenge (see {\em e.g.}~\cite{Isidori:2013cla,Brivio:2013pma,Ellis:2013lra,Liu:2019rce,Eboli:2021unw,Gomez-Ambrosio:2022qsi,Domenech:2022uud,Dawson:2023ebe}). The main obstacle to overcome in realising this program stems from ambiguities in the formulation of these EFTs; namely, the fact that the two can be partially mapped onto each other via field redefinitions, indicating that at least a fraction of the phenomenological differences emerging in an order-by-order comparison might be unphysical. In particular, some of these differences might be washed out upon (partially) resumming the infinite tower of EFT operators.

Geometrical formulations of these EFTs~\cite{Burgess:2010zq,Alonso:2015fsp,Alonso:2016oah,Helset:2018fgq,Corbett:2019cwl,Helset:2020yio} were introduced in recent years partly to address this issue, and have already proven very powerful in characterizing HEFT and SMEFT at a more fundamental level~\cite{Alonso:2017tdy,Hays:2020scx,Cohen:2020xca,Cohen:2021ucp,Alonso:2021rac,Alonso:2022ffe,Helset:2022pde,Talbert:2022unj,Helset:2022tlf}; see~\cite{Alonso:2023upf} for a pedagogical overview.
The core idea is that the (generalised) kinetic term of the scalar sector, by which we mean the set of all 2-derivative operators, could be seen as defining a metric on a real 4-manifold $M$ spanned (locally) by the four scalar fields. In this language, the invariance of physical scattering amplitudes under field redefinitions has been interpreted in terms of the invariance of geometric quantities (specifically, curvature invariants) of the field space under coordinate transformations. 
This approach allowed a more rigorous (and representation-independent) formulation of the differences between HEFT and SMEFT, and of the conditions under which each EFT is appropriate.
Because the scalar manifold geometry emerges as a remnant of the decoupling of heavy BSM particles, these conditions can also be related to specific properties of the UV sector~\cite{Alonso:2016oah,Cohen:2020xca,Banta:2023prj}. For instance, a SMEFT expansion cannot be constructed if the scalar geometry is singular at the point where the EW symmetry is restored. This can be associated to the presence of so-called `loryons', heavy particles that have been integrated out in the EFT but which take at least `half' of their mass from the Higgs mechanism~\cite{Banta:2021dek}.

Arguably, this geometric approach has two important limitations when applied to EFTs: (i) the scalar manifold geometry can only capture interaction terms with exactly 2 derivatives, and (ii) there is no geometric understanding of invariance under field redefinitions with derivatives, which, from the field theory perspective, are entirely reasonable `symmetries' of the path integral.
Point (i) is problematic because higher-derivative operators obviously play an important role in EFTs, and carry information that is in principle independent of that encoded in the 2-derivative terms. 
Point (ii) implies that geometric quantities are in fact basis-dependent. For instance, upon applying derivative field redefinitions, the curvature on $M$ changes, in general.
First steps towards obviating these issues were taken recently in~\cite{Cohen:2022uuw,Craig:2023wni}.

Of particular relevance to limitation (i), and to the present paper, Ref.~\cite{Craig:2023wni} recently introduced the idea of `Lagrange spaces' as a generalisation of the field space geometry, wherein directional derivatives of the field space coordinates are treated as independent coordinates. Lagrangian functions can be built with these coordinates, which encode a subset of operators with more than 2 derivatives in a covariant formalism -- in particular, because different spacetime derivatives $\partial_\mu$ cannot be distinguished, only operators with specific `flavour' symmetries can be realised using these derivative coordinates. The present paper is in a similar spirit, in that we also seek a geometric formulation of scalar field theories that extends to higher-derivatives. 
We approach the problem in a largely orthogonal direction, using, instead, pure geometry on `jet bundles' to formulate  scalar EFT Lagrangians.

\bigskip

In differential geometry, {\em jet bundles} provide a coordinate-free way to describe higher-derivatives of maps between manifolds; in the context of our EFTs, jet bundles are spaces on which derivatives of the scalar field are treated as extra coordinates in a consistent fashion. Roughly speaking, given a spacetime $\Sigma$, a field space $M$, and an integer $r$, one can construct an `$r$-jet bundle' which is itself a manifold, with local coordinates corresponding to spacetime coordinates $x^\mu$, field space coordinates $u^i$, and `$r$-derivative coordinates' $u^i_\mu$, $u^i_{\mu_1 \mu_2}$, ..., $u^i_{\mu_1 \dots \mu_r}$. This set of manifolds offers a promising arena for extending the geometric approach to EFTs to Lagrangians with more than 2 derivatives -- suggesting a route for overcoming limitation (i) described above, which we aim to explore in this paper.

\begin{figure}
    \centering
\includegraphics[width=0.9\textwidth]{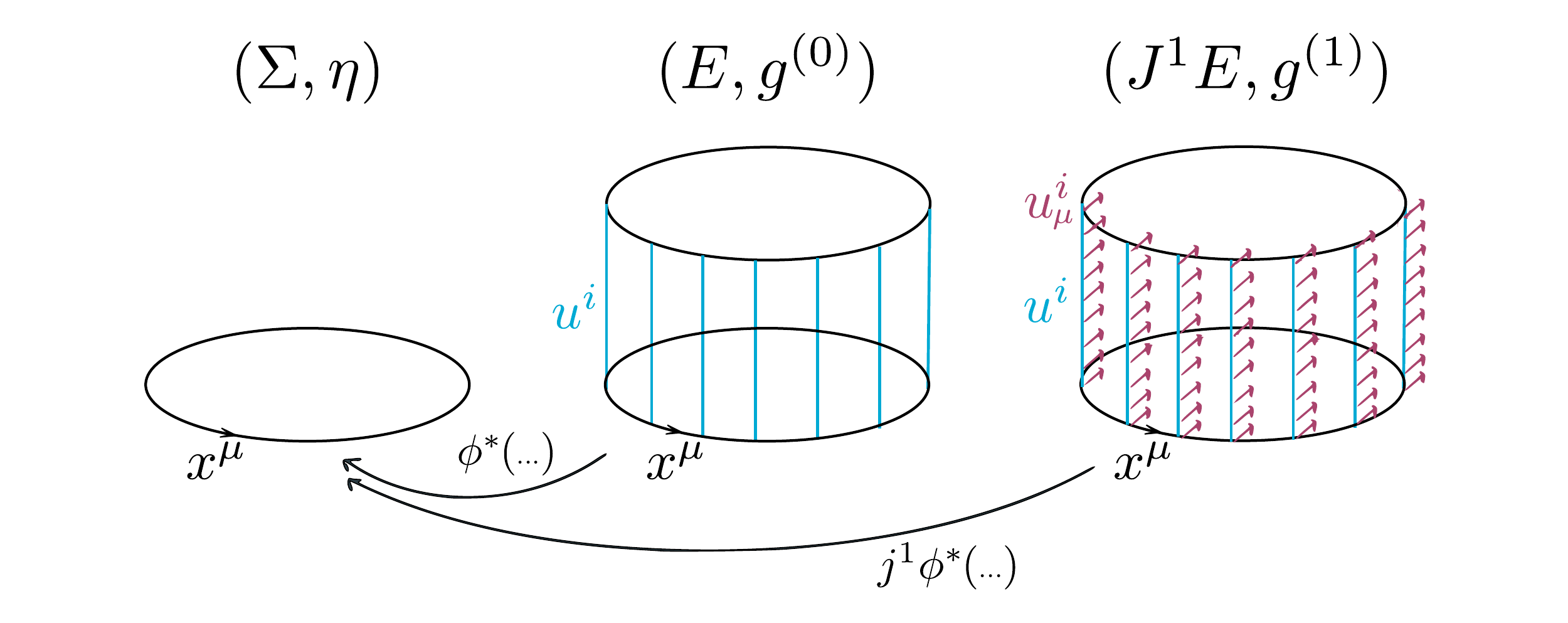}
    \caption{A schematic illustration of three key players in this paper: a spacetime manifold $(\Sigma, \eta)$ with local coordinates $x^\mu$ on a patch thereof; a `field space bundle' $E$ whose base space is $\Sigma$ and fibre is $M$, which has fibre coordinates $u^i$; and lastly the 1-jet manifold $J^1 E$ of the bundle $E$, which is itself a fibre bundle over both $E$ and over $\Sigma$. A `scalar field' configuration is a section $\phi$ of the bundle $E$, and the composition $u^i \circ \phi = \phi^i(x)$ returns its `field value', given the choice of local fibre coordinates $u^i$, at $x$. From $\phi$ one can construct a section $j^1 \phi$ of the bundle $j^1 E \to \Sigma$ by `prolongation'.
    The 1-jet bundle also admits a local fibred coordinate system $(x^\mu, u^i, u^i_\mu)$, where the extra $\{u_\mu^i\}$ are `derivative coordinates that, when evaluated on a section $j^1\phi$, agree with the first derivatives $\partial_\mu \phi(x)$. Finally, metrics $g^{(0)}$ and $g^{(1)}$ can be defined on $E$ and $J^1 E$ such that, upon pulling back to $\Sigma$ along $\phi$ or $j^1 \phi$ respectively, one can obtain the most general EFT Lagrangian for that scalar field theory with up to 2- or 4-derivative operators (and any number of fields).
    }
    \label{fig:drum}
\end{figure}

To that end, we formulate scalar field theory Lagrangians using geometry on the 1-jet bundle. The geometries of the various manifolds involved in our discussion, and maps between them, are illustrated in Figure~\ref{fig:drum}. Our strategy is to write down the most general metric on the 1-jet bundle consistent with the symmetries of the EFT, which are themselves naturally extended to (or, more precisely, `prolongated' to) the jet bundle. An invariant EFT Lagrangian function is then formed by pulling back this metric all the way to spacetime $\Sigma$ along a section of the jet bundle (which is itself obtained by prolonging a scalar field configuration $\phi$), and contracting with the inverse spacetime metric. One immediately obtains terms with more than 2 derivatives from this recipe; for example, if we pull back a metric component $\delta_{ij}\eta^{\mu \nu} du^i_\mu \, du^j_\nu$ along a prolongation of $\phi$, and then contract with $\eta^{\mu\nu}$, we obtain the 4-derivative term $(\partial_\mu \partial_\nu \phi)^2$. We show that, in the case of the SMEFT and HEFT, one obtains a {\em complete basis} of non-redundant EFT operators with 0-, 2-, and 4-derivatives in this way, starting from a 1-jet bundle geometry that is invariant under Poincar\'e and $O(4)$ custodial symmetry.

The conclusion, that one can construct a complete EFT basis with up to 4-derivatives from a jet bundle metric, remains unchanged upon allowing a breaking of custodial symmetry (\S \ref{sec:beyond-custard}), as occurs in the real-world electroweak theory. In fact, it generalises to scalar Lagrangians with arbitrary `flavour' structure, relying in no way on the assumption of an internal symmetry.
Nor is this an accident that occurs only at the 1-jet bundle order. We prove in Appendix~\ref{app:proof} that, following a similar recipe, the most general, Poincar\'e invariant $r$-jet bundle metric can be used to construct an EFT Lagrangian that contains a complete basis of operators with up to $2(r+1)$ derivatives and arbitrary number of field insertions. 
In this sense, we suggest that jet bundles offer a natural way of constructing generic scalar EFTs (barring only topological terms) using only geometry -- albeit on an increasingly high-dimensional space if we want to cover an increasingly large number of derivatives.

As well as extending the geometric formulation of scalar field theories to higher-derivatives, we try to develop an understanding of field redefinitions in this picture that includes derivative field redefinitions. An important realisation is that the scalar field $\phi(x)$ is not really identified with coordinates on the  field space; rather, as the notation suggests, $\phi(x)$ is the {\em map} from spacetime $\Sigma$ to field space $M$ (or, better still, $\phi$ is a local section of a bundle $\pi:E \to \Sigma$ with fibre $M$, which we take as our starting point in this work). The image of this map (or section) is of course evaluated in the coordinate chart, for each $x$, to return the `field value' at $x$. A field redefinition is then understood to be a change of map, or better a change of section, rather than just a change of coordinates (the latter leaves any metric $g$ trivially invariant, simply by virtue of $g$ being a tensor). Given  that the maps (sections) `know about' the spacetime source manifold $\Sigma$, it is not surprising that they can be differentiated with respect to $x$, whereas a coordinate on the target space cannot. Therefore a change of section can depend on derivatives of the original section, offering a more general understanding of field redefinitions -- and one that is, in essence, geometric.

In  certain cases, in particular those a physicist would call `non-derivative field redefinitions', the change of section can be equivalently described by doing a morphism on the bundle (or field space). All these notions lift to the 1-jet bundle formalism in a natural and consistent way. Lastly, derivative field redefinitions can, in a very limited range of examples, be replicated by (jet) bundle morphisms, but these are essentially accidents; indeed, there are mathematical arguments (see \S \ref{sec:lemma}) that suggest `derivative changes of section' cannot be induced by a (jet) bundle morphism. We try to elucidate these various points throughout this paper.

We remark that jet bundles have been recently used in the study of scalar field theories in Ref.~\cite{Gripaios:2021hsc} by Gripaios and Tooby-Smith, but in an altogether different context. There, the authors use jet manifolds to rigorously formulate so-called `inverse Higgs constraints'. The use of jet bundles is natural (and fruitful) in that context, because the inverse Higgs constraint equations relate fields to their derivatives, and so can be embedded as submanifolds of the 1-jet bundle that are invariant under the action of the symmetry. 
While we were completing this paper, we became aware of the forthcoming work~\cite{Craig:2023hhp} by Craig and Lee, that builds on~\cite{Craig:2023wni} and also utilizes the jet bundle formalism in the context of the Higgs EFTs.

\subsection*{Summary of results}

In this work we elucidate a geometric description of EFTs with up to four derivatives, in terms of geometry on the 1-jet bundle associated to our original field space maps. In doing so, here are some things we achieve (the various concepts and terminology we use will be duly explained in the main text):
\begin{enumerate}
\item We construct a map from the space of 1-jet bundle metrics $g$ to 4-derivative Lagrangians $\cL$, in a natural way. Namely, given a section $\phi$ of the field space bundle, {\em a.k.a.} a scalar field configuration, one pulls back the jet bundle metric $g$ to spacetime along the associated section $j^1 \phi$ of the jet bundle, and then contracts with the inverse metric $\eta^{-1}$ on spacetime to get an ordinary function or `Lagrangian': 
\begin{align}
\cL[g]
=\frac{1}{2} \langle \eta^{-1}, \, (j^1 \phi)^\ast g \rangle 
\end{align}
This map does not inject to the space of physically inequivalent Lagrangians, as one might expect given the target space does not know about {\em e.g.} integration by parts.
So, lastly (and somewhat na\"ively in our formalism), one can implement integration by parts (IBP) identities to reduce the Lagrangian to a particular basis.
\item We prove that this map surjects,
in the sense that every Lagrangian with up to 4-derivatives can be obtained by pulling back a metric on the 1-jet bundle. We emphasize that all terms with 0-, 2-, or 4-derivatives are obtained in this way, so that even the potential is captured `geometrically'.
We show this result explicitly in a number of toy examples, and we also prove that it holds at any order, for a map from the space of $r$-jet bundle metrics into the space of scalar Lagrangians with up to $2(r+1)$ derivatives.
\item Along the way, we discuss how the symmetries of, say, the Higgs EFTs, can be implemented in this (jet) bundle formalism. The internal $O(4)$ symmetry and the spacetime $SO(1,3)$ Poincar\'e symmetry are described by different kinds of bundle morphisms in this picture (depending on whether they move points in the base space or not). Both kinds of symmetry naturally carry over to the 1-jet bundle, and we enforce these symmetries on our metrics (hence Lagrangians).
\item We give a geometric meaning to the notion of general (including derivative) field redefinitions in terms of a (suitably smooth) change of section, which induces a change of section of the 1-jet bundle after prolongation.
\item In doing so, we also hope to clarify the geometric meaning of {\em non}-derivative field redefinitions. These are a special class of change of section that can be equivalently described by doing a bundle morphism (under which the metric changes), before pulling back to spacetime along the {\em same} section. Such an equivalent description is not available for changes of section that involve derivative maps.
\item 
Thanks to the components of the metric being smooth functions on the jet bundle, we are able to expand the metric around a point, which after specifying a section allows us to directly extract {\em $n$-point amplitudes}, here done for $n=2,\,3,\, 4$, in terms of products of momenta and the components of the metric evaluated at that point.
\end{enumerate}
Here are some things that we do {\em not} do in this paper and that are left for future work:
\begin{enumerate}
\item We do not describe how to implement gauging of a subgroup of the global symmetry in this jet bundle formulation.
\item We do not show how to couple this jet bundle geometry to fermions. For 2-derivative Lagrangian terms with two fermion insertions, this issue has been tackled recently in Refs.~\cite{Finn:2020nvn,Assi:2023zid,Gattus:2023gep}.
\item We do not go into any details concerning phenomenology. 
\end{enumerate}

The paper is structured as follows: in \S\ref{sec:preliminaries} we review and refine the formulation of 2-derivative EFT Lagrangians using field space geometry. In particular, we hope to clarify a little the notion of (non-derivative) field redefinitions.
In \S\ref{sec:fib-bundles} we recast this formalism using a field space bundle, which is more general (and motivated by locality). We again revisit the issue of field redefinitions, now equipped with the notion of a bundle morphism.
Jet bundles are introduced in \S\ref{sec:jets} and their relation to scalar Lagrangians is discussed in great detail in \S\ref{sec:geometry}. Section \ref{sec:examples} contains a few toy examples that will be useful to see how 1-jet bundle geometry captures operator bases including up to 4 derivatives; these examples also invite us to take a first look at the redundancies present in the description. In \S\ref{sec:HEFTandSMEFT} we turn to the SMEFT/HEFT case of four scalars respecting a custodial $O(4)$ symmetry. Finally, in \S\ref{sec:amplitudes} we comment on the interplay with scattering amplitudes, before concluding in \S\ref{sec:conclusions}.
 
\section{Preliminaries} \label{sec:preliminaries}

Our goal is to eventually describe the higher-derivative expansion of the Higgs EFTs, namely `HEFT' or `SMEFT', using geometry. We begin by reviewing the basic idea for geometrically formulating the Higgs EFTs.

The Higgs sector of the electroweak theory, by which we refer to the part of the Lagrangian involving {\em only} scalar fields, is described by a 4d sigma model with a target space $(M,g)$, which is a smooth manifold of real dimension 4, with Riemannian metric $g$. This means that, given also a 4d spacetime manifold with a (pseudo-)Riemannian metric, $(\Sigma,\eta)$, the degrees of freedom of the QFT are smooth ($C^\infty$) maps 
\be \label{eq:map}
\phi(x): \Sigma \to M\,,
\ee
with dynamics described by an action that is a particular functional $S[\phi]$. On a patch (open set) $\mathcal{F}_M$ of $M$ one can provide local coordinates $u^i :\mathcal{F}_M \to \R^4$, where $i=1,\dots, 4$, and on a patch $\mathcal{U}_\Sigma$ of $\Sigma$ one can provide local coordinates $x^\mu$. 
In this paper, we always take $\Sigma$ to be a fixed non-dynamical background manifold, with a fixed metric -- usually flat Minkowski space $\R^{3,1}$, with $\eta = \text{diag}(1,-1,-1,-1)$.\footnote{Extensions of scalar field theories of the kind we consider to include dynamical gravity, so-called `scalar-tensor theories', can also be described using a covariant geometric formalism~\cite{Finn:2019aip}. }

Of course this geometric picture is not reserved for the particular 4d theories that describe the Higgs, but equally describes scalar fields in any spacetime dimension $d$,\footnote{The source manifold $\Sigma$ need not be identified with spacetime, either; more generally it is the `worldvolume' of the theory. The most familiar example in which $\Sigma$ is not identified with spacetime occurs in string theory, where one usually identifies $\Sigma$ with the 2d string worldsheet, and the target space $M$ with the physical spacetime in which the string moves. A similar description can be used for any extended object.} 
with any target space geometry. In \S \ref{sec:toy}, for example, we consider a quantum mechanical example {\em i.e.} a scalar field theory in $0+1$ dimensions (related examples have been recently discussed in~\cite{Gripaios:2015pfa,Davighi:2019ffp}). An important class of examples in physics is those (pseudo-)scalars arising from spontaneous symmetry breaking $G \to H$, for $d>2$~\cite{Coleman:1973ci}. In this case the scalar fields, here pseudo Nambu Goldstone bosons (or pNGBs), are known to parametrize the coset space $M \cong G/H$. In addition to the chiral Lagrangian~\cite{Weinberg:1966fm} that describes QCD pions, arguably a pillar of theoretical particle physics, there exist physically important examples of such pNGB Lagrangians in condensed matter physics, {\em e.g.} describing fluids~\cite{Gripaios:2014yha,Dersy:2022kjd} and more exotic phases~\cite{Nicolis:2013lma,Nicolis:2015sra}, and even cosmology~\cite{Nicolis:2008in,Piazza:2017bsd}.

\paragraph{Example: Standard Model (SM).}
For the special case of the SM Higgs sector, we can take $M \cong \R^4$ with globally defined Cartesian coordinates $u^i$, and a flat metric.

\subsection*{Symmetries of SMEFT and HEFT} 

An essential bit of data required to define these Higgs EFTs is an {\em internal symmetry}, by which we mean (for now) that the target space {\em $M$ is equipped with the smooth action $\sigma$ by a Lie group $G$.} This induces a group action on the fields $\phi(x)$ which are (for now) maps into $M$, under which we require that the field theory action functional $S[\phi]$, or more precisely the phase $e^{iS[\phi]}$), be invariant. It is in this Lie group action where the difference between HEFT and SMEFT lies.
For both cases the group is the same, namely $G=O(4)$ in the custodial limit\footnote{Actually, in neither SMEFT nor HEFT do we know the global structure of the custodial symmetry group $G$, only that its Lie algebra is $\mathfrak{so(4)}$. We take $SO(4)$ to be enlarged by a parity symmetry {\em e.g.} $u^4 \to -u^4$ to an $O(4) \cong SO(4) \rtimes \Z/2$ symmetry; in other literature, the symmetry is taken to be the universal cover $SU(2)_L \times SU(2)_R \cong \Spin(4)$, which is a non-isomorphic central extension of $SO(4)$ also by $\Z/2$. This global ambiguity is unsurprising given that even the electroweak {\em gauge} symmetry is ambiguous, being either $SU(2)\times U(1)$ or $U(2)$ (see {\em e.g.}~\cite{Tong:2017oea,Davighi:2019rcd}).}
(which we assume throughout most of this paper),
but its action $\sigma$ on $M$ is different. 

This distinction is easy to express given a particular set of local coordinates on a patch, but is harder to formulate in a coordinate-free (or indeed field-redefinition-invariant) way. One of the successes of the geometric formulation of Higgs EFTs has been to provide coordinate-free criteria for detecting an $O(4)$ action that is decisively non-SMEFT~\cite{Falkowski:2019tft,Cohen:2020xca}.
For now let us use particular local coordinates to get started.

In the case of SMEFT, and for a particular set of local coordinates $u^i=h^i$ on some open set $\mathcal{F}_M\subset M$, the group action $\sS$ is simply
\begin{align} \label{eq:sS_action}
\sS: O(4) \times M \to M : \,\,
(O,h^i) \mapsto O h^i\, , 
\end{align}
where $O \in O(4)$ acts on 4-component real vectors by matrix multiplication.\footnote{We are being a little cavalier in expressing the group action, which is well-defined on all of $M$, only by its action on points in the patch $\mathcal{F}_M$ that is coordinatized by $\{h^i\}$. 
In principle, one should provide an atlas of coordinate charts covering $M$ and define the group action everywhere. Also note that the fixed patch $\mathcal{U}$ will itself be moved by the group action, and so there can be points `either side of the edge' of the patch for which Eq. (\ref{eq:sS_action}) cannot be used. We invite the reader to look past such subtleties here. } 
Evidently, this means that electroweak symmetry $SU(2)_L \times U(1)_Y \subset G$ acts {\em linearly}.
A consequence of this linear group action is that the patch $\mathcal{F}_M$ always contains a {\em fixed point} of the group action, namely the origin $o$ ($h^i=\vec{0}$), meaning that $\sS:(O,o) \mapsto o$ $\forall O\in O(4)$.  This is the electroweak symmetry preserving point.
As shown in~\cite{Alonso:2016oah}, the converse is also true, namely if $M$ contains an $O(4)$ fixed point then, in the vicinity of that point, there is always a coordinate chart $u^i = h^i$ in which $O(4)$ acts linearly on $h^i$.
Moving away from this fixed point, any other point on $\mathcal{F}_M$, written in the local coordinates $h^i$, preserves an $O(3) \subset O(4)$ subgroup. 
The vacuum of our Universe is such a symmetry breaking point $h^i \neq \vec{0}$, with length squared $\delta_{ij} h^i h^j = v^2$.

For HEFT, the symmetry breaking is made manifest by the choice of group action. 
Let us write our local coordinates in the form $u^i = (h, \pi^a/v)$, where $a=1, \dots, 3$.
HEFT is characterised by a {\em non-linear} group action $\sH$ in these local coordinates
\begin{align}
\sH: O(4) \times M \to M:\,\, 
(O, h, \pi^a/v) \mapsto \left(h, O \left(\pi^a/v, \sqrt{1-\pi\cdot\pi/v^2}\right)^T \right)\, .
\end{align}
In words, the radial mode $h$ is invariant under the $O(4)$ group action, while the Goldstone modes $\pi^a$ transform non-linearly, via the usual action of $O(4)$ on a unit vector spanning $S^3 \cong O(4)/O(3)$, the coset space associated with the electroweak symmetry breaking.

Spacetime $(\Sigma, \eta)$ is also equipped with a symmetry, which is a smooth action $\rho$ by the Poincar\'e group $\cong \R^{1,3} \rtimes O(1,3)$. For our coordinates $x^\mu$ on Minkowski spacetime $\R^{1,3}$, the (non-linear) group action is
\begin{equation}
    \rho: \text{Poinc} \times \Sigma \to \Sigma \,\, : \,\,
    (a^\mu, L^\mu_\nu,  x^\mu) \mapsto L^\mu_\nu x^\nu + a^\mu
\end{equation}
Throughout this paper we only consider Poincar\'e-invariant theories (including, but not only, SMEFT and HEFT). 

\subsection{Two-derivative Lagrangians from field space geometry} \label{sec:2-deriv}

The dynamics of such scalar field theories lend themselves to a geometric description. This means that certain terms in the Lagrangian can be built using geometric structures on the target space, which we have already stipulated is a Riemannian manifold (and sometimes, more basically, a topology alone is sufficient).

\medskip

The most important and natural geometric construction is of the kinetic term or, more generally, the Lagrangian terms featuring exactly two derivatives. This requires a geometry on $M$, in the form of the metric $g$, which recall is a symmetric and everywhere non-degenerate $(0,2)$-tensor on $M$. The two-derivative Lagrangian terms are then
\be \label{eq:2deriv_compact}
\cL_{2\partial}[\phi, g] = \frac{1}{2}\, \langle\eta^{-1} , \, \phi^\ast (g) \rangle \, .
\ee
That is, we pull back the metric $g$ along the map $\phi$,\footnote{We use the notation $f^\ast(\cdot)$, standard in differential geometry, to denote the pullback of an object along the map $f$. Likewise, pushforwards are denoted $f_\ast(\cdot)$. In order to avoid confusion, any notion of `complex conjugation' {\em e.g.} of a scalar Higgs field, will be denoted with a dagger, {\em viz.} $\phi^\dagger$.
}
to obtain a $(0,2)$ tensor on spacetime, which we contract with the inverse spacetime metric to get an ordinary function. Note that this formula for the Lagrangian is `coordinate-free'; the map $\phi:\Sigma \to M$ is defined irrespective of our choices of coordinate chart, and both $\eta^{-1}$ and $g$ are tensors. The Lagrangian will be moreover invariant under the symmetry of our theory if the metric $g$ is $\sigma$-invariant,\footnote{Note that the stipulation of `invariance' refers to the specific group {\em action}, which recall differs for HEFT {\em vs} SMEFT, and which we emphasize by writing `$\sigma$-invariant' rather than just `$G$-invariant'.} 
assuming also that $\eta$ is Poincar\'e invariant.
Our notation $\cL_{2\partial}[\phi, g]$ emphasizes that the Lagrangian is a functional of the map (field) $\phi$, given $g$. To obtain the action $S[\phi, g]$, we integrate $\cL_{2\partial}[\phi, g]$ over $\Sigma$ using the canonical volume form $\omega_\Sigma := \sqrt{|\eta|}d^4 x$, as we would for any Lagrangian `term' that is a function (rather than a form).

The compact, coordinate-free formula (\ref{eq:2deriv_compact}) can be written in a notation more familiar to physicists by using our local coordinates $x^\mu$ (on $\mathcal{U}_\Sigma \subset \Sigma$) and $u^i$ (on $\mathcal{F}_M \subset M$). In these local coordinates, the metrics are 
\begin{align}
\eta &=  \eta_{\mu\nu}\, dx^\mu \otimes dx^\nu\, , \qquad \text{(flat)}, \label{eq:eta} \\
g &=g_{ij}(u^i)\, du^i \otimes du^j \,, \label{eq:g} 
\end{align}
and then Lagrangian (\ref{eq:2deriv_compact}) is 
\be \label{eq:2deriv_long}
\cL_{2\partial}[\phi, g] = \frac{1}{2}\, \eta^{\mu\nu}\, g_{ij}\left(\phi^i(x)\right) \, \partial_\mu \phi^i(x) \partial_\nu \phi^j(x)\, ,
\ee
where 
\be
\phi^i(x) := (u^i \circ \phi)(x)\,.
\ee
In words, the quantity $\phi^i(x)$ returns the values of the local coordinates $u^i$ evaluated at the image of spacetime point $x$ under the map $\phi$ -- that is, the `field value' at $x$.

\paragraph{Example (ctd.): SM.}
Take $M=\R^4$ and $u^i$ to be global Cartesian coordinates thereon. The SM construction takes only the subset of Lagrangian terms that are renormalisable. With this extra condition, symmetry dictates that $g = \delta_{ij} du^i du^j$ is just the flat metric on $\R^4$, as anticipated above. This pulls back under $\phi(x)$ to give the usual Higgs kinetic term,
$$\cL_{\text{kin}} = \frac{1}{2} \eta^{\mu\nu} \delta_{ij} \partial_\mu \phi^i(x) \partial_\nu \phi^j(x)\, ,$$
upon evaluating (\ref{eq:2deriv_compact}).

\paragraph{Example: SMEFT.}
In the case of SMEFT the most general $\sS$-invariant metric can be expressed in terms of two arbitrary functions $A(z)$ and $B(z)$, via~\cite{Alonso:2015fsp,Alonso:2016oah} 
$$g_{ij}(u^i) = A\left(\frac{u \cdot u}{\Lambda^2} \right)\, \delta_{ij} + B\left(\frac{u \cdot u}{\Lambda^2} \right)\, \delta_{ik} \delta_{jl} \frac{u^k u^l}{\Lambda^2} \, ,$$ 
in our chosen SMEFT coordinate chart, where $v \cdot w := \delta_{ij} v^i w^j$. 
We have introduced appropriate factors of the formal power counting parameter $\Lambda$, which will be identified with the EFT cut-off scale in a physical theory. Here, $A(0) = 1$ and $B(0)=0$ in order to recover the SM in the limit $\Lambda \to \infty$.
Substituting this metric into (\ref{eq:2deriv_long}), or equivalently (\ref{eq:2deriv_compact}), yields the
most general set of $\sS$-invariant SMEFT operators with exactly two derivatives.

\subsection{Other terms in the scalar EFT} \label{sec:other_terms}

We emphasize that this procedure constructs only a subset of the non-renormalisable EFT Lagrangian, namely all operators with exactly two derivatives (but any number of field insertions), from the metric $g$ on field space $M$.
Other terms in the EFT are, in this picture, included `by hand', and can be considered as extra structures on $M$ beyond the metric. These other interactions fall into three classes:
\begin{enumerate}
\item The `potential' is the part of the Lagrangian with {\em zero derivatives}; it is the pullback of a $\sigma$-invariant function 
$V(\phi):M \to \R$ from $M$ to $\Sigma$, which is again integrated on $\Sigma$ using $\omega_\Sigma$ to obtain the action contribution; that is, $S[\phi] \supset \int_\Sigma \phi^\ast(V) \omega_\Sigma$.
\item There are all the local operators appearing in the EFT construction of Callan, Coleman, Wess, and Zumino~\cite{Coleman:1969sm,Callan:1969sn} with {\em more than two derivatives.} Capturing these operators systematically using geometry is a primary purpose of the present paper.
\item The action might also admit topological terms like the Wess--Zumino--Witten term~\cite{Wess:1971yu,Witten:1983tw}, that are the pullback of (possibly locally-defined) differential forms from $M$ to $\Sigma$, then integrated directly on $\Sigma$. These require only a topology on each space. Invariance under $\sigma$ is rather more subtle here than for local operators~\cite{Davighi:2018inx}, especially when the map $\phi$ is topologically non-trivial; a large class of such terms are given by $\sigma$-invariant differential cocycles~\cite{Davighi:2020vcm} on $M$, of degree $d+1$ where $d=\text{dim}(\Sigma)$ in general. Examples of such terms abound in Composite Higgs models (which are SMEFT)~\cite{Gripaios:2009pe,Davighi:2018xwn}, and {\em e.g.} in the condensed matter examples mentioned previously~\cite{Dubovsky:2011sk,Delacretaz:2014jka}.
\end{enumerate}
We largely neglect terms of type (3) in the rest of this paper, focussing rather on the non-topological ({\em i.e.} metric-dependent) sector of the EFT. We briefly comment on topological terms in the Higgs EFTs in \S\ref{sec:top-terms}.

\subsection{What is a (non-derivative) field redefinition?} \label{sec:field_redef_1}

One central feature of the geometric formulation of Higgs EFTs has been the identification of (some subset of) field redefinitions, which are symmetries of the partition function and thence of observables (see {\em e.g.}~\cite{Criado:2018sdb}), with coordinate transformations on field space $M$. 
Specifically, `non-derivative field redefinitions' of the form $\phi(x) \mapsto \psi(\phi(x))$ have been identified with changes of coordinates. It is not clear how to think about derivative field redefinitions, whereby  $\phi\mapsto\psi(\phi, \partial \phi, \dots)$, in this geometric picture, since the target space $M$ knows nothing about spacetime derivatives. 
At this stage, we want to revisit this established lore concerning non-derivative field redefinitions. Our discussion will be fairly pedantic, but in doing so will open the way to a simple geometric picture for more general smooth field redefinitions, even before we consider the jet bundle extension of field space in later Sections.

To begin, we emphasize that we have been careful, so far, to distinguish local coordinates $u^i$ on $\mathcal{F}_M$ from the {\em field variable} $\phi$, which is a map from $\Sigma$ into $M$. This is again distinguished from $\phi^i = u^i \circ \phi$, which evaluates the local coordinate values of the field for each spacetime point.
A `change of coordinates', {\em a.k.a.} a `coordinate transformation' or a `change of chart', is, when restricted to $\mathcal{F}_M$, a map $u^i \to u^{i\prime}(u^i)$.
This is not obviously related to a redefinition of the {\em field}, which we would more readily identify with some transformation on the map $\phi$ (inducing a transformation on the $\phi^i$). In this Subsection we contrast how these two kinds of transformation, namely a change in coordinates $u^i \to u^{i\prime}$ {\em vs.} a change in map $\phi \to \psi$, affect the Lagrangian that we construct from the metric on $M$ via (\ref{eq:2deriv_compact}). The main concepts presented in this Subsection are illustrated in Figure~\ref{fig:nonD_redefinitions}.

\subsubsection{Change of coordinates}
\label{subsec: change of coordinates}
The first key point is simply that the field space metric $g = g_{ij}(u^i) du^i \otimes du^j$ {\em does not change at all} under a change of coordinates, when evaluated at the same point $m$ in $M$. 
The components change via appropriate factors of the Jacobian matrix, {\em viz.} 
\be
g_{ij} =\frac{\partial u^{k\prime}}{\partial u^i} \frac{\partial u^{l\prime}}{\partial u^j}g_{kl}^\prime\,,
\ee 
which is compensated by the variation of the 1-forms 
\be
du^{\prime i} = \frac{\partial u^{\prime i}}{\partial u^k} du^k\,.
\ee
This is just the statement that the metric is a tensor, and so $g(m)$ evaluated at any point $m \in M$ is independent of the choice of the coordinate chart $\{u^i\}$. Once we pullback the metric along the map $\phi(x)$ and form the Lagrangian $\cL = (1/2) \langle\eta^{-1},\, \phi^\ast(g)\rangle$, one obtains exactly the same function of the field variable $\phi(x)$, regardless of having done the change of coordinates.

\begin{figure}[t]\centering
\includegraphics[height=3.53cm]{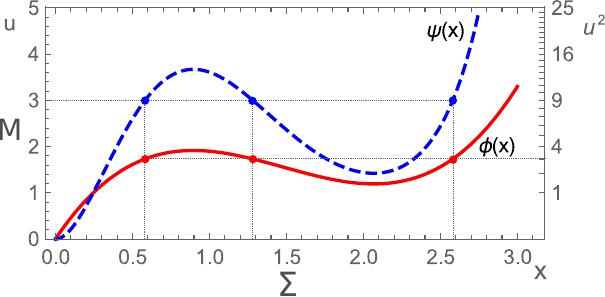}
\caption{Illustration of the notion of non-derivative field redefinitions in a toy example describing a single real scalar (take $M=\R$, with coordinate $u$) in 1d spacetime (take $\Sigma = \R$, with coordinate $x$). A field configuration is a map from $\Sigma$ to $M$; here $\phi$ (red-solid) and $\psi$ (blue-dashed) denote two such maps. A change of field space coordinates, such as $u \mapsto u'=u^2$, simply corresponds to reading off coordinate values using the right axis rather than the left axis. It does not move points in $M$, nor change the metric $g$ (or any other tensor) evaluated at any point $m \in M$; if we pull back $g$ to $\Sigma$ using a fixed map $\phi$ to form a Lagrangian, we obtain the same result regardless of whether we used the $u$ or $u'$ coordinate chart on $M$. A (non-derivative) field redefinition is rather a particular change of maps such that $u \circ \psi$ can be expressed as a function of $u \circ \phi$. In the figure, $u\circ\psi=(u\circ\phi)^2$, such that evaluating $\psi(x)$ is equivalent to first evaluating $\phi(x)$, then doing a diffeomorphism $f$ on $M$ that sends a point in $M$ with coordinate $u$ to a different point with coordinate $u^2$, expressed in a fixed coordinate chart.
Reading off the value of the curve $\psi(x)$ on the left is equivalent to reading off the value of $\phi(x)$ on the right.
The figure also illustrates that if $\phi$ sends multiple points on $\Sigma$ to the same point on $M$, the change of map can be equivalent to a diffeomorphism on $M$ only if $\psi$ preserves this feature. 
}\label{fig:nonD_redefinitions}
\end{figure}

If this argument sounds a little thin, we can make the point totally explicit with an example. Consider a 1-dimensional field space manifold $\R$, on which we define a local coordinate $\{u\}$. In this coordinate chart, consider a (flat) metric $g = du\otimes du$; in components, $g_{uu} = 1$. Choose a map $\phi:\Sigma \to \R$, namely a scalar field configuration. The Lagrangian formed as in Eq. (\ref{eq:2deriv_compact}) is then $\cL = \frac{1}{2} \partial_\mu \phi(x) \partial^\mu \phi(x)$, the kinetic energy for a real scalar.\footnote{We slightly abuse notation here and let $\phi(x)$ denote both the map itself and the coordinate value $u \circ \phi$; there is no component index because the target space is 1-dimensional.} Now, let us do a change of coordinates in the target space, and send 
\be
u \mapsto u^\prime = u^2\, .
\ee 
In the illustration of Fig.~\ref{fig:nonD_redefinitions}, this corresponds to reading off coordinates using the scale on the right side of the plot, rather than the left, for the \emph{same} points on $M$.
The metric expressed in the new coordinate system, but at the same point, is 
\be
g^\prime = \frac{1}{4u^\prime} du^\prime du^\prime (=g) \, ,
\ee 
and so its components are $g_{u' u'} = 1/(4u')$.

To form the `new' Lagrangian after doing the coordinate transformation, we then pullback the metric, in the new coordinates, along the same map $\phi(x)$. Note that the map $\phi(x)$ is defined independently of coordinates -- it takes points in $\Sigma$ to points in $M$ -- but that we evaluate its image in the given coordinate chart. We find the Lagrangian
\begin{align}
\cL[\phi, g] \mapsto \cL[\phi, g'] &= \frac{1}{2} \langle \eta^{-1} ,\, \phi^\ast(g^\prime)\rangle \nonumber \\
&= \frac{1}{2} \eta^{\rho\sigma} \left\langle{\partial_\rho} \otimes{\partial_\sigma},\, \frac{1}{4(u'\circ \phi)} \partial_\mu (u'\circ \phi) \partial_\nu (u'\circ \phi) \, dx^\mu \otimes dx^\nu \right\rangle  \nonumber \\
 &= \frac{1}{2} \eta^{\rho\sigma} \left\langle \partial_\rho \otimes \partial_\sigma ,\, dx^\mu \otimes dx^\nu \right\rangle \,\, \frac{1}{4\phi^2} \partial_\mu (\phi^2) \partial_\nu (\phi^2) \nonumber \\
 &= \frac{1}{2} \eta^{\mu\nu}\, \frac{1}{4\phi^2}\, (2\phi \partial_\mu \phi) (2\phi \partial_\nu \phi)  \nonumber \\
 &= \frac{1}{2} \partial_\mu \phi \partial^\mu \phi = \cL[\phi, g] \, ,  \label{eq:change-of-coords}
\end{align}
the same as before. The tensorial nature of the metric means that, in general, the Lagrangian (\ref{eq:2deriv_compact}) is not changed (as a functional of its argument) by any change of coordinates on $M$. The freedom to change coordinate charts is akin to a gauge redundancy in this geometric formulation of the theory.

\subsubsection{Change of maps} \label{sec:change-maps}

Let us instead consider what happens to the Lagrangian (\ref{eq:2deriv_compact}) when we change field space maps, from $\phi(x)$ to a different map $\psi(x):\Sigma \to M$. We assume the change in map $\phi \to \psi$ is smooth. Recall that $\phi^i (x)= (u^i \circ \phi)(x)$ denotes the local coordinates of the image of point $x$ under the original map $\phi$. A new map $\psi$ sends each $x$ to a different point $\psi(x) \in M$, for which $\psi^i (x)= (u^i \circ \psi)(x)$ are its local coordinates. The two different maps are indicated in Fig.~\ref{fig:nonD_redefinitions} by the red-solid and blue-dashed curves.

We again use the simple example of a real scalar field for illustration, with field space $M=\R$, local coordinate $u$, and flat metric $g=du\otimes du$. 
Consider the particular change of map, defined in terms of coordinate values (in fixed chart $\{u\}$) of field space points, 
\be \label{eq:eg_map_change}
\phi(x) \mapsto \psi(x) \quad | \quad u \circ \psi = (u \circ \phi)^2\, .
\ee
(If we continue to abuse notation as above, we would write this simply as $\psi=\phi^2$, denoting both the map and the coordinate value by the same symbol $\phi$.)
Now, when we pull back the metric $g$ along this new map $\psi$, the Lagrangian (\ref{eq:2deriv_compact}) is not necessarily the same (even though the metric $g$ is unchanged) because $\phi^\ast(g) \neq \psi^\ast(g)$. We obtain
\begin{align}
\cL[\phi, g] \mapsto \cL[\psi, g] &= \frac{1}{2} \langle \eta^{-1},\, \psi^\ast(g)\rangle \nonumber \\
&= \frac{1}{2} \eta^{\rho\sigma}\left\langle \partial_\rho \otimes \partial_\sigma ,\, dx^\mu \otimes dx^\nu \right\rangle \,\,\partial_\mu (u\circ \psi) \partial_\nu (u\circ \psi) \nonumber \\
 &= \frac{1}{2} \eta^{\mu\nu}\,\, \partial_\mu (\phi^2) \partial_\nu (\phi^2) = 2\phi^2 \partial_\mu \phi \partial^\mu \phi \, .  \label{eq:change-of-map}
\end{align}
When expressed in terms of the original field variable, the functional form of the Lagrangian has indeed changed; the key difference to the manipulations in (\ref{eq:change-of-coords}) is that there is no compensating factor of $1/4\phi^2$ coming from the Jacobian. 

Intuitively, the reason we get a different answer is because we pulled back the metric from {\em different points in field space}, namely those in the image of $\psi$ rather than $\phi$; this is not true for a simple change of coordinates, which does not move points in $M$ but only relabels them.
The transformation (\ref{eq:change-of-map}) of $\cL$ under change of section $\phi \to \psi$ is what one would usually call a `field redefinition'.

\subsubsection{Field space diffeomorphisms}

The change of maps $\phi \to \psi$ that we have performed is, in this case, precisely equivalent to doing a {\em (local) field space diffeomorphism} $f$ before pulling back the transformed metric $f^\ast(g)$ along the same section $\phi$. 
Recall that a diffeomorphism is a smooth invertible function between manifolds $M$ and $N$:
\be
f: M \to N\, .
\ee
In words, every point in the manifold $N$ can be related uniquely to a point in the manifold $M$; if $f$ moves the points hit by $\phi$ to the points hit by $\psi$, then it exactly replicates the effect of changing maps. To relate two maps $\phi$ and $\psi$ we try to construct a diffeomorphism $f$ to fill in the commutative diagram:
\be
\begin{tikzcd}
	\Sigma \\
	(M,g) & (N,g)
	\arrow["{\phi }"', from=1-1, to=2-1]
	\arrow["{\psi }", from=1-1, to=2-2]
	\arrow["f"', from=2-1, to=2-2]
\end{tikzcd}
\ee
so that
\be \label{eq:change-section=f}
\psi = f \circ \phi\,.
\ee
When considering field redefinitions, as we do here, we can take $M$ and $N$ to be the same manifold, and so $f$ becomes a map 
from $M$ to itself.

It is clear that such a diffeomorphism $f$ cannot be found for {\em any} pair $\phi$ and $\psi$ of field space maps. So, when can a change of maps ({\em a.k.a.} a field redefinition) be realised as a field space diffeomorphism? 
Consider a map $\phi$ that sends two spacetime points $x$ and $y$ to the same point in field space, {\em viz.} $\phi(x) = \phi(y) = m_x$. Then, via (\ref{eq:change-section=f}), we can only change to a new map $\psi$ that satisfies $\psi(x) = f(\phi(x)) = f(m_x)$ and $\psi(y) = f(\phi(y)) = f(m_x)$, thus $\psi(x)$ and $\psi(y)$ must also agree (see Fig.~\ref{fig:nonD_redefinitions} for illustration). This means that $\psi$ must be a function of $\phi$.
Explicitly, evaluating Eq. (\ref{eq:change-section=f}) on our coordinate chart $\{u^i\}$ on $M$, 
\be
u^i \circ \psi = u^i \circ f \circ \phi = u^i \circ f(\phi)\, ,
\ee
thus the particular change of map induced by a smooth map $f$ can be expressed as
\be \label{eq:non-deriv-FR}
\phi^i \mapsto \psi^i = f^i(\phi^j)\, .
\ee
This is precisely what physicists call a `non-derivative field redefinition', and applies to the example (\ref{eq:eg_map_change}) considered above. If we additionally can write
\be \label{eq:non-deriv-diffeo-FR}
\psi^i \mapsto \phi^i = (f^{-1})^i (\psi^i)
\ee
then the change of maps is a diffeomorphism and the theories are equivalent at the level of the manifolds on which they live.

Indeed, if we reconsider the example considered in the previous Subsection, we can see explicitly how one recovers the shift (\ref{eq:change-of-map}) in the Lagrangian that we obtained by changing map $\phi \to \psi$, by instead doing a diffeomorphism $f:M \to M$ before pulling back along the same map $\phi$.
The required diffeomorphism\footnote{If the manifold contains a point $m \in M$ with $u(m)=0$, then this map is only a {\em local} diffeomorphism, {\em i.e.} a diffeomorphism on a patch that does not include this point, because $f^{-1}: u \to \sqrt{u}$ is not differentiable at $m$. Alternatively, the map $f$ can be `promoted' to a proper diffeomorphism by instead taking $M = \R_{>0}$, the positive half-line. \label{foot:R+}}
is one that sends field space point $m \in \R$ with local coordinate $u$ to the point $m'$ with coordinate $u^2$ (written in the same coordinate chart); in local coordinates, we might simply write this as $f:u \mapsto u^2$.
In Fig.~\ref{fig:nonD_redefinitions}, this can be visualised by interpreting the right axis of the plot as the image of the left axis under $f$, such that the two axes represent \emph{different} sets of points on $M$.
Under this diffeomorphism the {\em metric changes}, as
\begin{equation} \label{eq:diffeo-of-field-space-metric}
g = du \otimes du \mapsto f^\ast g = \bigg(\frac{\partial u \circ f}{\partial u} \bigg)^2 du \otimes du = 4 u^2 du \otimes du\, .
\end{equation}
Pulling back $f^\ast g$ along $\phi$ then gives the transformed Lagrangian
\begin{align}
    \mathcal{L}[\phi, g] \mapsto \cL[\phi, f^\ast (g)] &= \frac{1}{2} \langle\eta^{-1},\, \phi^\ast (f^\ast(g))\rangle \nonumber \\
    &= \frac{1}{2} \eta^{\mu\nu} 4(u \circ \phi)^2 \partial_\mu (u \circ \phi) \partial_\nu (u \circ \phi) 
    \nonumber \\
    &= 2 \phi^2 \partial_\mu \phi \partial^\mu \phi = \cL[\psi, g]\, ,
\end{align}
matching the result (\ref{eq:change-of-map}) that we obtained by changing maps.

In a similar way, any non-derivative field redefinition of the form (\ref{eq:non-deriv-FR}) can be implemented as a smooth map on the field space.
The transformed Lagrangian can be equivalently expressed in terms of the transformed map $\psi$, or in terms of the smooth map $f$: 
\be
\cL \mapsto  \frac{1}{2} \langle \eta^{-1},\, \psi^\ast(g)\rangle = 
\frac{1}{2} \langle\eta^{-1},\, \phi^\ast (f^\ast(g))\rangle \, .
\ee
At risk of being overly pedantic, we wish to distinguish a diffeomorphism on $M$ from a change of coordinates, although the two notions are intimately related. 
A diffeomorphism involves a change of coordinates {\em plus} a pushforward to the new point; equivalently, a change of coordinates means doing a diffeomorphism, but then pulling back the tensor to be evaluated at the original point. 

Finally, we stress that a change of maps $\phi(x) \to \psi(x)$ is more general than the special case (\ref{eq:non-deriv-FR}) that can be captured by field space diffeomorphisms (or even general smooth maps on field space). Differentials of the smooth map $\phi$ can be used to construct, at least infinitesimally, what one would call `derivative field redefinitions'. 

\subsubsection{Local field space diffeomorphisms, and HEFT {\em vs.} SMEFT}

There is an important subtlety in this discussion that we have so far glossed over, that was hinted at in footnote~\ref{foot:R+}. Namely, it may be the case that the diffeomorphism $f$ is {\em only defined locally}, meaning in words that not all points in $M$ correspond to points in $N$. A particularly relevant physics example of this situation occurs for the map from SMEFT to HEFT, as follows.
Let us start with a coordinate system $\{u^1,u^2,u^3,u^4\}$ on a patch $\mathcal{F}_M \subset M$ in which the metric is
\begin{equation}
    g_{ij} = \begin{pmatrix}
        1 & 0 & 0 & 0\\
        0 & 1 & 0 & 0\\
        0 & 0 & 1 & 0\\
        0 & 0 & 0 &1\\
    \end{pmatrix}\, .
\end{equation}
Pulling this back to $\Sigma$ along a section $\phi$ whose components in the chart are $u^i\circ\phi = \phi^i(x)$, we find the familiar 2-derivative Lagrangian
\begin{equation}
\label{kinetic Lagrangian SMEFT}
    \mathcal{L} = \frac{1}{2} \langle \eta^{-1}, \phi^* g \rangle = \frac{1}{2} \delta_{ij} \partial_\mu \phi^i \partial^\mu \phi^j\, 
\end{equation}
of a canonically normalised set of Higgs fields.
 
Now consider the map $f:\mathcal{F}_M \to \mathcal{F}_N \subset N$ defined by

\begin{equation}
\label{SMEFT to HEFT manifold level}
    f: \begin{pmatrix}
        u^1\\
        u^2\\
        u^3\\
        u^4\\
    \end{pmatrix}
    \to \left(1+\frac{u^4}{v}\right) \begin{pmatrix}
       u^1\\
        u^2\\
        u^3\\
        \sqrt{v^2 - (u^1)^2 - (u^2)^2 - (u^3)^2}
    \end{pmatrix}\
\end{equation}
expressed
in the same coordinate chart $\{u^i\}$. After doing this smooth mapping, we compute the new metric to be

\begin{equation}
   f^* g_{ij} = \begin{pmatrix}
        \left(1+\frac{u^4}{v}\right)^2 \left(1 + \frac{(u^1)^2}{v^2-\vec{u}^2} \right) &  
        \left(1+\frac{u^4}{v}\right)^2 \frac{u^1 u^2}{v^2-\vec{u}^2} &  \left(1+\frac{u^4}{v}\right)^2\frac{u^1 u^3}{v^2-\vec{u}^2} & 
        0 \\[2mm]
        \left(1+\frac{u^4}{v}\right)^2\frac{u^1 u^2}{v^2-\vec{u}^2} & 
        \left(1+\frac{u^4}{v}\right)^2\left(1 + \frac{(u^2)^2}{v^2-\vec{u}^2} \right)  &  
        \left(1+\frac{u^4}{v}\right)^2\frac{u^2 u^3}{1-\vec{u}^2}  & 
        0 
        \\[2mm]
        \left(1+\frac{u^4}{v}\right)^2 \frac{u^1 u^3 }{v^2-\vec{u}^2} &  
        \left(1+\frac{u^4}{v}\right)^2\frac{u^2 u^3}{1-\vec{u}^2}  &  
        \left(1+\frac{u^4}{v}\right)^2 \left(1 + \frac{(u^3)^2}{1-\vec{u}^2} \right) & 
        0 
        \\[2mm]
        0 &  0 &  0 & 1
    \end{pmatrix}\, ,
\end{equation}
where $\vec{u} = (u^1, u^2, u^3)$.
Once again we can pullback along a section $\phi$ and contract with $\eta^{-1}$ to obtain the Lagrangian function
\begin{nalign}
    \mathcal{L} &= \frac{1}{2} \partial_\mu \phi^4 \partial^\mu \phi^4 
    + \frac12\left(1+\frac{\phi^4}{v}\right)^2
    \bigg(\partial_\mu \phi^1 \partial^\mu \phi^1 + \partial_\mu \phi^2 \partial^\mu \phi^2 + \partial_\mu \phi^3 \partial^\mu \phi^3 \bigg)\\
    &+ \frac{(1+\phi^4/v)^2}{v^2 - (\phi^1)^2 - (\phi^2)^2 - (\phi^3)^2} \bigg( \phi^1 \partial_\mu \phi^1 \phi^2 \partial^\mu \phi^2 +  \phi^1 \partial_\mu \phi^1 \phi^3 \partial^\mu \phi^3 +  \phi^2 \partial_\mu \phi^2 \phi^3 \partial^\mu \phi^3 \bigg)\\
    &+\frac12\frac{(1+\phi^4/v)^2}{v^2 - (\phi^1)^2 - (\phi^2)^2 - (\phi^3)^2} \bigg( (\phi^1)^2 \partial_\mu \phi^1 \partial^\mu \phi^1 + (\phi^2)^2 \partial_\mu \phi^2 \partial^\mu \phi^2 + (\phi^3)^2 \partial_\mu \phi^3 \partial^\mu \phi^3 \bigg)\, .
\end{nalign}

Introducing the notation $\vec{\phi} = (\phi^1,\phi^2,\phi^3)$, this can be written in the more compact form:

\begin{nalign} \label{eq:HEFT-lag-phi}
    \mathcal{L} &= \frac{1}{2} \partial_\mu \phi^4 \partial^\mu \phi^4 
    + \frac12\left(1+\frac{\phi^4}{v}\right)^2\left[ \partial_\mu \vec{\phi} \cdot \partial^\mu \vec{\phi} 
    + \frac{1}{v^2 - \vec{\phi}^2} \bigg( \vec{\phi} \cdot \partial_\mu \vec{\phi} \bigg)^2
    \right]\,.
\end{nalign}
If we identify $\phi^4$ with the physical Higgs boson field and the other $\phi^i$ with the Goldstones, we see that the Higgs has a canonical kinetic term and does not enter any of the scalar products. Eq. (\ref{eq:HEFT-lag-phi}) is in agreement with the results in Refs.~\cite{Gomez-Ambrosio:2022qsi,Alonso2016:smn}. The map from SMEFT to HEFT is known to always exist and be smooth at the level of the Lagrangian.

On the other hand, the inverse map (that would take one from HEFT to SMEFT) may have a singularity.
The inverse map $f^{-1}$ would be

\begin{equation}
    f^{-1} : \begin{pmatrix}
        u^1\\
        u^2\\
        u^3\\
        u^4\\
    \end{pmatrix}
    \to    \frac{v}{\sqrt{\mathbf{u}^2}}\begin{pmatrix}
      u^1\\
      u^2\\
      u^3\\
    \frac{\sqrt{\mathbf{u}^2}}{v} \left(\sqrt{\mathbf{u}^2} - v\right)
    \end{pmatrix}\,,
\end{equation}
where $\mathbf{u} = (u^1,u^2,u^3,u^4)$. This is clearly only well defined if $\mathbf{u}^2 \neq 0$.

The map from this Lagrangian back to (\ref{kinetic Lagrangian SMEFT}) may have a singularity at the origin, in which case $f$ is simply a smooth map on the whole manifold and only becomes a diffeomorphism (which must be invertible) when restricted to a patch that does not contain the origin (the $O(4)$ fixed point). Mathematically speaking, this means that the two manifolds on which SMEFT and HEFT are formulated are indistinguishable {\em away} from this point.
When SMEFT is not enough~\cite{Cohen:2020xca}, we really only have a `smooth map' $f$ from SMEFT to HEFT, which cannot be inverted.

\subsection{Geometrizing amplitudes} \label{sec:amplitudes-part1}

As we saw in \S \ref{subsec: change of coordinates}, a change of coordinates does not change the Lagrangian (whereas, we reiterate, a smooth map does -- and the latter is what we mean by a non-derivative field redefinition). The coordinate invariance of the Lagrangian nevertheless remains a useful redundancy to exploit, because it means we are free to pick any valid coordinate system on a given local patch.

In particular, this opens up the choice of `normal coordinates' in the vicinity of some point $m$ in patch $\mathcal{F}_M \subset M$ of field space, defined by the following properties:\footnote{Normal coordinates always exist under the conditions we assume. This is because, for any pseudo-Riemannian manifold $(M,g)$ and any point $m \in M$, there exists a neighbourhood $U \subset M$ and a neighbourhood $V \subset T_m M$ such that the map
\begin{equation}
    \exp_m: V \to U
\end{equation}
is a diffeomorphism. Using a torsion free connection (such as the Levi-Civita connection) we can construct an isomorphism $T_m M \sim \mathbb{R}^n$ allowing us to consider the triple $\{ \exp_m^{-1}, V, U \}$ as a coordinate chart, which we refer to as normal coordinates (see {\em e.g.}~\cite{Petersen:2006riemann}). The chart is unique up to the choice of isomorphism.
\label{foot:NC}}
\begin{enumerate}
    \item For all $1 \leq i,j \leq n$ we have 
    \be \label{eq:NCs_1}
    g_{ij}(m) = \bar{g}_{ij}, 
    \ee 
    which is a diagonal matrix whose entries are either $+1$ or $-1$, depending on the signature of the metric; 
    \item For all $1 \leq i,j,k \leq n$ we have 
    \be \label{eq:NCs_2}
    \Gamma_{ij}^k(m) = 0;
    \ee
    \item For all $1 \leq i,j,k \leq n$ we have 
    \be \label{eq:NCs_3}
    \partial_k g_{ij}(m) = 0.
    \ee
\end{enumerate}
Using this system of normal coordinates, the metric tensor $g_{ij}$ admits the following expansion around point $m \in \mathcal{F}_M$
\begin{equation}
    g = g_{ij}(u) du^i \otimes du^j = \bigg(\bar{g}_{ij} + \frac{1}{3} R_{iklj}(m) u^k u^l + \frac{1}{6}R_{iklj;r} (m) u^k u^l u^r + \dots \bigg) du^i \otimes du^j
\end{equation}
where the normalization of the first term in the expansion naturally leads to a canonical kinetic term once we fix the signature of the metric.

The choice of normal coordinates is  useful because the expansion then naturally organizes itself based on the number of fields, thus allowing for an easy relation to be established between components of the Riemann tensor and $n$-point amplitudes~\cite{Cohen:2021ucp,Nagai2019:nc}:
\begin{itemize}
    \item $\delta_{ij} du^i \otimes du^i$ enters in the propagator;
    \item $\frac{1}{3} R_{iklj}(m) u^k u^l du^i \otimes du^j$ enters in four point amplitudes;
    \item $\frac{1}{6}R_{iklj;r} (m) u^k u^l u^r  du^i \otimes du^j$ enters in five point amplitudes.
\end{itemize}
In this field space picture, a potential (0-derivative) contribution to the Lagrangian is added to the Lagrangian by hand (see \S \ref{sec:other_terms}), independently of the field space geometry. Choosing coordinates $u^i$ on $\mathcal{F}_M$ such that $u^i \circ m_{\text{vac}} = \vec{0}$, where $m_{\text{vac}} \in M$ is the vacuum point, we have
\begin{nalign}
    \mathcal{L} &=  \bigg(\delta_{ij} + \frac{1}{3} R_{iklj}(0) \phi^k \phi^l + \frac{1}{6}R_{iklj;r} (0) \phi^k \phi^l \phi^r + \dots \bigg) \partial_\mu \phi^i \partial^\mu \phi^j\\  &- \bigg(\frac{1}{2} \frac{\partial^2 V}{\partial \phi^k \partial \phi^l}(0) \phi^k \phi^l + \frac{1}{6} \frac{\partial^3 V}{\partial \phi^k \partial \phi^l \partial \phi^r}(0) \phi^k \phi^l \phi^r + \dots \bigg)
\end{nalign}
It can be seen that the contribution to the three point amplitude at this point comes entirely from the potential as a result of the vanishing of the Christoffel symbols~\cite{Cohen:2021ucp}.    
When we eventually pass to the 1-jet bundle, we will see (\S \ref{sec:amplitudes}) how to adapt this story to incorporate higher-derivative terms (and also how, in the jet bundle picture, the potential contribution also comes from components of the metric).

\section{From Field Space to Bundles} \label{sec:fib-bundles}

Before we get into the main business of this paper, namely our consideration of EFT terms with $>2$ derivatives, 
we find it helpful to first generalise slightly the above account of EFTs in terms of field space geometry.
The generalisation, which is motivated by locality, is to replace the description in terms of maps (\ref{eq:map}) to a field space $M$ by one in terms of fibre bundles, which is almost equivalent but a bit more general.\footnote{JD is grateful to Ben Gripaios and Joseph Tooby-Smith for past discussion and correspondence related to ideas in this Section. See also Ref.~\cite{Gripaios:2021hsc}.}
While the generalisation to bundles might seem like a technicality here, we will describe several advantages over the field space picture in this Section; moreover, the bundle picture provides the starting point for passing to jet bundles when we turn to describe higher-derivative Lagrangians.

\subsection{Fields as sections of fibre bundles}

The idea is simple, but first let us recap some basic concepts. Recall that a fibre bundle is a triple $(E,\Sigma, \pi)$, where
\begin{itemize}
\item $E$ is called the {\em total space} of the bundle; 
\item $\Sigma$ is called the {\em base space}. Both $E$ and $\Sigma$ are smooth manifolds; 
\item The {\em projection map} $\pi:E \to \Sigma$ is a surjective submersion; 
\item {\em Local trivializations:} for any point $x\in \Sigma$ there exists an open neighbourhood $\mathcal{U}_x$ and a diffeomorphism $\varphi: \pi^{-1}(\mathcal{U}_x) \to \mathcal{U}_x \times F$, where $F$ is itself a manifold called the `fibre', such that the projection $\pi$ agrees with projection from $\mathcal{U}_x \times F$ onto its first factor. In other words, locally the total space $E$ looks like a direct product space, but this need not be true globally.
\end{itemize}
A {\em section} of the fibre bundle is then a smooth map $\phi:\Sigma \to E$ such that $\pi \circ \phi = \text{id}_\Sigma$.  A section is like an `inverse' of the projection map $\pi$, taking you from the base space into the total space of the bundle, while preserving the base space point. Identifying the base $\Sigma$ with spacetime, what a physicist calls a `field' is thus a section of some bundle over $\Sigma$.
Sections will moreover play a starring role in defining jet bundles. 
We denote the set of all sections of $\pi$ by $\Gamma(\pi)$, and the set of {\em local} sections whose domains include a point $x\in \Sigma$ by $\Gamma_x(\pi)$.
With these definitions, let us return to scalar field theory on target space $M$.

The sigma model map $\phi(x): \Sigma \to M$ that we encountered before trivially defines a {\em section} $\phi$ of a particular fibre bundle $(\Sigma \times M,\Sigma,\pi)$ with fibre $F=M$ coinciding with the `field space' of the previous picture, and
where the projection is trivially $\pi(x,m) = x$ $\forall m\in M$, $x\in \Sigma$. This bundle can moreover be equipped with a product metric $\eta \oplus g$, given the metrics $\eta$ and $g$ on $\Sigma$ and $M$ respectively (\ref{eq:eta}, \ref{eq:g}).
Such a bundle that is globally a direct product is called a trivial bundle, for which $\phi$ is a global section.

\subsubsection*{Locality}

This field space formulation described in \S \ref{sec:preliminaries} implicitly assumes that $\phi$ is a global section of a trivial bundle; but this restriction is unecessarily strong.
Given that we, as local physicists exploring our patch of the Universe, can only know about the structure of field space on some open set $\mathcal{U}_\Sigma\subset \Sigma$, there is no reason to enforce that the field space bundle, which {\em locally} has a product structure $\cong \mathcal{U}_\Sigma \times M$, has a direct product structure $=\Sigma \times M$ globally.
More generally, therefore, one can (and should) allow fibre bundles $(E,\Sigma,\pi)$ with fibre $F=M$ that are topologically non-trivial.  

\begin{figure}[t]\centering
\includegraphics[height=3.7cm]{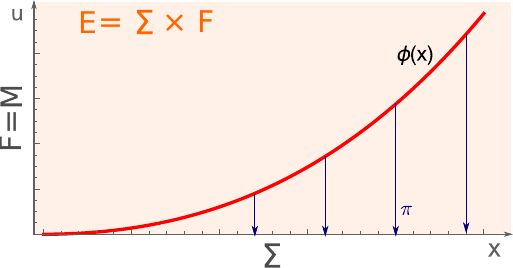}
~~
\includegraphics[height=3.7cm]{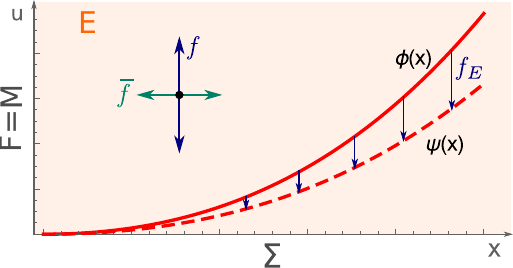}
\caption{Left: Illustration of a fibre bundle $(E, \Sigma, \pi)$ with fibre $M$, which here we take to be a trivial bundle ($E \cong \Sigma \times M$) for ease of illustration, on which we have a fibred coordinate system $(x,u)$. This toy example describes a single real scalar ({\em e.g.} take $M=\R$) in 1d spacetime ({\em e.g.} take $\Sigma=\R$).  The field $\phi$ is a \emph{section} of such a fibre bundle, which in general need not be a trivial bundle.
Right: bundle morphisms are pairs of maps $(f,\bar f)$ that generically move points on $E$ and $\Sigma$ respectively, in a way that is compatible with the bundle projection maps. A field redefinition is a change of section. A non-derivative field redefinition is equivalent to a bundle morphism with $\bar f={\rm id}_{\Sigma}$, as indicated in our sketch by the map $f_E$, for which the arrows are directed `vertically'. }\label{fig:bundle}
\end{figure}

We should therefore pass from a description in which the field $\phi(x)$ is a map from $\Sigma$ to $E$ to a slightly more general picture, in which:\footnote{This viewpoint is emphasized in Ref.~\cite{Gripaios:2021hsc}, which adopts an even more general setup, again motivated by locality arguments, whereby the fibre bundle is replaced by a {\em fibred manifold}. For a fibred manifold, the fibres $\pi^{-1}(x)$ above each spacetime point $x$, which the reader can think of as the `field space' associated to the point $x$, are not even required to be diffeomorphic to one another. We are happy to limit ourselves to plain old fibre bundles in the present work.%
} 
\be \label{eq:scalar-field-section}
\text{A scalar field $\phi\in \Gamma_x(\pi)$ is a local section of a fibre bundle $(E,\Sigma, \pi)$ with fibre $M$} .
\ee 
Given a patch $\mathcal{U}_\Sigma \subset \Sigma$ with local coordinates $x^\mu$, one can introduce a local fibred coordinate system $(x^\mu, u^i)$ on $\pi^{-1}(\mathcal{U}_\Sigma) \subset E$, that matches our previous choice. 
To guide the reader, we illustrate the basic `local product' structure of a fibre bundle and the notion of a section in Fig.~\ref{fig:bundle} (left), for a toy example in which the total space $E$ has dimension two.

\subsubsection*{Lower derivative Lagrangian terms from geometry}

In these coordinates, and starting from metrics $\eta$ and $g$ on $\Sigma$ and $M$ respectively (in the field space picture), a natural choice of geometry on $E$ is to take the metric on $E$ to be locally the product metric, $g_E =\eta \oplus g$. Pulling $g_E$ back to $\Sigma$ along a section $\phi$, and contracting with the inverse metric $\eta^{-1}$ on $\Sigma$, via essentially the same formula as (\ref{eq:2deriv_compact}), we get the same 2-derivative Lagrangian (\ref{eq:2deriv_long}) that we obtained from the metric $g$ in the `field space maps' picture of \S \ref{sec:preliminaries}, up to a constant shift coming from pulling back the $\eta_{\mu\nu}dx^\mu dx^\nu$ part of $g_E$. 

Even though this geometry is a natural choice, one can consider more general geometries on the field space bundle; in particular, consider a metric on the field space bundle $E$ of the form
\begin{equation} \label{eq:metric-0jet-A}
    g_E = -\frac{2}{d}V(u)\eta_{\mu\nu}\,  dx^\mu \otimes dx^\nu + g_{ij}(u) du^i \otimes du^j\, ,
\end{equation}
where $d$ is the spacetime dimension (say, $d=4$ for the Higgs EFTs).
Upon pulling this metric back to spacetime $(\Sigma,\eta)$, which we emphasize is still equipped with the flat geometry of Minkowski space, and contracting with the inverse spacetime metric, we get the Lagrangian
\begin{equation}
    \cL[\phi, g_E] = \frac{1}{2}\, \langle\eta^{-1} , \, \phi^\ast (g_E) \rangle
    =\frac{1}{2} g_{ij}(\phi) \partial_\mu \phi^i \partial^\mu \phi^j - V(\phi)\, .
\end{equation}
We thus recover the most general scalar EFT Lagrangian with {\em up to} 2-derivatives, but including also the 0-derivative potential terms, by passing from geometry on a target space $M$ to a geometry on the fibre bundle $E$. As an example, this is already enough to capture the whole (renormalisable) Higgs sector of the SM Lagrangian via geometry. We will revisit this feature when we pass to jet bundles later in the paper. 

\subsubsection*{Symmetries on the field space bundle}

For the case of SMEFT or HEFT, in this fibred coordinate system $(x^\mu, u^i)$ there is also a natural implementation of the symmetries described in \S \ref{sec:preliminaries} at the level of field space bundles $(E,\Sigma,\pi)$. The group action on the bundle is, at least locally, simply
\begin{equation}
    (a^\mu, L^\mu_\nu, O; \, x^\mu, u^i) \mapsto (L^\mu_\nu x^\nu + a^\mu, \sigma(O,u^i))\, ,
\end{equation}
where one can insert the group action $\sigma = \sS$ or $\sH$ for SMEFT or HEFT respectively.
It is, however, instructive to still think of the spacetime and internal symmetries as two distinct group actions, both in this case of Higgs EFTs and more generally. As we will see in \S \ref{sec:symms-on-bundle}, we can think of these symmetries as bundle morphisms, and
the identification of a symmetry as `internal' can be given a natural definition in these terms.

\subsection{Field redefinitions are changes of section}

In addition to the motivation by locality that we have described, plus the straightforward inclusion of lower-derivative Lagrangian terms, passing to this `field space bundle' formulation of scalar EFTs affords a technical benefit when it comes to describing field redefinitions geometrically. 
Recall from \S \ref{sec:field_redef_1} that, in the `maps to field space' picture, a non-derivative field redefinition (or change of map $\phi^i \mapsto \psi^i(\phi^j)$) is equivalent to doing a local diffeomorphism on the target space.

Regarding a scalar field to instead be a section $\phi$ of the bundle $(E,\Sigma,\pi)$, a general field redefinition will be implemented as a smooth change of (local) section 
\be
\Gamma_x(\pi)  \ni \phi \mapsto \psi  \in \Gamma_x(\pi)
\ee
This is itself a geometric notion of what it means to do a general field redefinition. A simple example of this is given in \S \ref{sec:phi4} below. For derivative field redefinitions, however, we are not guaranteed an equivalent description in terms of morphisms of the target space (or bundle), as we study more carefully in the next Section.

\subsection{Morphisms on the field space bundle} \label{sec:bundle-morphism}

In this `field space bundle' picture, a more general class of field redefinitions can be described in terms of morphisms of the bundle, than we saw in the case of field space smooth maps, above.
To unpack this statement requires a more-or-less formal definition of bundle morphisms, which also provide a natural language for describing {\em e.g.} symmetries of the theory.

Suppose we have a fibre bundle $\pi:E\to \Sigma$, in the notation of \S \ref{sec:fib-bundles}, and $\phi$ is a (local) section of that bundle.
Given another bundle $\rho: F \to \Omega$, a bundle morphism is a pair of maps $f: E \to F$ and $\bar{f}: \Sigma \to \Omega$ such that 
\be \label{eq:CD-bundle}
\begin{tikzcd}
	E & F \\
	\Sigma & \Omega
	\arrow["\pi"', from=1-1, to=2-1]
	\arrow["\rho", from=1-2, to=2-2]
	\arrow["f", from=1-1, to=1-2]
	\arrow["{\bar{f}}"', from=2-1, to=2-2]
\end{tikzcd}
\ee
is a commutative diagram. 
(We could, more generally, consider only a {\em local} bundle morphism, whereby every element in the commutative diagram is replaced by open submanifolds, but such generality will not add much to the discussion.)

To describe both symmetries and field redefinitions, we are interested in fixing both bundles to be the same, $(E,\Sigma,\pi)$. Fig.~\ref{fig:bundle} (right) illustrates how a bundle morphism acts in this case, in a 1d example. We are also sometimes interested in fixing the map from the base space to itself to be 
\begin{equation} \label{eq:idSigma}
\bar{f} = \text{id}_\Sigma\, .
\end{equation}
(The obvious exception to this will be when we discuss spacetime symmetries below.)
The map $f_E$ along the top is an automorphism from the bundle $(E,\Sigma,\pi)$ to itself, $f_E: E \to E$. The commutative diagram becomes
\be
\label{eq:CD-bundle-morph-E}
\begin{tikzcd}
	E & {} & E \\
	{} & \Sigma
	\arrow["f_E", from=1-1, to=1-3]
	\arrow["\pi"', from=1-1, to=2-2]
	\arrow["\rho", from=1-3, to=2-2]
\end{tikzcd}
\ee
We thus have the relation of maps
\begin{equation}    \label{eq:bundle_morphism_relation}
    \pi = \rho \circ f_E\, .
\end{equation}
This, which follows from assumption (\ref{eq:idSigma}), means that $f_E$ does not move points in the base space, but only moves points in the fibre. In our local fibred coordinate system $(x^\mu, u^i)$ on $E$, we have
\be \label{eq:fE-coords-b-p}
f_E: (x^\mu, u^i) \mapsto (x^\mu, f^i(x^\mu,u^j))\, .
\ee
The map $f^i(x^\mu,u^j)$ on the fibre coordinates can be regarded as a family of diffeomorphisms on fibres $\{M_x\}$, for each value of the base point $x$, that is free to vary (smoothly) with $x$. 

A pair of local sections $\phi \in \Gamma_x(\pi)$ and $\psi \in \Gamma_x(\rho)$ for these two bundles will also be related by the bundle morphism $f_E$:
\begin{equation}    \label{eq:bundle_morhpism_relation_sections}
    \psi = f_E \circ \phi\, .
\end{equation}
In our fixed local fibred coordinate system $(x^\mu, u^i)$, Eq. (\ref{eq:bundle_morhpism_relation_sections}) implies
\be \label{eq:fi-x}
\psi^i(x) = f^i (x, \phi^j(x))\, .
\ee
This is clearly a generalisation of (\ref{eq:non-deriv-FR}) because now the functions $f_x^i$ can vary continuously (and smoothly) with the base point $x$.
However, it is perhaps not an especially useful generalisation, given our primary interest is in describing Poincar\'e-invariant quantum field theory actions -- the more general change of section expressed in (\ref{eq:fi-x}) allows for field redefinitions that have explicit spacetime dependence, which typically violate Poincar\'e symmetry. Next, we discuss how this notion of bundle morphisms applies first to symmetries, and then turn to field redefinitions.

\subsubsection{Internal {\em vs} spacetime symmetries} \label{sec:symms-on-bundle}

We already offered a description of symmetries in the `field space bundle' formulation of Higgs EFTs, in terms of a group action by $\text{Poinc} \times O(4)$ on the bundle $E$. Generally, a group action of Lie group $G$ on a manifold $E$ is equivalent to specifying a diffeomorphism of $E$ for each group element $g \in G$. A symmetry that can be described via such a group action can thus be naturally interpreted in terms of bundle morphisms; moreover, the bundle structure offers an obvious notion of what we mean by saying that a symmetry is `internal' or otherwise. 

In this language, we might say rather abstractly that a {\em symmetry} is a particular bundle morphism $(f, \bar{f})$ such that the Lagrangian  $\cL = (1/2) \langle \eta^{-1}, \phi^\ast (g) \rangle$ shifts by at most a total derivative:\footnote{We also implicitly assume an invariance condition~\cite{Davighi:2018inx} is satisfied for the topological term, if present.} 
\be \label{eq:symm-abstract}
\langle \eta^{-1}, \phi^\ast (f^\ast -1)g \rangle = \partial_\mu K^\mu\,.
\ee
Such a broad definition would account for the fact that  (i) there can be many redundancies in the map from metric to Lagrangian (as will become more obvious when we pass to 1-jet bundles, especially in the examples of \S \ref{sec:examples} and \S \ref{sec:HEFTandSMEFT}), and (ii) we wish to identify Lagrangians that differ by total derivatives. Both these facts mean that requiring invariance of the metric itself, {\em viz.}
\begin{equation} \label{eq:symm-metric}
    f^\ast g = g\, ,
\end{equation}
is too strong a condition for a symmetry. Nonetheless, the more general condition (\ref{eq:symm-abstract}) is rather nasty to implement in practice -- indeed, it is not even obvious why the set of morphisms satisfying (\ref{eq:symm-abstract}) should form a group,\footnote{Moreover, a definition of `symmetry' like Eq. (\ref{eq:symm-abstract}) is still far from being comprehensive; indeed, finding all symmetries can be an almost overwhelming task even for very simple theories, if a symmetry is broadly defined to be any `thing that does not change observables'. This would include transformations that act completely trivially, such as the change of coordinates considered in \S \ref{subsec: change of coordinates} above as well as more general redundancies of description such as gauge symmetries, right through to the huge symmetry under field redefinitions that is not seen at the Lagrangian level, but only after doing the path integral. The symmetries we consider here, namely symmetries of the metric on $E$ that can be captured by bundle morphisms, lie somewhere between these two extremes. Finally, we remark that in recent years yet further `generalised' notions of symmetry have been appreciated, which are not seen by their action on any local operators but by their action on extended objects like Wilson lines (see~\cite{Bah:2022wot,Cordova:2022ruw} for recent surveys). 
}
whereas (diffeo)morphisms satisfying (\ref{eq:symm-metric}) clearly do form a group.
In practice, the particular symmetries that we consider will be implemented by bundle morphisms that are symmetries of the metric (isometries), satisfying (\ref{eq:symm-metric}), so we are happy to proceed with this definition of (a certain class of) symmetry.

Continuing, an {\em internal symmetry} is then a symmetry that is described by a `base-point preserving' bundle morphism $(f,\bar{f})$ of type (\ref{eq:CD-bundle-morph-E}), namely for which $\bar{f}=\text{id}_\Sigma$, which means the group action does not move points in spacetime. More precisely, it moves points in a given fibre $M_x$ to other points in that same fibre $M_x$. In the case of SMEFT, say, the internal symmetry is unsurprisingly the $O(4)$ group action:
\begin{equation}
    \sS: O(4) \times E \to E : (O, x^\mu, u^i) \mapsto (x^\mu, O u^i), 
\end{equation}
in our local fibred coordinates. This defines a set of base-point preserving bundle morphisms satisfying both the commutative diagram (\ref{eq:CD-bundle-morph-E}) and the group multiplication law. In general, an internal symmetry described by a Lie group action by $G$ on the bundle $E$ corresponds to the set
\be
\{(f_g, \bar{f} =\text{id}_\Sigma) \text{~a bundle morphism} \,\, \forall g \in G \, \, | \,\, f_{g_1} \circ f_{g_2} = f_{g_1 g_2} \,\, \forall g_1, \, g_2 \in G\}\, .
\ee
In our SMEFT example, the morphisms are
\be
f_O: (x^\mu, u^i) \to (x^\mu, O u^i)\, \, \forall O \in O(4)\, .
\ee
On the other hand, a {\em spacetime} symmetry is a map $\bar{f}$ such that $(f,\bar{f})$ is a bundle morphism that is a symmetry. In our HEFT and SMEFT examples, which have $(3+1)$-d Poincar\'e invariance, we can take this spacetime symmetry to be defined by the group action on the bundle
\be \label{eq:Poincare-on-E}
\rho: \text{Poinc} \times E \to E : (a^\mu, L^\mu_\nu, x^\mu, u^i) \mapsto (L^\mu_\nu x^\nu + a^\mu, u^i)\, .
\ee
Again this defines a collection of bundle morphisms, where the important thing is the map on the base of the bundle, $\bar{f}_{a,L}:x^\mu \mapsto L^\mu_\nu x^\nu + a^\mu$ for every $a^\mu$ and $L^\mu_\nu$. 

It is interesting that, as a collection of bundle morphisms, the internal symmetry is naturally implemented {\em locally}. The general formula for an `internal morphism' in  the bundle picture is $\sigma: (x^\mu,u^i) \mapsto (x^\mu, O(x).u^i)$, where $O(x)$ is an $O(4)$-valued function on spacetime. This suggests there should be a natural notion of gauging in this language (indeed, gauge fields are perhaps the most familiar example of quantum fields being sections on bundles over spacetime). We save a proper treatment of the gauging of internal symmetries, which is of course a crucial element in the electroweak theory, for future work.

\subsubsection{Non-derivative field redefinitions (again)}

Having discussed symmetries, we now return to the issue of field redefinitions, and ask when such transformations can be described by the bundle morphisms defined in \S \ref{sec:bundle-morphism} above.

Firstly, if we take the maps $f^i(x^\mu, u^j)$ in {\em e.g.} Eq. (\ref{eq:fE-coords-b-p}) to be independent of $x$, then we have that $u^i \circ \psi = u^i \circ f(\phi)$, equivalent to doing the same `field space diffeomorphism' everywhere in spacetime, {\em i.e.} at every fibre. 
This special case of bundle morphisms can describe any non-derivative field redefinition, as considered in \S \ref{sec:change-maps}. Later, in \S \ref{sec:last-time} and \S \ref{sec:qm_1d_redef}, we will see how this is implemented in the more general jet bundle context.

\subsubsection{More general field redefinitions?} \label{sec:towards-deriv}

Consider again the general (base-point preserving) bundle morphism described by Eq.~\eqref{eq:fi-x}. We now examine, for completeness, what kind of field redefinitions can be captured by such bundle morphisms, and whether we get `anything more' than non-derivative field redefinitions. Recall that
we already have a geometric description of arbitrary derivative field redefinitions as arbitrary changes of section $\phi \to \psi$. We see an explicit example of this below in \S \ref{sec:phi4}.

To study this, let us start from the more general class of Lagrangian captured by our field space bundle geometry, including 0-derivative (potential) terms. We take 
\begin{equation} \label{eq:metric-0jet}
    g_E = -\frac{2}{d}V(u)\eta_{\mu\nu}\,  dx^\mu \otimes dx^\nu + g_{ij}(u) du^i \otimes du^j\, 
\end{equation}
as in (\ref{eq:metric-0jet-A}), which, pulling back along a section $\phi:\Sigma \to E$, gives the general 2-derivative Lagrangian $\cL[\phi, g_E] = \frac{1}{2}g_{ij}(\phi)\partial_\mu \phi^i \partial^\mu \phi^j - V(\phi)$.
If we do an `internal diffeomorphism' on the bundle of the kind (\ref{eq:CD-bundle-morph-E}), which acts as $f_E:(x^\mu,u^i) \to (x^\mu, f^i(x,u))$ in our local fibred coordinate system, then the metric $g_E$ transforms as
\begin{align} \label{eq:gE-shift}
    g_E \to f_E^\ast g_E &= \,\,-\frac{2}{d}V(u \circ f_E)\eta_{\mu\nu}\, dx^\mu dx^\nu +g_{ij}(u \circ f_E) \frac{\partial (u^i \circ f_E)}{\partial u^k} \bigg|_{x,u}\frac{\partial (u^j \circ f_E)}{\partial u^m} \bigg|_{x,u} du^k du^m \nonumber \\
    &\qquad\qquad+ g_{ij}(u \circ f_E)\frac{\partial (u^i \circ f_E)}{\partial x^\mu} \bigg|_{x,u}\frac{\partial (u^j \circ f_E)}{\partial x^\nu} \bigg|_{x,u} dx^\mu dx^\nu 
\end{align}
Upon pulling back along the (same) section $\phi$ and forming the new Lagrangian, we get
\begin{align} \label{eq:change-of-L-bundle-morph}
    &\cL[\phi, g_E] \mapsto \cL[\phi, f^\ast_E g_E] \\
    &=\frac{1}{2} g_{ij}(f(x,\phi)) \left[\partial_\mu f^i\bigg|_{x,u=\phi} \partial^\mu f^j\bigg|_{x,u=\phi} +  \frac{\partial f^i}{\partial u^k} \bigg|_{x,u=\phi} \frac{\partial f^j}{\partial u^m}\bigg|_{x,u=\phi}  \partial_\mu \phi^k \partial^\mu \phi^m  \right] -V(f(x,\phi)) \nonumber 
\end{align}
While formulae (\ref{eq:gE-shift}) and (\ref{eq:change-of-L-bundle-morph}) may look elaborate, they are essentially the result of applying the chain rule. 

The result (\ref{eq:change-of-L-bundle-morph}) generalises a little the effect of `field space smooth maps' considered in \S \ref{sec:preliminaries}. There are two main differences: first, the possible dependence of the transformed $g_{ij}$ and $V$ functions on spacetime coordinates $x^\mu$; second, the first term on the RHS which comes from partial differentiation of the functions $f^i$ with respect to spacetime. If we restrict to bundle morphisms that respect Poincar\'e symmetry, then neither of these generalisations are relevant. 

If we further assume the morphism is a `perturbation' around the trivial morphism, {\em i.e.} that it takes the form
\be \label{eq:f-expansion}
f^i(x^\mu,u^j) = u^i + \epsilon \lambda^i(x^\mu,u^j)\, , \qquad \epsilon \ll 1\, ,
\ee
then we can expand the various functions $g_{ij}(f)$ and $V(f)$ around their `old values'. For the potential, for example, we can do an ordinary Taylor expansion
\begin{equation}
    V(u) \mapsto V(f(x,u)) = V(u) + \epsilon \lambda^i \frac{\partial V}{\partial u^i} + \frac{1}{2} \epsilon^2 \lambda^i \lambda^j \frac{\partial^2 V}{\partial u^i \partial u^j} + \dots\, ,
\end{equation}
while the various Jacobian factors can also be expanded
\begin{equation}
     \frac{\partial f^i}{\partial x^\mu} = \epsilon \frac{\partial \lambda^i}{\partial x^\mu}, \qquad 
     \frac{\partial f^i}{\partial u^j} = \delta^i_j + \epsilon \frac{\partial \lambda^i}{\partial u^j}\,,
\end{equation}
before substituting into (\ref{eq:change-of-L-bundle-morph}).

One might wonder whether such a base-point preserving bundle morphism of this kind can ever capture the effects of a derivative field redefinition, which is a change of section $\phi \to \psi$ such that $u\circ \psi$ is a function of $u \circ \phi$ and its spacetime derivatives.
The first point to note is that, starting from the 2-derivative Lagrangian $\cL[\phi, g_E]$ above, the transformation (\ref{eq:change-of-L-bundle-morph}) cannot generate terms with more than 2-derivatives. So, we straightaway learn that this kind of bundle morphism cannot generically capture a derivative field redefinition; 
at best, one might be able to reproduce the effects of a derivative field redefinition if we truncate the transformed Lagrangian also at two derivatives. We next consider a concrete example, for which such a diffeomorphism can capture the effects of a derivative field redefinitions -- but we stress that this example is far from generic.

\subsubsection{Example: $\phi^4$ theory} \label{sec:phi4}

Consider a 4d theory of a single real scalar field. In the bundle formulation, we take a trivial bundle $E = \R^{1,3} \times \R_u \xrightarrow{\pi} \R^{1,3}$, on which we specify an initial fibred coordinate system $(x^\mu, u)$. We take the metric 
\be
g_E = -\frac{1}{2} \frac{y\, u^4}{4!} \eta_{\mu \nu} dx^\mu\, dx^\nu + du\, du, \qquad y \in \R
\ee
which is flat in the field space coordinate $u$, corresponding to a canonical 2-derivative kinetic term. The real parameter $y$ is a coupling constant. Pulling back along a section $\phi:x^\mu \to (x^\mu, \phi(x))$, our initial Lagrangian is simply 
\be
\cL[\phi] = \frac{1}{2} \partial_\mu \phi \partial^\mu \phi - \frac{y}{4!} \phi^4 \, ,
\ee
that of `$\phi^4$ theory' in 4d, with potential $V(\phi) = (y/4!) \phi^4$.

Now, consider a derivative field redefinition that is defined by a change of section $\phi \to \psi$ where, in our coordinate system,\footnote{We choose a field redefinition that is consistent with a $\text{Poinc} \times O(1)_u \cong \text{Poinc} \times \Z_2$ symmetry. The field redefinition is consequently second-order in derivatives and `preserves the number of fields'. This kind of field redefinition, which shifts $\phi$ by `$\Box \phi$', can be used to eliminate redundancies due to the equations of motion (EOMs), and will play an important role in the examples of higher-derivative Lagrangians that we discuss in \S \ref{sec:examples}. } 
\begin{align} \label{eq:deriv-change-of-section}
x^\mu \circ \psi = x^\mu \, , \qquad 
u \circ \psi = \left(1-\epsilon \partial_\mu \partial^\mu\right)(u \circ \phi)\, , 
\end{align}
where $1 \gg \epsilon \in \R$.
The first condition ensures the change of section only acts `internally' on the fibres, as is our assumption throughout. Abusing notation as before and letting $\psi$ ($\phi$) also denote the component values of the section $\psi$ ($\phi$) in the local fibre coordinate, and letting $\Box := \partial_\mu \partial^\mu$ as usual, we would write this (in notation more familiar to the physicist) as
\be
\phi \mapsto \psi =\phi - \epsilon\Box\phi\, . 
\ee 
Pulling back the metric $g$ along $\psi$ rather than $\phi$, we obtain the transformed Lagrangian,
\begin{align} \label{eq:deriv-change-L}
\cL[\psi, g] &= \frac{1}{2}\eta^{\rho\sigma} \langle\partial_\rho \otimes \partial_\sigma, \, dx^\mu\otimes dx^\nu \rangle
\partial_\mu (u \circ \psi) \partial_\nu (u \circ \psi) - V(u \circ \psi) \nonumber \\
&= \frac{1}{2} \partial_\mu (\phi - \epsilon \Box \phi) \partial^\mu (\phi - \epsilon \Box \phi) - V(\phi - \epsilon \Box \phi)
\\
&= \frac{1}{2} (\partial \phi)^2 - \epsilon \partial_\mu \Box \phi \partial^\mu \phi + \frac{1}{2}\epsilon^2 \partial_\mu \Box \phi \partial^\mu \Box \phi - V + \epsilon \Box \phi \frac{dV}{d\phi} + \frac{1}{2} \epsilon^2 (\Box \phi)^2 \frac{d^2 V}{d\phi^2}
+\mathcal{O}(\epsilon^3) \nonumber
\end{align}
exactly as one would expect from a derivative field redefinition, where we have here kept the potential $V(\phi)$ general. We stress that there is no obstacle to considering derivative field redefinitions in the geometric picture, once we consider a field redefinition to be a change of section (along which the unchanged metric is pulled back to form the Lagrangian).

Knowing that our bundle geometry construction can only capture Lagrangians with up to 2-derivatives, it is instructive to also expand this only up to 2-derivative terms:
\begin{align} \label{eq:deriv-redef-0jet}
\cL[\phi,g_E] \mapsto \cL[\psi,g_E] &= \frac{1}{2} (\partial \phi)^2\left(1 - 2\epsilon V^{\prime\prime}(\phi) \right)  - V(\phi)
+\mathcal{O}(4\partial)\, \nonumber \\
&= \frac{1}{2} (\partial \phi)^2\left(1 - \epsilon y\phi^2 \right)  - \frac{y}{4!}\phi^4
+\mathcal{O}(4\partial)\, ,
\end{align}
where we have also used IBPs, and the relation 
\be
\partial_\mu F(\phi) = \partial_\mu \phi \frac{dF}{d\phi}\, .
\ee
The result, truncating to 2-derivative order, is a $\phi$-dependent shift of the kinetic term.

Truncated to this order, the effect of doing this derivative field redefinition can be realised instead via a diffeomorphism on the bundle $E$. Because the field redefinition we seek to replicate is a perturbation around doing nothing, we assume the bundle morphism is also a perturbation around the trivial morphism, of the form (\ref{eq:f-expansion}).
By Poincar\'e invariance we also assume the bundle morphism has no explicit dependence on $x^\mu$, so we take
\be
(x^\mu, u) \mapsto (x^\mu, u+\epsilon \lambda(u))\, .
\ee 
Subbing into (\ref{eq:change-of-L-bundle-morph}), we get
\begin{nalign} 
    \cL[\phi, g_E] \mapsto \cL[\phi, f_E^\ast g_E] = \frac{1}{2}(\partial \phi)^2 \left(1+ 2\epsilon \lambda^\prime \right) - V - \epsilon \lambda V^\prime  + \mathcal{O}(\epsilon^2)  \, .
\end{nalign}
Thus, in this example, we can reproduce the effects (\ref{eq:deriv-redef-0jet}) of the derivative field redefinition, truncated to operators with 2 derivatives and up to 4 field insertions, by choosing the function $\lambda = -y\phi^3/6$. 

This is far from a generic example; for more general (2-derivative) Lagrangians, derivative field redefinitions cannot be described by base-point preserving bundle morphisms of the kind we have described -- but they can always be described as changes of section of the bundle. 
A relevant example here was discussed in Appendix E of~\cite{Cohen:2020xca}; there, starting from a theory with zero Ricci scalar on the field space, a derivative field redefinition in the EFT was performed (motivated by a simple non-derivative rotation in the UV), that was found to 
introduce a singularity in the Ricci scalar; this could not occur if the metric is pushed forward under a diffeomorphism (since a diffeomorphism must act invertibly on tensors).

\subsubsection{A limiting lemma} \label{sec:lemma}

It is likely not even possible to remedy this by allowing for {\em general} bundle morphisms $(f, \bar{f})$, as described by (\ref{eq:CD-bundle-morph-E}), which have a non-trivial morphism on the base manifold $\Sigma$. That is, allowing totally general bundle morphisms (even, in desperation, allowing for generic Poincar\'e violation) is still not enough to fully capture any old `derivative change of section'. A big clue to this is the following lemma, which is admittedly proven for the case of vector bundles (\cite{saunders1989geometry}, Lemma 2.2.9):
\begin{lemma} \label{lemma}
If $(E,\pi,\Sigma)$ and $(H,\rho,\Sigma')$ are vector bundles and $(f, \bar{f})$:  $(E,\pi,\Sigma) \to (H,\rho,\Sigma')$ is a vector bundle morphism such that $\bar{f}$ is a diffeomorphism, then $\tilde f: \Gamma((E,\pi,\Sigma)) \to \Gamma((H,\rho,\Sigma'))$ defined by $\tilde f (\phi) = f \circ \phi \circ \bar{f}^{-1}$ is a module homomorphism over the ring isomorphism $\bar{f}^{-1 *}: C^\infty (\Sigma) \to C^\infty (\Sigma')$
\end{lemma}
Crucially, a `derivative change of section' such as $\phi(x) \mapsto \psi = \partial_\mu \phi(x)$ (abusing notation in the now-familiar way), or indeed the specific change of section (\ref{eq:deriv-change-of-section}) considered in the previous example, is not a module homomorphism over the ring of smooth functions, because differentiation doesn't commute with multiplication. The lemma then implies it is not possible to find any bundle morphism $(f,\bar{f})$, where $\bar{f}$ is a diffeomorphism, such that $\phi(x) \mapsto \partial_\mu \phi(x)$, since that would require the map to be a homomorphism.

\bigskip

In this and the previous Section we have carefully reviewed and, we hope, refined the geometric picture of the 2-derivative term in scalar EFTs, and how one can think about field redefinitions geometrically as simply a change of section. In doing so we have laid the groundwork for extending this geometric approach to higher derivative terms. The key objects, mathematically, will be {\em jet bundles} of the original field space bundle $E$, and metrics that we will define thereon. Since jet bundles might be unfamiliar territory to most effective field theorists, in the next Section we review the concept and necessary definitions.
Our main reference is the textbook of Saunders~\cite{saunders1989geometry}.

\section{Jet Bundles} \label{sec:jets}

Starting from the `field space bundle' $(E,\Sigma, \pi)$, one can construct a sequence of associated manifolds of higher dimension, all themselves fibres bundles, called jet bundles. These jet bundles are indexed by an integer; there is a `1-jet bundle' denoted $J^1 E$, a `2-jet bundle' $J^2 E$, and so on; the `0-jet bundle' is just $E$ itself. 
For our humble physics purposes, $r$-jet bundles with higher $r$ will be needed to describe EFTs with more and more derivatives; the 1-jet bundle shall suffice to describe all terms (at least in SMEFT and HEFT) with {\em up to four derivatives}.\footnote{The 1-jet bundle also has well-known mathematical physics applications in formal treatments of Lagrangian mechanics. See {\em e.g.}~\cite{saunders1989geometry,de1998lie,Sardanashvily:2009br}.
}

Jets are built up from equivalence classes of sections, as follows. Starting from a fibre bundle $(E, \Sigma, \pi)$, with a local fibred coordinate system $(x^\mu,u^i)$ as introduced in \S \ref{sec:fib-bundles}, we say that two local sections $\phi, \psi \in \Gamma_x(\pi)$ are $1$-equivalent at the base point $x$ if both their values and the values of their first derivatives agree there -- but $\phi$ and $\psi$ can disagree in their higher derivatives. Recalling our notation $\phi^i(x) := (u^i \circ \phi)(x)$ for the `components' of a section (given local fibre coordinates $u^i$), then $\phi \sim \psi$ iff
\begin{align}
\phi^i(x) =\psi^i(x) \, , \qquad \text{and} \qquad
\frac{\partial \phi^i(x)}{\partial x^\mu} = \frac{\partial \psi^i(x)}{\partial x^\mu} \, .
\end{align}
It is easy to show that this equivalence relation is independent of the choice of coordinate chart on our patch of $E$.
The 1-equivalence class containing a representative section $\phi$ is called the {\em 1-jet of $\phi$ at $x$}, denoted $j_x^1 \phi$. In a similar way, two local sections are $r$-equivalent at $x \in \Sigma$, for some $r \in \Z_{\geq 0}$, if their first $r$ derivatives agree at $x$; the $r$-equivalence class containing $\phi$ is the $r$-jet of $\phi$ at $x$, denoted $j_x^r \phi$.

The 1-jet bundle $J^1 E$ is then defined to be the set of 1-equivalence classes of sections,
\be
J^1 E = \{j^1_x \phi \, | \, x \in \Sigma, \phi \in \Gamma_x(\pi) \}\, ,
\ee
which moreover has a natural structure as a differentiable manifold.
Continuing in a similar fashion, the $r$-jet bundle $J^r E$ is the set of $r$-equivalence classes of sections -- or, more prosaically, is the space of field configurations that agree in their first $r$ derivatives. The $0$-jet bundle is defined to be $J^0 E = E$ itself.

Focussing our attention on the 1-jet bundle $J^1 E$, 
from its definition one can define two (ordinary) fibre bundles with total space $J^1 E$. One is a bundle over the original base space $\Sigma$, and the other is a bundle over the original bundle $E$:\footnote{More generally, the $J^r E$ bundle can always be modelled as a fibre bundle over base space $J^{r-1} E$, or indeed as a fibre bundle over any $J^{q <r} E$. (There is natural inductive limit of the sequence $\{J^r E\}$ of jet bundles, that defines a space $J^\infty E$~\cite{Sardanashvily:2009br}. One can model $J^\infty E$ by sections that agree as $C^\infty$ maps -- we will not discuss this infinite order jet bundle here.) 
}
\begin{align}
\pi_1 &: J^1 E \to \Sigma \,: \,\,\, j^1_x \phi \mapsto x\, , \\ \label{eq:p1-bundle}
\pi_{1,0} &: J^1 E \to E : \,\,\, j^1_x \phi \mapsto \phi(x)\, .
\end{align}
These projections are sometimes referred to as the `source' and `target' bundles respectively, for obvious reasons.
These projections are compatible with the original bundle structure of $E$, in that they clearly fit into a commutative diagram
\begin{center}
\begin{tikzcd}
	& {J^1E} \\
	E && \Sigma
	\arrow["{\pi_1}", from=1-2, to=2-3]
	\arrow["\pi"', from=2-1, to=2-3]
	\arrow["{\pi_{1,0}}"', from=1-2, to=2-1]
\end{tikzcd}
\end{center}
involving the original projection $\pi:E\to \Sigma$. Thus, $\pi_1 = \pi \circ \pi_{1,0}$.

\subsection*{Local coordinates on the 1-jet bundle}

Recall that we start from local coordinates $(x^\mu, u^i)$ on the fibre bundle $E$, or more precisely on an open set $\pi^{-1}(\mathcal{U})\subset E$, and that a particular section $\phi:\Sigma\to E$ maps a point $x\in \Sigma$ to the point in $E$ with coordinates $(x^\mu, \phi^i(x^\mu))$.
The 1-jet bundle $J^1 E$ then admits a set of induced local coordinates
\be  \label{eq:J1Y-coords}
(x^\mu, u^i, u^i_\mu)\, .
\ee
The value of these coordinates at a point $j^1_x\phi \in J^1 E$ are
\begin{align}
x^\mu\circ j^1_x\phi &= x^\mu \, ,  &&\text{(spacetime points)}\\
u^i \circ j^1_x\phi &= \phi^i(x)\, ,  &&\text{(field values)} \\
u^i_\mu \circ j^1_x\phi &=\left. \frac{\partial \phi^i}{\partial x^\mu}\right|_{x}\,  &&\text{(first derivatives)}. \label{eq:deriv-coords}
\end{align}
Thus, the extra fibre coordinates in the 1-jet bundle $(J^1 E, \Sigma, \pi_1)$ encode the first derivatives of sections passing through that base point $x$ -- which is intuitive given the definition of a point $j^1_x \phi \in J^1 E$ as the collection of all sections passing through $x$ with the same value and first derivative there. 

One can model the fibres of the bundle $(J^1 E, E, \pi_{1,0})$, on which $u^i_\mu$ provide fibre coordinates, as vector spaces~\cite{Sardanashvily:2009br}
\be
\pi_{1,0}^{-1}(x\in \Sigma , m\in M_x) \cong T^\ast_x (\Sigma) \otimes T_m (M)\, .
\ee
This structure is encoded in the placement of indices of the derivative coordinate $u^i_\mu$; it transforms as a vector with respect to its $i$ index, and a co-vector with respect to its $\mu$ index. 
We will often call these extra coordinates $u^i_\mu$ `derivative coordinates'.

Summarising, the induced fibre coordinates on $(J^1 E, \Sigma, \pi_1)$, which evaluate on a section to $(\phi^i, \partial_\mu \phi^i)$, provide the familiar ingredients that we use to write down Lagrangians in field theory. The 1-jet bundle provides a coordinate-free description of this data. The geometry of $\Sigma$, $J^0 E = E$ and $J^1 E$, all of which play a big role in the EFT formalism developed in this paper, is summarized schematically in Figure~\ref{fig:drum}.

Lastly, it is easy to find the dimension of the 1-jet manifold. Starting with a general fibre bundle $E \to \Sigma$, with $\text{Dim} \Sigma = d$ and $\text{Dim} E - \text{Dim} \Sigma = n$ (the $\R$-dimension of the fibre, equivalently the number of real scalar fields), the $\{u^i_\mu\}$ provide $dn$ extra coordinates. Then
\be
\text{Dim} (J^1 E) = \text{Dim} E + dn = d+n+dn\, .
\ee
Let's count the dimensions of $J^1 E$ for some examples:
\begin{itemize}
\item Real scalar field in $d$ dimensions: $\text{Dim} (J^1 E)  = 2d + 1$ (see \S \ref{sec:4dRscalar});
\item HEFT or SMEFT: $\text{Dim} (J^1 E)  = 24$ (see \S \ref{sec:HEFTandSMEFT}).
\end{itemize}
Going further, the dimensions of the jet manifolds $J^r E$ quickly blow up when $d = \text{Dim} \Sigma$ is large; in general, it is easy to see that 
\be
\text{Dim} (J^r E) = d + n\frac{(d+r)!}{d! r!}\, .
\ee
For SMEFT or HEFT, we have $(J^2E) = 64$ for the 2-jet bundle.

\section{Higher-derivative EFTs from Jet Bundle Geometry} \label{sec:geometry}

The 1-jet bundle, introduced in \S \ref{sec:jets}, defines a bundle over spacetime with local fibre coordinates that we identify with scalar fields and their first derivatives. While these degrees of freedom are treated independently on the jet bundle, we always eventually pull back objects (like metrics) along sections, whereupon the relation between the field and its derivative is enforced ({\em i.e.} upon pullback $u^i_\mu$ will `remember' that it is a derivative of $u^i$). 

The key idea, from the EFT perspective, is that we can consider more general geometric structures involving higher-derivatives of fields in this larger space $J^1E$. In particular, we will see how defining different metrics on $J^1 E$, equivalently `measures of distance', encodes different scalar EFT Lagrangian with up to 4 spacetime derivatives acting on the fields. In this Section we outline how this works, for a general scalar EFT. In Sections~\ref{sec:examples} and~\ref{sec:HEFTandSMEFT}, we will apply this to specific examples including the Higgs EFTs, which is our primary interest.

\subsection{Defining geometry on the 1-jet bundle} \label{sec:natural-jet-geometry}

The subject of {\em geometry} on jet bundles is not so widely explored in the mathematical literature.
There are nonetheless `natural metrics' that are induced on jet bundles $J^r E$, given a metric on the original bundle $E \xrightarrow{\pi} \Sigma$ that we take to be, locally, a product metric $\eta \oplus g$, with $\eta$ flat (ignoring any potential contribution $V(u)$ for now) and $g(u)$ possibly curved, as described in \S\S \ref{sec:2-deriv} and \ref{sec:fib-bundles}. To give the reader a flavour for geometry on jet bundles, we begin with a brief sketch of such a natural geometry on $J^1 E$,\footnote{{\em Contact forms} provide another source of natural geometries on jet bundles, which can also be used to define a class of 4-derivative EFT Lagrangians using the ideas in this paper.} 
and see how it offers a starting point for constructing Lagrangians with 4-derivatives.

We want to define a Riemannian metric on $J^1 E$ considered as a manifold. We already saw that $J^1 E$ is itself a vector bundle $(J^1 E, E, \pi_{1,0})$ with fibre $T^\ast_x (\Sigma) \otimes T_m (M)$. Defining a metric on $J^1 E$ means equipping the tangent bundle $T(J^1 E)$ with an inner product 
\be
g^{(1)} : T (J^1E) \times_{J^1E} T (J^1E) \to \R
\ee
that is symmetric and non-degenerate, and that varies smoothly with the base coordinate $j_x^1 \phi$ (in other words, it is a tensor field on $J^1 E$).  We use a notation whereby $g^{(r)}$ denotes a metric on the $r$-jet bundle $J^r E$. 
Now, the tangent bundle $T (J^1E)$ is isomorphic to the pullback bundle\footnote{To show (\ref{eq:bundle_iso}), we need some theorems concerning the structure of the tangent bundle of such a vector bundle. There is a short exact sequence of smooth vector bundles on $E$, namely $0 \to VE \to TE \to HE \to 0$, where  the sub-bundle $VE := \text{ker~} d \pi$ is called the `vertical bundle', and the quotient bundle $HE = TE/VE$ is called the `horizontal bundle'. Both these bundles are isomorphic to pullback bundles along $\pi^\ast$; we have $VE \cong \pi^\ast E$, and $HE \cong \pi^\ast TY$, which gives rise to the isomorphism (\ref{eq:bundle_iso}).
}
\be \label{eq:bundle_iso}
T (J^1 E) \cong \pi^\ast (J^1 E \oplus TE)\, .
\ee
We can therefore build a (fibre-wise) inner product on $T (J^1 E)$ ({\em i.e.} a metric on $J^1 E$) out of (fibre-wise) inner products on the bundles $J^1 E$ and $TE$. The latter is just the metric on $E$, which we already stipulated is (locally) the product metric $\eta\oplus g$.
The former is not much more complicated, given the fibre is $T^\ast_x (\Sigma) \otimes T_\phi (X)$; we define the inner product via
$\eta^{-1}$ and $g$. Note the {\em inverse} spacetime metric appears here, because we are contracting co-vectors with respect to spacetime, but vectors with respect to the target space.

Putting things together, we can express this `natural' choice of metric, which recall is the one built from the metrics $\eta$ and $g$ that we already have at our disposal, in our local coordinate system:
\be \label{eq:first_g1}
g^{(1)} = \Lambda^4\, \eta_{\mu\nu} \,dx^\mu \otimes dx^\nu + g_{ij}(u)\, du^i \otimes du^j + \frac{1}{\Lambda^2}\eta^{\mu\nu} g_{ij}(u)\, du^i_\mu \otimes du^j_\nu\, .
\ee
We have introduced appropriate factors of $\Lambda$ to give everything consistent dimensions.

\subsection{Our first four-derivative Lagrangian}

Now, starting from this metric on the jet bundle $J^1 E$, which we intrepret as a `derivative-generalisation of field space', one can construct a Lagrangian using exactly the same recipe that we described for the 2-derivative Lagrangian in \S \ref{sec:2-deriv}.
The only difference is that we now need to pull the metric $g^{(1)}$ all the way back to spacetime, and so we should use a section $j^1\phi$ of the fibre bundle  $(J^1 E, \Sigma, \pi_1)$ introduced above. 

In particular, given an original section $\phi$ of $E$ ({\em i.e.} a scalar field configuration), let 
\be
j^1 \phi \in \Gamma_x(\pi_1)
\ee
be its `prolongation' to the 1-jet bundle, defined as the section of $\pi_1$ that passes through the 1-jet $j^1_x\phi$.\footnote{Note that not every section of the 1-jet bundle $\pi_1:J^1E \to \Sigma$ is a prolongation of a section $\phi$ of $\pi:E\to \Sigma$, where the latter is for us identified with a scalar field configuration. We are only interested in sections obtained in this way. }
Evaluating on the coordinate chart, its components are simply
\begin{align} \label{eq:j1phi-coords}
x^\mu\circ j^1\phi = x^\mu \, ,  \qquad 
u^i \circ j^1\phi = \phi^i(x)\, ,  \qquad
u^i_\mu \circ j^1\phi &=\left. \frac{\partial \phi^i}{\partial x^\mu}\right|_{x}\,  ,
\end{align}
from (\ref{eq:deriv-coords}).

We define the Lagrangian
\be \label{eq:4deriv_compact}
\cL_{4\partial}[\phi, g^{(1)}] := \frac{1}{2}\, \left\langle \eta^{-1} ,\, (j^1\phi)^\ast g^{(1)}\right\rangle \, ,
\ee
which we regard as a functional of the original section $\phi$.
Assuming the system of local coordinates set out in Eqs. \eqref{eq:J1Y-coords}--\eqref{eq:deriv-coords}, this gives the Lagrangian
\be \label{eq:4deriv_long_example}
\cL_{4\partial}[\phi, g^{(1)}] = \frac{1}{2} \eta^{\mu\nu} g_{ij}(\phi(x)) \partial_\mu \phi^i \partial_\nu \phi^j  + \frac{1}{2\Lambda^2} \eta^{\rho\sigma} \eta^{\mu\nu} g_{ij}(\phi(x)) \partial_\rho\partial_\mu \phi^i \partial_\sigma\partial_\nu \phi^j \, ,
\ee
up to a cosmological constant term coming from the first term in (\ref{eq:first_g1}) that we can neglect when doing flat-space quantum field theory (although its natural appearance seems intriguing nonetheless).

Let us pause to make some further comments. First and foremost, we have shown that defining a geometry on the 1-jet bundle naturally `refines' our original 2-derivative Lagrangian, as constructed from field space geometry in \S \ref{sec:2-deriv}, by a particular 4-derivative `correction' term that is also geometric in origin. This is already, to our knowledge, a novel result in the context of EFT.

We can try to unpack the particular Lagrangian that we have obtained, which is clearly not yet general, in the case of the Higgs EFTs. Let us therefore take $\Sigma$ to be flat 4d spacetime and the target space $M$ is a real 4-manifold equipped with the $SO(4)$ group action $\sS$ for SMEFT, say.
We can, for illustration, keep only terms in the Lagrangian (\ref{eq:4deriv_long_example}) that have {\em two field} insertions. This is equivalent to taking only the leading term in the expansion of the original $\sS$-invariant metric,
\be
g_{ij}(u^i) = \delta_{ij} + \dots\,
\ee
In this limit, and contracting the various indices, the Lagrangian (\ref{eq:4deriv_long_example}) becomes
\be
\cL = \frac{1}{2} \partial_\mu \phi \cdot \partial^\mu \phi  + \frac{1}{2\Lambda^2} \partial_\mu\partial_\nu \phi \cdot \partial^\mu\partial^\nu \phi  + \mathcal{O}(\phi^4)\, .
\ee
Interestingly, we recover the most general SMEFT Lagrangian with {\em up to two fields and up to four derivatives.}
The SM can, as for the purely 2-derivative case, be obtained by taking the cut-off scale $\Lambda \to \infty$, which decouples all the non-renormalisable operators from the EFT.

\subsection{General geometries for general Lagrangians}

In the previous two Subsections, we considered a metric on $J^1 E$ that is `natural', mathematically, given the geometry we started with on the field space bundle $E$. We found that this metric delivered a 4-derivative EFT Lagrangian, via the `obvious map' to Lagrangians specified by Eq. (\ref{eq:4deriv_compact}).
Unsurprisingly, if we start from such a naturally induced metric on $J^1 E$, this will only ever get us a 4-derivative term whose form is determined by the 2-derivative terms already present in the action. For a general EFT description, we of course want to consider 4-derivative terms that are independent of the 2-derivative terms. This suggests that our geometry on the 1-jet bundle $J^1 E$ should introduce {\em new} structures not already present in the geometry of the field space bundle $E$.

This poses no problem, and in a sense simplifies the picture.
One can shortcut the mathematical niceties of \S \ref{sec:natural-jet-geometry}, and instead take a more pedestrian approach that we are used to as physicists;  to wit, we should consider the {\em most general geometry on the 1-jet bundle that is consistent with symmetry}. This is, after all, the most natural approach to take in the spirit of EFT!

\subsubsection*{General Poincar\'e invariant 1-jet metrics}

We will, however, make rather stringent assumptions about the structure and symmetry of spacetime, to simplify our discussion (and specialise to those EFTs we are most interested in as high-energy physicists). As previously, we fix spacetime to be $\Sigma = \R^{1,3}$, flat Minkowski in $d$ spacetime dimensions with metric $\eta=\eta_{\mu\nu}\,dx^\mu \otimes dx^\nu=dt^2 - d\vec{x}\cdot d\vec{x}$.\footnote{Lest there is any confusion, the scalar field theory Lagrangian will be defined in terms of a geometry on the 1-jet bundle $\pi_1:J^1 E \to \Sigma$, and we equip $J^1 E$ with a geometry ({\em i.e.} a metric $g^{(1)}$) that is independent of the geometry on spacetime. In particular, the metric components in the `base space directions' of $J^1 E$ need not agree with those components of $\eta$, the metric on $\Sigma$.} 
We moreover enforce Poincar\'e symmetry on the jet bundle geometry. In practice, this means we must contract all $\mu$ indices (that derive from both the base space coordinates $x^\mu$ and the derivative coordinates $u^i_\mu$) to make Lorentz invariants, but we do not form contractions with $x^\mu$ vectors since that would break translations (nor can any explicit $x$ dependence appear in the metric components). We will discuss symmetries on the jet bundle again in \S \ref{sec:symmetries-1jet-bundle}, using the language of bundle morphisms (building on \S \ref{sec:bundle-morphism}).

So, let us write down the most general such metric, in our local coordinates. We keep all terms, including `cross-terms' that mix different types of jet bundle coordinates:
\begin{align}
\label{eq:general_g1}
g^{(1)}=&\begin{pmatrix} 
dx^\mu & du^i & du^i_\mu
\end{pmatrix} [g^{(1)}]
\begin{pmatrix}
dx^\nu\\ du^j\\ du^j_\nu
\end{pmatrix},
 &
 [g^{(1)}] &= 
 \begin{pmatrix}
  \Lambda^4 g_{\mu\nu}(\dots)\; & \Lambda^2 g_{\mu j}(\dots)& g^\nu_{\mu j}(\dots)
  \\[2mm]
  \Lambda^2 g_{\nu i}(\dots) & g_{ij}(\dots)& \dfrac{1}{\Lambda} g_{ij}^\nu(\dots)
  \\[2mm]
  g^\mu_{\nu i}(\dots)& \dfrac{1}{\Lambda} g_{ij}^\mu(\dots)& \dfrac{1}{\Lambda^2} g_{ij}^{\mu\nu}(\dots)
 \end{pmatrix}\,.
\end{align}
Because the derivative coordinates themselves have {\em pairs} of indices, one upstairs and one downstairs, it is convenient to use the same index notation for the components of the metric; thus, a component such as $g_{ij}^\mu$ or $g_{ij}^{\mu\nu}$ still has two downstairs jet bundle indices ({\em i.e.} all metric components are of course components of a $(0,2)$ tensor, never a 3- or 4-tensor), but taken with respect to one (or two) derivative coordinates respectively.
The $(\dots)$ record the fact that, in  general and before imposing any symmetry, each metric component is a functions of {\em all} the jet bundle coordinates -- including the `derivative coordinates'. Symmetry will of course restrict the form of these components dramatically.

Upon pulling back a general metric of the form (\ref{eq:general_g1}) along a section $j^1 \phi$ of the jet bundle $(J^1 E, \Sigma, \pi_1)$ and forming the Lagrangian (\ref{eq:4deriv_compact}), one obtains the following Lagrangian terms:
\begin{align} \label{eq:4deriv_long}
\cL[\phi, g^{(1)}] \,=\,\, &\eta^{\rho\sigma} \bigg[
\frac{1}{2}\Lambda^4 \delta^{\mu}_\rho\delta^\nu_\sigma\, g_{\mu\nu}(\phi,\partial \phi) 
+ \Lambda^2 \delta^{\mu}_\rho g_{\mu j}(\phi,\partial \phi)\, \partial_\sigma \phi^j
+ \Lambda \delta^{\mu}_\rho g_{\mu j}^\nu(\phi,\partial \phi)\,\de_\rho \partial_{\nu}\phi^i \nonumber \\
 +\, &\textcolor{magenta}{\frac{1}{2}
 g_{ij}(\phi,}\partial \phi\textcolor{magenta}{)\, \partial_\rho \phi^i \partial_\sigma \phi^j} 
+ \frac{1}{2\Lambda}  g_{ij}^\mu(\phi,\partial \phi)\, \partial_\rho \partial_\mu \phi^i \partial_\sigma \phi^j
+ \frac{1}{2\Lambda}  g_{ij}^\nu(\phi,\partial \phi)\,  \partial_\rho \phi^i \partial_\sigma \partial_\nu \phi^j
\nonumber\\
\,+& \frac{1}{2\Lambda^2} g_{ij}^{\mu\nu}(\phi,\partial \phi)\, 
\partial_\rho\partial_\mu \phi^i \partial_\sigma\partial_\nu  \phi^j\,
\bigg]\,. 
\end{align}
Let us unpack this general form of the Lagrangian a little.
First, we highlight in \textcolor{magenta}{magenta} the piece of this Lagrangian that is captured by the usual field space geometry picture, as in Eq. (\ref{eq:2deriv_long}). In fact, this $g_{ij}$ term is {\em already more general} than the 2-derivative Lagrangian (\ref{eq:2deriv_long}), because the component function $g_{ij}(\phi,\partial \phi)$ is itself a function of first derivatives, and so this term already includes a subset of 4-derivative EFT operators (and indeed operators with $>4$ derivatives), not just 2-derivative operators.

Another important remark to make is that we also capture terms with {\em fewer} than 2-derivatives from this geometry, coming from the first two terms on the RHS of (\ref{eq:4deriv_long}) -- we already saw this feature in \S \ref{sec:fib-bundles} when describing the 2-derivative EFT using the field space bundle picture (which is equivalent to formulating the theory geometrically on the `0-jet bundle' $J^0 E = E$).
In particular, the first term 
\be
\cL \supset \eta^{\mu\nu} \Lambda^4 g_{\mu\nu}(\phi,\partial \phi)
\ee
can encode an arbitrary scalar potential contribution; taking $g_{\mu\nu}=\frac{1}{d}\eta_{\mu\nu} V(\phi/\Lambda) + \dots$, and Taylor expanding, the preceding equation becomes
\be
\label{eq:potential}
\cL \supset \Lambda^4 \sum_{n=0}^{\infty} V_n \left(\frac{\phi}{\Lambda}\right)^n \, .
\ee
In the case of SMEFT or HEFT, when we impose $O(4)$ symmetry and so forbid terms with odd powers of $\phi$, this contains the (super-)renormalisable terms $c + \mu^2 \phi^2 - \lambda \phi^4$ appearing in the SM, where $c=\Lambda^4 V_0$, $\mu^2 = \Lambda^2 V_2$ and $\lambda = -V_4$.\footnote{Of course, tiny values are required for $V_0$ and $V_2$, given any viable EFT cutoff scale $\Lambda$; this manifests the (tree-level) cosmological constant and electroweak hierarchy problems.}
We emphasize that, in the usual geometric approaches, the scalar potential does not have a geometric origin but rather is added as an extra ingredient in defining the EFT -- see our discussion in \S \ref{sec:other_terms}.

Let us return to non-renormalisable operators, with more derivatives. 
Because all of the metric components in (\ref{eq:4deriv_long}) are arbitrary functions of $\phi$ and $\partial\phi$, the Lagrangian (\ref{eq:4deriv_long}) contains operators with arbitrary numbers of derivatives -- but not all such terms -- in a similar way to the usual geometric Lagrangian (\ref{eq:2deriv_long}) containing operators with arbitrary numbers of fields. The more precise statement here is that the 1-jet bundle geometric Lagrangian (\ref{eq:4deriv_long}) yields operators with
\be \label{eq:DvN_1jet}
D \leq N+2
\ee
where $D$ is the number of derivatives in the operator, and $N$ is the number of fields. 
The reason for the $+2$ on the RHS of (\ref{eq:DvN_1jet}) is the same reason that the ordinary field space geometry (equivalently, the $J^0 E$ case) describes 2-derivative terms, namely because there are already two derivatives due to $g$ being a $(0,2)$ tensor. 

We do not claim, in this fully general EFT without any symmetries, that we capture {\em all} operators satisfying (\ref{eq:DvN_1jet}); to answer this would require careful counting of non-redundant operators with unbounded mass dimension (since $N$ can be taken arbitrarily large), which is clearly difficult. Adapting Hilbert series techniques (see {\em e.g.}~\cite{Henning:2015daa,Henning:2017fpj}) to jet bundle `field space' might provide helpful tools for tackling such completeness questions (should they be interesting).
A more practical approach is to truncate the general Lagrangian (\ref{eq:4deriv_long}) and count {\em e.g.} all operators with a maximum number of derivatives. For the 1-jet bundle, we are able to count operators in the examples considered in \S \ref{sec:examples} and \S \ref{sec:HEFTandSMEFT}, including for HEFT and SMEFT, accounting for all integration by parts (IBP) redundancies; in all cases, we find the complete basis of operators with up to 4-derivatives is covered by the  Lagrangian (\ref{eq:4deriv_long}). In fact, we prove in general that by pulling back a 1-jet bundle metric one obtains a complete non-redundant basis of 0-, 2-, and 4-derivative operators for any scalar EFT. See Appendix~\ref{app:proof}.

Passing to higher jet bundles, one captures more and more operators (with more and more derivatives, relative to the number of fields); specifically, for an $r$-jet bundle geometry, one can obtain operators with
\be
D \leq rN+2
\ee
number of derivatives; those operators with $(r-1)N < D-2 < rN$ would {\em not} have already been captured by the $(r-1)$-jet bundle geometry.
We thus see how going to higher jet bundles systematically captures more and more operators in the EFT, and so provides a natural organising principle for the EFT expansion. Indeed, one can prove that the $r$-jet bundle geometry can capture a complete basis of operators in a general scalar EFT with up to $2(r+1)$ derivatives -- again, see Appendix~\ref{app:proof}.

We will use 1-jet bundles to study several examples of scalar field theories in \S \ref{sec:examples}, culminating in a geometric formulation of the Higgs EFTs,  including all operators containing up to 4 derivatives, in \S \ref{sec:HEFTandSMEFT}. Before we get into the nitty gritty of these examples, in the remainder of this Section we want to dig a bit deeper into how field redefinitions might be described in this context, and how the various notions of `bundle morphism' described in \S \ref{sec:bundle-morphism} lift to the 1-jet bundle description. This will also allow a more general formulation of symmetries on the jet bundles.

\subsection{Field redefinitions are (prolonged) changes of section} \label{sec:1jet-change-section}

To recap, in the `bundle formulation' of a scalar field theory that we described in \S \ref{sec:fib-bundles}, we identified scalar fields with local sections $\phi \in  \Gamma_x(\pi)$ of bundles $(E, \Sigma, \pi)$ over spacetime $\Sigma$ (\ref{eq:scalar-field-section}). A field redefinition, in general, was identified as a change of section $\phi \mapsto \psi \in \Gamma_x(\pi)$. Furthermore, for the special class of field redefinitions that are `non-derivative', this is equivalent to doing a base-point preserving bundle morphism $(f, \bar{f}) = (f_E:E\to E, \text{id}_\Sigma:\Sigma \to \Sigma)$.

Upon passing to the 1-jet bundle $(J^1, \Sigma, \pi_{1})$, which we suggest provides the right geometric ingredients for constructing EFT Lagrangians with more derivatives, the section $\phi$ is promoted to its prolongation $j^1 \phi \in \Gamma_{x}(\pi_1)$, a section of the 1-jet bundle. This is uniquely determined by the original section $\phi$, simply by $j^1 \phi(x) = j^1_x\phi$, the 1-jet of $\phi$ at $x$. The natural notion of a `field redefinition' in this picture is therefore sending
\be
j^1 \phi \mapsto j^1 \psi\, ,
\ee
the result of which is also a section of the 1-jet bundle 
$\pi_1$. 
We can think of this as a `prolongation' of the field redefinition $\phi \to \psi$. Since the Lagrangian will be defined by pulling back objects (a metric) from $(J^1 E, \Sigma, \pi_1)$ along a section, this tells us how we can implement the field redefinition at the level of the Lagrangian -- it is straightforward to check that the Lagrangian transforms in the `right way' (see \S \ref{sec:qm_1d_redef} for some more-or-less trivial examples). We believe this can accommodate any (possibly derivatively-dependent, but sufficiently smooth) field redefinition.

\subsection{Morphisms on the jet bundle} \label{sec:symmetries-1jet-bundle}

The notion of bundle morphisms that we gave in \S \ref{sec:bundle-morphism}, and used to describe non-derivative field redefinitions and (internal and spacetime) symmetries, can be promoted to a notion of morphism on the 1-jet bundle (which extends suitably to higher $r$-jet bundles), which we describe in this Section.

\subsubsection{Prolongation of bundle morphisms}

We start with the notion of bundle morphism $(f, \bar{f})$ introduced in Eq. (\ref{eq:CD-bundle}), where we straightaway fix both bundles to have the same base and total space.
Going to the 1-jet bundle $J^1 E$, the prolongated morphism $j^1 f$ is defined by its action on points in the jet bundle $J^1 E$, which recall are jets $j^1_x \phi$, by~\cite{saunders1989geometry}
\be \label{eq:prolong_f_general}
j^1 (f, \bar{f})\, (j^1_x \phi) = j^1_{\bar{f}(x)} (f \circ \phi \circ \bar{f}^{-1})\, 
\ee
where $\bar{f}^{-1}$ is the inverse of $\bar{f}$, which we assume to be a (local) diffeomorphism.
The commutative diagram (\ref{eq:CD-bundle}) is extended to
\begin{equation} \label{eq:CD-prolonged-general}
    \begin{tikzcd}
	{J^1E} && {J^1 E} \\
	E && E \\
	\Sigma && \Sigma
	\arrow["{\bar{f}}", from=3-1, to=3-3]
	\arrow["f", from=2-1, to=2-3]
	\arrow["{j^1 f}", from=1-1, to=1-3]
	\arrow["{\pi_{1,0}}", from=1-1, to=2-1]
	\arrow["{\rho_{1,0}}", from=1-3, to=2-3]
	\arrow["\pi", from=2-1, to=3-1]
	\arrow["\rho", from=2-3, to=3-3]
\end{tikzcd}
\end{equation}
Specialising to the case of bundle morphisms that act trivially on the base (\ref{eq:CD-bundle-morph-E}), {\em i.e.} taking $\bar{f} = \text{id}_\Sigma$, 
Eq. (\ref{eq:prolong_f_general}) simplifies to
\be \label{eq:prolong_f}
j^1 (f)\, (j^1_x \phi) = j^1_{x} (f \circ \phi)\, .
\ee
The commutative diagram (\ref{eq:CD-prolonged-general}) then simplifies to
\begin{equation} \label{eq:CD-prolonged}
    \begin{tikzcd}
	{J^1E} && {J^1E} \\
	E && E \\
	& \Sigma
	\arrow["\pi"', from=2-1, to=3-2]
	\arrow["\rho", from=2-3, to=3-2]
	\arrow["f", from=2-1, to=2-3]
	\arrow["{\pi_{1,0}}", from=1-1, to=2-1]
	\arrow["{\rho_{1,0}}", from=1-3, to=2-3]
	\arrow["{j^1 f}", from=1-1, to=1-3]
\end{tikzcd}
\end{equation}
Recall $\pi_1 = \pi_{1,0} \circ \pi$ and likewise $\rho_1 = \rho_{1,0} \circ \rho$.
For two sections $j^1 \phi \in \Gamma_x(\pi_1)$ and $j^1 \psi \in \Gamma_x(\rho_1)$, the `prolonged' commutative diagram implies the relation
\begin{equation}
\label{bundle morhpism relation sections jet} 
    j^1 \psi = j^1 f \circ j^1 \phi\, 
\end{equation}
between sections, analogous to (\ref{eq:bundle_morhpism_relation_sections}).

We can obtain the coordinate representation of the prolongated diffeomorphism $j^1 f$ by taking the composition with the various coordinate functions. For the original base and `field space coordinates' $x^\mu$ and $u^i$, we get
\begin{align}
    x^\mu \circ j^1 f &= x^\mu \circ \rho_1 \circ j^1 f = x^\mu, \label{eq:pro-x}\\
    u^i \circ j^1 f &= u^i \circ \rho_{1,0} \circ j^1 f = u^i \circ f \, . \label{eq:pro-u}
\end{align}
That is, the prolongated morphism $j^1 f$ acts exactly as $f$ on the $x^\mu$ and $u^i$ components (remembering our restriction that the bundle morphism $f$ acts trivially on the base space).
Lastly, for the derivative coordinates $u_\mu^i$, we can deduce their transformation under $j^1 f$ by considering the action on a particular jet $j^1_x\phi$:\footnote{We reiterate that all formulae here are for the specific subset of bundle morphisms that act trivially on the base.} 
\begin{nalign} \label{eq:pro-ux}
    (u_\mu^i \circ j^1 f) (j^1_x \phi) &=  \partial_\mu \left(u^i \circ (f \circ \phi)\right) \big|_x \\
    & = \partial_\mu (u^i \circ f) \big|_x + \partial_\mu (u^j \circ \phi) \big|_x \frac{\partial(u^i \circ f)}{\partial u^j} \bigg|_x \\
    & = \partial_\mu (u^i \circ f) \big|_x + u_\mu^j \,\, \frac{\partial(u^i \circ f)}{\partial u^j} \bigg|_x\, .
\end{nalign}
One can think of this as a `total derivative' of the transformed coordinate functions $u^i \circ f$; it follows essentially from applying the chain rule. To summarize, we can write the prolongated bundle morphism as:
\begin{equation} \label{eq:prolonged-morphism}
    j^1 f : (x^\mu, u^i, u_\mu^i) \mapsto (x^\mu, f^i, \partial_\mu f^i + u_\mu^j \partial_j f^i)\, ,
\end{equation}
in our local fibred coordinates, where $f^i(x^\mu, u^j) = u^i \circ f$ as in \S \ref{sec:bundle-morphism}.

The construction we have described for prolonging a bundle morphism from $\pi:E\to \Sigma$ to its 1-jet bundle $\pi_1:J^1\to \Sigma$ can be extended straightforwardly to the higher-jet bundles. Since on the physics side we focus on $J^1 E$ in this paper, we omit details of this construction here.

\subsubsection{Prolonging internal and spacetime symmetries to the 1-jet bundle} \label{sec:prolonging-symmetres-1jet-bundle}

In \S \ref{sec:symms-on-bundle} we defined notions of internal and spacetime symmetries as bundle morphisms satisfying particular conditions. Namely, all symmetries leave the Lagrangian  invariant; internal symmetries are base-point preserving bundle morphisms, which in local coordinates can be expressed as $f_g: (x^\mu, u^i) \mapsto (x^\mu, f^i_g(u^i, x^\mu))$; spacetime symmetries have a non-trivial action on the base space $\Sigma$ of the bundle. We discussed the particular cases of the Higgs EFTs in \S \ref{sec:symms-on-bundle}.

How do these symmetries act on the 1-jet bundle? Having phrased symmetries in the above language of bundle morphisms, there is an obvious way to `lift' them to the 1- and higher-jet bundles using the notion of prolongation, that we have introduced in this Section. 

\subsubsection*{Prolonging an internal symmetry}

For an internal symmetry, we can directly apply the formula (\ref{eq:prolonged-morphism}) for the prolongation of a base-point preserving bundle morphism, which we have for each group element $g \in G$, our symmetry Lie group. The set of prolonged symmetry morphisms are\footnote{As usual, all morphisms and prolongations thereof are understood to be only {\em local} here; the formulae we write are valid only in certain open submanifolds coordinatized by certain charts; in other words, all manifolds appearing in commutative diagrams such as (\ref{eq:CD-prolonged}) should be understood as open submanifolds of the objects written. }
\begin{equation} 
    j^1 f_g : (x^\mu, u^i, u_\mu^i) \mapsto (x^\mu, f_g^i, \partial_\mu f_g^i + u_\mu^j \partial_j f_g^i)\, , \quad \forall g \in G
\end{equation}
For the SMEFT/HEFT symmetries, we can, at least in a coordinate patch, write $f_{O\in O(4)}^i = \sigma(O, u^i)$, for $\sigma = \sS$ or $\sH$ in the case of SMEFT or HEFT respectively. Let us unpack this further in the case of SMEFT, for illustration, in which the group action in our coordinate chart of choice is linear:
\begin{equation}
    f^i = O^i_j u^j\, ,
\end{equation}
where $O^i_j$ is a constant $O(4)$ matrix in the fundamental representation. Notice that, in addition to being an {\em internal} symmetry, it is also a {\em global} symmetry, in the usual physicist's sense, which is a restricted kind of base-point preserving morphism. (It is the same restriction we have often imposed when viewing field redefinitions as morphisms, where we appealed to Poincar\'e invariance as our justification.)
Thus, $\partial_\mu f^i = 0$ while $\partial_j f^i = O^i_j$. The prolongation of the internal $O(4)$ symmetries to the 1-jet bundle is thus given by the set of morphisms:
\begin{equation} 
    \{j^1 f_O : (x^\mu, u^i, u_\mu^i) \mapsto (x^\mu, O^i_j u^j,  O^i_j u_\mu^j)\qquad \forall O \in O(4)\, \}\, .
\end{equation}
exactly as one would na\"ively expect.
In practice, this means that one forms $O(4)$-invariants (say, when constructing invariant metrics on the 1-jet bundle, and thence invariant Lagrangians) by contracting the upper case $i$ indices wherever they appear both in the $u^i$ coordinates {\em and} in the derivative coordinates $u^i_\mu$. The simplicity of this group action, which could hardly have been otherwise, will be frequently invoked in \S \ref{sec:HEFTandSMEFT} when we finally turn to constructing the 4-derivative Higgs EFT Lagrangians.

\subsubsection*{Prolonging a spacetime symmetry}

Starting from the field space bundle $(E, \Sigma, \pi)$, a spacetime symmetry is a collection of bundle morphisms $(f, \bar{f})$ for which $\bar{f}$ is a non-trivial diffeomorphism on the base space. The Poincar\'e group action for our Lorentz-invariant EFTs is still specified by Eq. (\ref{eq:Poincare-on-E}). The prolonging of a spacetime symmetry requires the more general version shown in Eqs. (\ref{eq:prolong_f_general}, \ref{eq:CD-prolonged-general}). 
This requires the generalisation of Eqs. (\ref{eq:pro-x}--\ref{eq:pro-ux}) to the case where $\bar{f}$ is non-trivial. (We refer the interested reader to Section 4.2 of the book~\cite{saunders1989geometry} for these more general formulae, which we only use explicitly here). The prolongation of the Poincar\'e symmetry morphisms (\ref{eq:Poincare-on-E}) gives:
\begin{equation} \label{eq:prolonged-Poincare-formula}
    \{j^1 f_{a,L} : (x^\mu, u^i, u_\mu^i) \mapsto (L^\mu_\nu x^\nu + a^\mu, u^i, (L^{-1})^\nu_\mu u_\nu^i)\qquad \forall (a, L) \in O(1,3)\, \}\, .
\end{equation}
Notice that while the Poincar\'e action is non-linear on the $x^\mu$ coordinates, including the effects of translations, this does not appear in the group action on the derivative coordinates upon prolonging the symmetry. Moreover, notice that it is the inverse matrix $L^{-1}$ that appears multiplying the derivative coordinates. Again, as for the $O(4)$ internal indices, the straightforward action of the prolonged spacetime symmetry on the derivative coordinates means that we can treat the $\mu$ index exactly as we would any other Lorentz index when forming Lorentz-invariant contractions; namely, we should contract with the Minkowski metric $\eta^{\mu\nu}$ on spacetime, as we do everywhere in the examples that follow in \S \ref{sec:examples} and \S \ref{sec:HEFTandSMEFT}.

\subsubsection{Non-derivative field redefinitions (last time!)} \label{sec:last-time}

Again, analogous to the situation we described in \S \ref{sec:bundle-morphism}, {\em not any} change of prolonged section $j^1 \phi \to j^1 \psi$ can be obtained by prolonging a bundle morphism, as in (\ref{eq:CD-prolonged}). 
Indeed, the same considerations of \S \ref{sec:bundle-morphism}, noting in particular the lemma discussed in \S \ref{sec:lemma}, suggest that a morphism $j^1 f$ on the 1-jet bundle cannot reproduce the effects of a derivative field redefinition (except under special circumstances, analogous to those considered in Example \ref{sec:phi4}),\footnote{Later in \S \ref{sec:4dRscalar}, we will perform an analogous exercise to that in \S \ref{sec:phi4} ; for a scalar field theory at the 4-derivative level, which is described by geometry on the 1-jet bundle, we ask when (if ever) a prolonged bundle morphism can reproduce the effects of a {\em derivative} field redefinition. }
but can reproduce the effects of non-derivative field redefinitions. It is important to check, as we do in this Subsection, that non-derivative field redefinitions are still implemented consistently by prolonging the bundle morphism $f$ to $j^1 f$.

Recall that a general Lagrangian (with up to 4 derivatives) is, in this formulation, obtained by pulling back a metric $g^{(1)}$ from the 1-jet bundle to $\Sigma$ along a section $j^1 \phi \in \Gamma_x(\pi_1)$, given an initial section $\phi \in \Gamma_x(\pi)$, before contracting with the inverse metric on spacetime:
$$\cL[\phi, g^{(1)}] = \frac{1}{2}\langle \eta^{-1}, (j^1 \phi)^\ast g^{(1)}\rangle\, .$$
Doing a prolongated bundle morphism $j^1 f: J^1 E \to J^1 E$ on the 1-jet bundle, obtained from a morphism $f:E \to E$ via (\ref{eq:CD-prolonged}), which is itself a diffeomorphism on $J^1 E$ considered as a manifold, the metric changes via the usual transformation law
\begin{equation} \label{eq:prolong-morph-metric}
    g^{(1)} = g^{(1)}_{IJ}({\bf x}) d{\bf x}^I \otimes d{\bf x}^J \mapsto (j^1 f)^\ast g^{(1)} = g^{(1)}_{KL}({\bf x} \circ j^1f) \frac{\partial {\bf j^1 f}^K}{\partial {\bf x}^I} \frac{\partial {\bf j^1f}^L}{\partial {\bf x}^J} d{\bf x}^I \otimes d{\bf x}^J\, ,
\end{equation}
where we employ a compact notation in which 
\be
\{{\bf x}^I\} = \{x^\mu, u^i, x_\mu^i\}
\ee
runs over all the 1-jet bundle coordinates, and here ${\bf j^1 f}^I = {\bf x}^I \circ j^1 f$. When we pull back the transformed metric along the {\em original} section $j^1 \phi$, we get 
\begin{nalign}
\label{Transformation of metric under diffeomorphism}
    (j^1 \phi)^\ast (j^1 f)^\ast (g^{(1)}) 
    = g^{(1)}_{KL}({\bf x} \circ j^1f \circ j^1 \phi) &\frac{\partial ({\bf x}^K \circ j^1f \circ j^1 \phi)}{\partial {\bf x}^I} \frac{\partial ({\bf x}^L \circ j^1f \circ j^1 \phi)}{\partial {\bf x}^J} \\ 
    &\times d({\bf x}^I \circ j^1\phi) \otimes d({\bf x}^J \circ j^1 \phi) \, .\nonumber
\end{nalign}
This can be expanded out in the various pieces, analogous to the expansion (\ref{eq:4deriv_long}), using Eqs. (\ref{eq:prolonged-morphism}) and (\ref{eq:j1phi-coords}) to evaluate objects like ${\bf x} \circ j^1f \circ j^1 \phi$, but the result would be very cumbersome and so we do not write it explicitly.

\subsubsection{Example: free particle quantum mechanics} \label{sec:qm_1d_redef}

To see a more explicit expression, it is helpful to lose this full generality and consider a specific example. We consider the quantum mechanics of a single real scalar field, described by a $1$d sigma model on $M=\R$ (we will treat this example in full generality in \S \ref{sec:toy}), for which we view a field as a section $\phi$ of a bundle $(E\cong \R_t \times \R_u, \R_t, \pi:(t,u)\mapsto t)$. The 1-jet bundle is just 3-dimensional, on which we specify an initial fibred coordinate system $(t, u, u_t)$. We start from the simplest viable Lagrangian, just $\cL = \frac{1}{2}\dot\phi^2$ that describes a free particle. As mentioned at the end of \S \ref{sec:toy}, one valid choice of metric is\footnote{One cannot take $g^{(1)} = dt dt + du du$ on the 1-jet bundle, because that is not invertible and so does not define a metric.} 
\begin{equation} \label{eq:QM-free-metric}
g^{(1)}=u\, dt\, du_t + 2\, du\, du  \, .  
\end{equation}
To recap, given a section $\phi:t \mapsto (t, \phi(t))$ which we prolong, this gives the Lagrangian 
\be
\cL[\phi, g^{(1)}] = \frac{1}{2}\langle \eta^{-1}, (j^1 \phi)^\ast g^{(1)} \rangle =  \frac{1}{2}\phi \ddot\phi + \dot\phi^2 \sim \frac{1}{2}\dot\phi^2
\ee
after using IBPs. 

Now, consider the base-point preserving bundle morphism
\begin{equation}
    f:(t,u) \mapsto (t,u^2)\, ,
\end{equation}
which, similar to the field theory example considered in \S \ref{sec:field_redef_1}, implements a non-derivative field redefinition sending our initial section $\phi$ to a section $\psi$ whose components satisfy $\psi = \phi^2$.
Applying Eq. (\ref{eq:prolonged-morphism}), the prolongation of this morphism to the 1-jet bundle is
\begin{equation}
    j^1 f:(t,u,u_t) \mapsto (t,u^2, 2u u_t)\, ,
\end{equation}
as expected from the chain rule.
Now apply this morphism to the metric $g^{(1)}$, using (\ref{eq:prolong-morph-metric}), to obtain
\begin{equation}
    g^{(1)} \mapsto (j^1 f)^\ast g^{(1)} = 2u^2 u_t dt \, du + 2u^3 dt \, du_t + 8 u^2 du \, du \, . 
\end{equation}
Finally, pulling this back and forming the transformed Lagrangian gives 
\be
\cL[\phi, (j^1 f)^\ast g^{(1)}] = \frac{1}{2}\langle \eta^{-1}, (j^1 \phi)^\ast (j^1 f)^\ast g^{(1)} \rangle =  \phi^2\Dot{\phi}^2 + \phi^3 \Ddot{\phi} + 4 \phi^2 \Dot{\phi}^2 
\buildrel {\text{IBPs}}\over\sim 
2 \phi^2 \Dot{\phi}^2 \, ,
\ee
after doing IBPs. We see that the formulae for prolonging the bundle morphism to the 1-jet, where the original morphism corresponds to a particular non-derivative field redefinition, gives the consistent transformation of the Lagrangian when computed using a 1-jet metric. We expect the same applies for any non-derivative field redefinition, which we would write as $\phi \mapsto \psi=\phi^2$ at the level of coordinates implemented via the general formula (\ref{eq:prolonged-morphism}), and for general Lagrangians (with up to 4-derivatives).

\subsubsection*{Derivative field redefinitions}

We emphasize that, mirroring our discussion in \S \ref{sec:bundle-morphism} of 2-derivative Lagrangians using the field space bundle picture, one can always implement more general derivative field redefinitions just by pulling back the metric $g^{(1)}$ along a different section $j^1 \psi$, as described in \S \ref{sec:1jet-change-section}; but, in this more general situation, one loses the equivalent description in terms of a prolongated morphism on the jet bundle.

We can exhibit this by considering an explicit derivative field redefinition, still in the context of this QM example for simplicity. 
Consider a change of section $\phi \to \psi$ which, in components, is defined by $\phi \to \psi = \phi - \epsilon\ddot \phi$, analogous to the 
field theory example (\ref{eq:deriv-change-of-section}).
Starting from the same metric (\ref{eq:deriv-change-of-section}) describing the free particle (when pulled back along $j^1 \phi$), 
but pulling back along the different section $\psi$, we get
\begin{align}
    \cL[\phi, g^{(1)}] \mapsto \cL[\psi, g^{(1)}] &= \frac{1}{2}\langle \eta^{-1}, (j^1 \psi)^\ast g^{(1)} \rangle \\
    &=  \frac{1}{2} \langle \partial_t \otimes \partial_t, dt\otimes dt \rangle \left((u \circ j^1 \psi)\partial_t(u_t \circ j^1 \psi) + 2 \partial_t(u \circ j^1 \psi)\partial_t(u \circ j^1 \psi)  \right)\nonumber \\
    &=  \frac{1}{2} (\phi-a\ddot\phi)\partial_t(\dot\phi-a\dddot\phi) + \partial_t(\phi-a\ddot\phi)^2  \nonumber\\
    &= \frac{1}{2}\left(\phi \ddot\phi - a\ddot\phi^2 - a\phi \ddddot\phi + a^2\ddot\phi\ddddot\phi \right) + \left(\dot\phi - a\dddot\phi \right)^2 \buildrel {\text{IBPs}}\over\sim \frac{1}{2}\left(\dot\phi - a\dddot\phi \right)^2\, .\nonumber
\end{align}
As for the example considered above in \S \ref{sec:phi4}, the formulae for implementing this field redefinition are essentially trivial. Changing section corresponds to `subbing in' ${\bf x} \circ \psi$ everywhere in place of ${\bf x} \circ \phi$, which are precisely the manipulations one would do to `change fields' at the level of the Lagrangian.

\section{Examples}\label{sec:examples}

In this Section we consider concrete examples of scalar field theories, to better acquaint the reader with the jet bundle formalism described in \S \ref{sec:jets}, and the notion of geometry thereon (\S \ref{sec:geometry}). The examples start simple and increase in complexity.

\subsection{Quantum mechanics on the line} \label{sec:toy}

 We begin with an example that could not be much simpler, namely the quantum mechanics of a point particle moving on the real line $M \cong \R$, which is described by a $(0+1)$-dimensional sigma model on $\R$. The 1-jet bundle is here only 3-dimensional, meaning it is feasible to write down metrics explicitly and thereby get a feel for the general formalism.

It is worth remarking that, in $0+1$ dimensions, the scalar field $\phi$ has negative (classical) mass dimension, and so operators with more insertions of the field $\phi$ become more and more relevant (in contrast to the story we are used to in $3+1$ dimensions). Therefore, the `EFT expansion' we consider in this Section should be regarded as a formal construction, in which we expand in both number of fields and number of derivatives, not to be interpreted as an expansion in increasingly irrelevant operators.

The sigma model field is simply a real function
$\phi: \R \to \R: \, t \mapsto \phi(t)$.
For our symmetries, we take a pair of $O(1) \cong \Z_2$ factors that act as inversions on both the source and target lines:
\begin{align}
T &: t \mapsto -t,  &&\text{(time reversal)}\, , \\
P &: \phi \mapsto -\phi,  &&\text{(parity)}\, .
\end{align}
This defines a simple toy quantum field theory to which we can apply the jet bundle formalism, which nonetheless is structurally similar to the Higgs EFTs; the $O(1)_t \times O(1)_u$ symmetry, like the $O(3,1)_\Sigma \times O(4)_{M}$ of the HEFT or SMEFT, allows only terms with even numbers of derivatives, and even numbers of fields. 

The first step is to pass from the formulation in terms of maps to field space $M \cong \R$, to the formulation in which the scalar field is a section of a bundle $E$.
For simplicity, let us take the field space bundle here to be (globally) a product manifold, $E = \R_t \times \R_u \to \R_t$, on which our field 
\be
\phi:t \mapsto (t, \phi(t))
\ee
is a section.
The 1-jet bundle $J^1 E$ can then be modelled as the real 3-manifold
\be
J^1 E \cong T^\ast \R_t \times \R_u\, ,
\ee
with local fibred coordinate system
\be \label{eq:toy-coords}
y^I = \{t, u, u_t\}\, .
\ee
Along the section $j^1\phi$ of $J^1 E \to \Sigma$, these coordinates evaluate to
\begin{align}
t \circ j^1 \phi &= t, \\
u \circ j^1 \phi &= \phi(t), \\
u_t \circ j^1 \phi &= \dot\phi(t)\, .
\end{align}
Following our general formalism,
an invariant metric on this 1-jet bundle can be used to define a general 4-derivative Lagrangian for this system. 

Before writing down the 1-jet metric, it is helpful to know where we want to end up: using the usual procedure, let us first write down the most general Lagrangian for this toy EFT with up to 4-derivatives in a non-redundant basis, consistent with our pair of $O(1)$ symmetries.
At the 2-derivative level, there are two independent operators $\phi \ddot \phi$ and $\dot\phi^2$, one of which can be removed (we remove $\phi \ddot \phi$) by an IBP relation. At the 4-derivative level, there are 5 operators and 3 IBP relations. We choose to remove the operators 
$\ddddot\phi$ and $\dddot\phi \dot\phi$, which cannot in fact be written using the 1-jet bundle (they contain too many derivatives acting on a single field), plus the operator $\ddot \phi \dot\phi \dot\phi$, to leave a non-redundant basis spanned by the 4-derivative operators $\dot\phi^4$ and $\ddot\phi^2$. In general all these operators (including the 0-derivative constant piece) are multiplied by arbitrary even functions of $\phi$.

Now let us see how to recover this general Lagrangian using the 1-jet bundle described above. We write down the most general metric
\be
g = g_{IJ} dy^I \otimes dy^J
\ee
on $J^1 E$ that is
consistent with our $O(1)_t \times O(1)_u$ global symmetry. We truncate each metric component to include all contributions that will pull back to Lagrangian terms with up to and including four time derivatives.
We have:
\begin{align}
g_{u_t u_t} &= A(u), \label{eq:metricA} \\
g_{u u_t} &= u_t u B(u), \\
g_{uu} &=  C(u) + u_t^2 D(u), \\
g_{t u_t} &= u E(u) + u u_t^2 F(u), \\ 
g_{t u} &= u_t G(u) + u_t^3 H(u) , \\
g_{tt} &= -2V(u) + u_t^2 J(u) + u_t^4 K(u) , \label{eq:metricZ}
\end{align}
where $A(u)$, ... $V(u)$ are even functions of $u$, as dictated by our symmetry. Written as a symmetric matrix of components:
\be
\label{general metric toy model}
[g_{IJ}] = 
\begin{pmatrix}
-2V+u_t^2J+u_t^4K & \quad u_t G + u_t^3 H& \quad u E + u u_t^2 F \\
\cdot & C+u_t^2 D & u u_t B \\
\cdot & \cdot & A
\end{pmatrix}\, ,
\ee
in our local fibred coordinates (\ref{eq:toy-coords}).

To form a Lagrangian, we pull back this metric along the section $j^1\phi$, and contract with the inverse metric on the 1d spacetime:
\be \label{eq:cLofg}
\cL[\phi, g] = \frac{1}{2}\left\langle \eta^{-1},\, (j^1\phi)^* g \right\rangle\, .
\ee
Here, the inverse metric is just 
$$\eta^{-1} = \partial_t \otimes \partial_t.$$ 
The resulting object is a scalar function on the worldline, {\em i.e.} a function of $t$, that is our Lagrangian. We get
\begin{align} \label{eq:toy-general}
\cL[\phi, g] = 
\frac{1}{2}\left\langle \eta^{-1},\, (j^1\phi)^* g \right\rangle = &
\frac{1}{2}g_{tt} + g_{tu}\dot{\phi} + g_{tu_t}\ddot{\phi} + \frac{1}{2}g_{uu} \dot{\phi}^2
+ g_{uu_t} \dot{\phi}\ddot{\phi} + \frac{1}{2}g_{u_tu_t}\ddot{\phi}^2
\\
= &-V(\phi) +\frac{1}{2} \dot\phi^2 (C+2G+J) + \ddot\phi \phi E \\ \nonumber
&+ \frac{1}{2}\dot\phi^4 (D+2H+K) + \ddot\phi \dot\phi^2 \phi(B+F)+\frac{1}{2}\ddot\phi^2 A\, .
\end{align}
where $V, A,B\dots K$ are implicitly understood to be even functions of the field: $V(\phi), A(\phi)$, etc.
As anticipated, we do not obtain the operators $\ddddot\phi$ and $\dddot\phi \dot\phi$ with more than two derivatives on a single field, but this is not an issue because those operators can be removed from the Lagrangian using IBPs. 

We do, however, obtain 2 (3) operator structures with 2 (4) derivatives, only 1 (2) of which are independent, as discussed.
The extra operators can be removed via IBP:
\begin{align}
\ddot\phi\phi E &= - \dot\phi^2E-\dot\phi\phi \,\de_tE 
=- \dot\phi^2E - \dot\phi\phi\,2\phi\dot\phi\frac{dE}{d(\phi^2)}
=-\dot\phi^2(E+2\phi^2 E')
\,,
\\
\ddot\phi\dot\phi^2\phi (B+F) &= -\frac{1}{3}\dot\phi^4\left(B+F+2\phi^2(B'+F')\right)\,,
\end{align}
where $E'=dE/d(\phi^2)$ and analogously for $B',F'$. With this notation, $B',E',F'$ are even functions of $\phi$.
The Lagrangian (\ref{eq:toy-general}) can now be written in a non-redundant basis:
\begin{align} \label{eq:toy-reduced}
\cL[\phi,g] =& \text{IBP} \circ \frac{1}{2}\left\langle \eta^{-1},\, (j^1\phi)^* g \right\rangle \\
= & -V + \frac{1}{2}\dot\phi^2 (C+2G+J-2E -4\phi^2 E') 
\\
&+ \dot\phi^4 \left(\frac{1}{2}(D+K)+H-\frac{1}{3}(B+F)-\frac{2}{3}\phi^2(B'+F') \right) +\frac{1}{2}\ddot\phi^2 A \,.\nonumber
\end{align}
In fact we know a little more; the 2-derivative term must be non-vanishing, and its leading order piece is fixed by canonical normalisation of $\phi$. We require
\be
C(0) + 2G(0) +J(0) - 2E(0) = 1\, .
\ee
This defines a map from metric $g$ to $\cL[\phi,g]$, the Lagrangian functional (of $\phi$), in a particular basis with no residual IBP redundancies.
The map is, importantly, surjective, meaning that the most general Lagrangian with up to 4-derivatives can be captured by pulling back a metric on the 1-jet bundle.

The map $g \mapsto \cL[\phi,g]$ is not, however, injective.
This is a general feature that will recur in every example: there are redundancies in the map from 1-jet metrics to Lagrangians, in the sense that many different metrics map to the same Lagrangian. A natural next step is to try to define equivalence classes of metrics that map to the same Lagrangian (up to a total derivative), in order to fix up a map $\cL[\phi,g]$ that is both surjective and injective. Generally, two metrics $g$ and $\tilde g$ are equivalent iff
\be
\eta^{-1} (j^1 \phi)^\ast (g-\tilde g) = \partial_\mu K^\mu\, .
\ee 
However, it is not so straightforward to make a consistent choice of representative for each such equivalence class, as is made apparent even in this toy example where we can explicitly `read off' the redundancies. For example, one might try to define representative metrics by setting the functions $G$, $J$, $E$ to zero (capturing the 2-derivative term via $C$), and setting $H$, $K$, $B$, and $F$ to zero also (capturing the $\dot\phi^4$ term via $D$). But this can lead to pathologies when trying to describe certain (physically reasonable) Lagrangians, such as the free theory $\cL = \frac{1}{2}\dot\phi^2$; with the `representative' above, the `metric' would have components $[g_{IJ}] = \text{diag}(0,1,0)$, which is singular and so not a metric at all. A valid metric that does map to this Lagrangian, accounting for IBPs in the manner described above, would be $[g_{IJ}] = \text{diag}(u/2,2,u/2)$.

\subsection{Real scalar in 4d} \label{sec:4dRscalar}

We now consider the case of a single real scalar field in Minkowski spacetime. We consider the scalar field $\phi$ to be a section of the bundle $\pi: E=\R^{1,3}\times \R_u\to\R^{1,3}$:
\begin{equation}
    \phi: x^\mu \mapsto \left(x^\mu,\phi(x^\mu)\right)\,.
\end{equation}We require Poincar\'e invariance, but do not impose any internal symmetries on the field. The 1-jet bundle $J^1 E$ is a real 9-manifold
\begin{equation}
    J^1 E \cong T^*\R^{1,3} \times \R_u\,,
\end{equation}
with local fibred coordinates
\begin{equation}
    y^I = \{x^\mu, u, u_\mu\}\,.
\end{equation}
Along the section $j^1\phi$ of $J^1E\to \Sigma$, they evaluate to
\begin{align}
x^\mu\circ j^1\phi &= x^\mu,\\
u \circ j^1\phi &= \phi(x^\mu),\\
u_\mu\circ j^1\phi &= \de_\mu\phi(x^\mu)\,.
\end{align}
Following the same steps as in the previous example, we first write down the Lagrangian we wish to obtain. 

In this case, we are after a complete Green's basis for a scalar EFT, {\em i.e.} a set of operators that is non-redundant under IBP (but can in general contain redundancies under Equations-of-Motion),  with up to 4 derivatives and arbitrarily large number of field insertions. We also require no more than 2 derivatives acting on each field and that boxes $\square=\de_\mu\de^\mu$ are absent. 
The former condition is necessary, because it is not possible to obtain such structures from a 1-jet bundle metric.
The complete absence of operators with boxes is not strictly required by the metric structure, but turns out to be a very convenient choice: the Lagrangian obtained pulling back the most general 1-jet bundle metric contains \emph{all} box-less operators with up to 4 derivatives, but only a subset of those with boxes, see~\S\ref{app:proof}. 
Restricting to box-less bases guarantees that the scalar Lagrangian will be manifestly matched by pulling back the 1-jet bundle metric, without further manipulations and without introducing cumbersome evaluations of which operators with boxes can or cannot be obtained.
The fact that all operators with boxes can always be removed from a Green's basis is not obvious a priori,  but can be proven explicitly for any scalar theory at any derivative order, see \S\ref{app:proof}.

In the case of a single real scalar with interactions with up to 4 derivatives, the conditions listed above identify unambiguously:\footnote{With $N\geq2$ field insertions, there are 2 operator structures with 2 derivatives and 1 IBP. With $N=2$ (3) fields, one can write 4 (6) structures with 4 derivatives, related by 3 (4) IBP. With $N\geq 4$ fields, there are 7 structures with 4 derivatives and 4 IBP. The case $N=1$ is never relevant as any such operator is automatically a total derivative. The number of independent operators for each field multiplicity is automatically reflected in the Lagrangian~\eqref{eq.toy-1scalar}. }
\begin{align}\label{eq.toy-1scalar}
\mathcal{L} =&\, \frac{1}{2}\de_\mu\phi \de^\mu\phi \mathcal{F}_0(\phi) - \mathcal{V}(\phi)
\\
&
+ \frac{1}{\Lambda^2}(\de_\mu\de_\nu\phi\de^\mu\de^\nu\phi)\,\mathcal{F}_1(\phi)
+ \frac{1}{\Lambda^3}(\de_\mu\de_\nu\phi\de^\mu\phi\de^\nu\phi)\,\mathcal{F}_2(\phi) 
+ \frac{1}{\Lambda^4}(\de_\mu\phi\de^\mu\phi)^2 \,\mathcal{F}_3(\phi) \,.\nonumber
\end{align}
where $\mathcal{F}_0(0)=1$ to have a canonical kinetic term and in general $\mathcal{F}_i(0)=c_i$ for $i=1,2,3$, accounting for arbitrary Wilson coefficients.\footnote{We will see in \S \ref{sec:amplitudes} that the Wilson coefficient $\mathcal{F}_1(0)$ is {\em not} completely arbitrary, but  must satisfy a positivity bound (\ref{eq:positivity-F_1}). In  particular, its sign is negative~\cite{Brivio:2014}. In the context of our jet bundle geometry, this translates to a condition on the signature of the metric. }
The operator bases in~Eqs.~\eqref{eq.toy-1scalar} and~\eqref{eq.toy-1scalar_eomreduced} below were cross-checked with {\tt BasisGen}~\cite{Criado:2019ugp}.

The Lagrangian~\eqref{eq.toy-1scalar} can be recovered from a 1-jet bundle metric which is formally similar to the one in the 1d example. The metric components, truncated such that only terms with up to 4 derivatives are generated when pulling back to the Lagrangian, are
\begin{align}
\label{eq.1scalar_metric_first}
g_{uu}^{\mu\nu} &= \eta^{\mu\nu} A(u)\,,
\\
g_{uu}^\mu &= \frac{u^\mu}{\Lambda^2} B(u)\,,
\\
g_{uu} &= C(u) +\frac{u_\rho u^\rho}{\Lambda^4} D(u)\,,
\\
g_{\mu u}^\nu &= \delta^\nu_\mu E(u) + \frac{u^\nu u_\mu}{\Lambda^4} F_1(u) + \delta^\nu_\mu \frac{u_\rho u^\rho}{\Lambda^4} F_2(u)
\,,
\\
g_{\mu u}&= \frac{u_\mu}{\Lambda^2} G(u) + \frac{u_\mu\, u_\rho u^\rho}{\Lambda^6} H(u)\,,
\\
g_{\mu\nu} &= -\frac{\eta_{\mu\nu}}{2} V(u)
+ \left(\frac{u_\mu u_\nu}{\Lambda^4} +\frac{\eta_{\mu\nu}}{4} \frac{u_\rho u^\rho}{\Lambda^4}\right)\frac{J(u)}{2} 
+\left(\frac{u_\mu u_\nu}{\Lambda^4} +\frac{\eta_{\mu\nu}}{4} \frac{u_\rho u^\rho}{\Lambda^4}\right)\frac{u_\sigma u^\sigma}{\Lambda^4} \frac{K(u)}{2}
\,,
\label{eq.1scalar_metric_last}
\end{align}
where the coordinates $u^\mu$ are defined by raising indices with the spacetime metric $u^\mu = \eta^{\mu\nu} u_\nu$ and we have restored the powers of $\Lambda$ to make the dimensionalities manifest. The functions $V,A\dots K_i$ are defined to be dimensionless, and therefore functions of $(u/\Lambda)$.
Moving from 1d to 4d Minkowski spacetime, we find a larger number of independent tensor structures: the terms proportional $F_1$ and $F_2$ were degenerate in the QM example, but become distinguishable in this case. 
In Eq.~\eqref{eq.1scalar_metric_last} the functions $J$ and $K$ multiply a combination of two Lorentz structures: they were grouped together because they yield the same result upon  contracting with $\eta^{\mu\nu}$:
\begin{equation}
  \eta^{\mu\nu} u_\mu u_\nu = u_\mu u^\mu = \frac{\eta^{\mu\nu}\eta_{\mu\nu}}{4}  u_\rho u^\rho\,.
\end{equation}
Pulling back along the section $j^1\phi$ and contracting with the inverse metric:
\begin{align}
\mathcal{L}[\phi,g] &= 
\frac{1}{2}\left\langle \eta^{-1},\, (j^1\phi)^* g \right\rangle
&
\text{where}\quad
\eta^{-1} &= \eta^{\rho\sigma} \de_\rho \otimes \de_\sigma\,,
\end{align}
we get
\begin{align}
\mathcal{L}[\phi,g] =&\,  \frac{\Lambda^4}{2} \eta^{\mu\nu} g_{\mu\nu} + \Lambda^2 g_{\mu u} \de^\mu\phi
+\Lambda g_{\mu u}^\nu \de^\mu\de_\nu\phi + \frac{1}{2}g_{uu} \de_\mu\phi\de^\mu\phi
\label{eq.lag_frompullback_1scalar}
\\
&
+ \frac{g_{uu}^\mu}{\Lambda} \de^\nu\phi\de_\mu\de_\nu\phi 
+ \frac{1}{2}\frac{g_{uu}^{\mu\nu}}{\Lambda^2} \de_\rho\de_\mu\phi\de^\rho\de_\nu\phi
\nonumber
\\
=&\,
-\Lambda^4 V
+ \frac{1}{2}(\de_\mu\phi\de^\mu\phi)\left(C+2G+J\right)
+ \Lambda \square\phi E 
\\
&
+ \frac{(\de_\mu\de_\nu\phi\de^\mu\de^\nu\phi)}{\Lambda^2} \frac{A}{2}
+ \frac{\de_\mu\de_\nu \phi  \de^\mu\phi\de^\nu\phi}{\Lambda^3}\left(B+F_1\right)
+ \frac{(\square\phi)(\de_\mu\phi\de^\mu\phi)}{\Lambda^3} F_2
\nonumber
\\
&
+\frac{(\de_\mu\phi\de^\mu\phi)^2}{\Lambda^4} \frac{D+2H+K}{2}
\,,\nonumber
\end{align}
where $V,A\dots K$ are understood to be functions of $\phi/\Lambda$.

The terms proportional to $E$ and $F_2$ contain boxes. Both arise from pulling back the $g_{\mu u}^\nu$ metric element, which, at the 1-jet level, is the only one that allows the contraction of two derivatives acting on the same field. As is evident in Eq.~\eqref{eq.lag_frompullback_1scalar}, box contractions can't be implemented in any of the other terms. Following the reasoning above and the argument proved in \S\ref{app:proof}, all operators with boxes obtained from the 1-jet bundle are redundant with the box-less ones, and can be removed using IBPs. Let us do this explicitly: 
if $E$ is a constant, the corresponding term is a total derivative and therefore can be removed. If $E$ is at least linear in $\phi$, then it can be recast as:
\begin{equation}
\square\phi E = - \de_\mu\phi \de^\mu E =
 - \frac{\de_\mu\phi \de^\mu \phi}{\Lambda}\frac{d E}{d(\phi/\Lambda)}\,.
\end{equation}
The term proportional to $F_2$ can always be removed via IBP, even if $F_2$ is a constant:
\begin{equation}
(\square\phi)\de_\mu\phi\de^\mu\phi F_2 = - 2\de_\mu\de_\nu\phi\de^\mu\phi\de^\nu\phi F_2 - 
\frac{\de_\mu\phi\de^\mu\phi\, \de_\nu\phi\de^\nu\phi }{\Lambda}\frac{dF_2}{d(\phi/\Lambda)}
\end{equation}
No further manipulation is required in order to obtain a non-redundant Lagrangian, that manifestly matches Eq.~\eqref{eq.toy-1scalar}:
\begin{align}
 \mathcal{L}[\phi,g] &=\, {\rm IBP} \circ\frac{1}{2}\left\langle \eta^{-1},\, (j^1\phi)^* g \right\rangle
 \\
 &=
 \,
- \Lambda^4V
+ \frac{1}{2}(\de_\mu\phi\de^\mu\phi)\left(C+2G+J-2E'\right)
\nonumber
\\
&
+ \frac{(\de_\mu\de_\nu\phi\de^\mu\de^\nu\phi)}{\Lambda^2} \frac{A}{2}
+ \frac{\de_\mu\de_\nu \phi  \de^\mu\phi\de^\nu\phi}{\Lambda^3}\left(B+F_1-2F_2\right)
\nonumber\\
&
+\frac{(\de_\mu\phi\de^\mu\phi)^2}{\Lambda^4} \frac{D+2H+K-2F_2'}{2} \,,
\nonumber
\end{align}
where now $E'=dE/d(\phi/\Lambda)$ and analogously for $F_2$.
A canonical kinetic term requires
\begin{equation}
\label{canonical kinetic term requirement}
    C(0) + 2G(0)+J(0)-2 E'(0) = 1\,.
\end{equation}
The results obtained for a real scalar in Minkowski spacetime share several features with the case of quantum mechanics on the line: the map from metric to Lagrangian is surjective, and a degeneracy is observed between the same groups of functions, namely $(B,F_i)$, $(C,G,J,E)$, $(D,H,K)$. The presence of $E'$ alone in this example, rather than both $E$ and $E'$ as in the 1d case, is simply due to the different internal symmetry assumed. 

Finally, we comment on the conditions for $g^{(1)}$ with elements~\eqref{eq.1scalar_metric_first}-\eqref{eq.1scalar_metric_last} to be invertible, which is a required to have a valid metric. The necessary condition to have $\det [g^{(1)}]\neq 0$ is that at least one of the following 9 polynomials is non-vanishing:\footnote{
Note that, in general, this requirement does not preclude the possibility of coordinate singularities, i.e. conditions in which the metric is singular only at specific values of the coordinates $(u,u_\mu)$, which should be examined case by case.}
\begin{align}
\label{eq.invertible_metric_first}
&C (2 E^2+A V)\,,
\\
&D (8F_2^2-A K)\,,
\\
&B (4 F_1 G-B J)-2C F_1^2\,,
\\
&
B (4 F_1 H-B K)-2D F_1^2\,,
\\
&
4D(2E^2+AV) + C( 16E F_2- A J)\,,
\\
&
C(8F_2^2 -A K) +D(16 E F_2-A J)\,,
\\
&
4E(BG-CF_1)+A(CJ- 2G^2)+B^2 V\,,
\\
&
16E(BH-DF_1) +16F_2(BG-CF_1)
+4A(D J+C K -4  G H) -B^2 J\,,
\\
&16F_2(BH-DF_1)+4A(DK- 2H^2)-B^2 K\,.
\label{eq.invertible_metric_last}
\end{align}
In the generic case where  4-derivative operators with non-zero Wilson coefficients are present in the Lagrangian, the conditions~\eqref{eq.invertible_metric_first}--\eqref{eq.invertible_metric_last} leave large freedom to choose a possible metric that pulls back to Eq.~\eqref{eq.toy-1scalar}. A minimal example is:
\begin{align} \label{eq:1R-scalar-G-basis}
g^{\mu\nu}_{uu} &= 2\eta^{\mu\nu} \mathcal{F}_1(u)\,,
&
g_{uu}^\mu &= \frac{u^\mu}{\Lambda^2}\mathcal{F}_2(u)\,,
&
g_{uu} &= \mathcal{F}_0(u) +2\frac{u_\mu u^\mu}{\Lambda^4} \mathcal{F}_3(u)\,,
\\
g_{\mu u}^\nu &=0 \,,
&
g_{\mu u}&= 0\,,
&
g_{\mu\nu} &= -\frac{1}{2}\eta_{\mu\nu} \frac{\mathcal{V}(u)}{\Lambda^4} \,. \nonumber
\end{align}
If we restrict the Lagrangian to the 2- and 0-derivative terms, setting to zero all functions except $V,C,E$, then the conditions for invertibility collapse into $2CE^2\neq 0$, which implies that both $C$ and $E$ must be non-vanishing. This indicates that, even though it produces a physically redundant operator, the $E$ term is necessary for the jet-bundle metric to be non-singular for any value of the 4-derivative Wilson coefficients, consistently with what observed in \S\S\ref{sec:qm_1d_redef} and \ref{sec:toy}. Restricting further to 0-derivatives only is incompatible with both physics (as the kinetic term is removed) and jet-bundle geometry, as the metric would be inevitably singular.

\subsubsection*{Equation-of-Motion redundancies}
The  operator basis considered up to this point is non-redundant under IBP, i.e. it is a Green's basis for operators with up to 4 derivatives and arbitrary number of fields. It is interesting to note that the 1-jet bundle geometry automatically accounts for all IBP among operators with up to 4 derivatives, except the residual $E$ and $F_2$ function redundancies.\footnote{The fact that no redundant box-less operators are produced by the jet bundle metric is due to the fact that, in this particular case, there are exactly as many operators with boxes as independent IBP. Therefore the set of \emph{all} box-less operators with up to 4 derivatives forms a complete and non-redundant Green's basis.}

When constructing EFTs, we are typically interested in removing Equation-of-Motion (EOM) redundancies as well. 
EOM relations express invariance properties of the $S$-matrix. Because this information is not contained in the geometric formulation, they need to be imposed as additional conditions on top of the Lagrangian in Eq.~\eqref{eq.toy-1scalar}.

Formally, the removal of EOM redundancies in an EFT is performed by applying \emph{derivative} field redefinitions on the Lagrangian. Therefore, based on the discussion in \S\ref{sec:geometry},
two jet bundle metrics pulling back to EOM-equivalent Lagrangians are in general not related by a morphism. In practice, we do not expect to find an operation on the jet bundle metric that is mathematically equivalent to basis reduction via EOM.
Nevertheless, it is instructive to consider the EOM-reduced Lagrangian for the toy example under study, and ask what classes of jet bundle metrics would pull back to it. We will come back to the inadequacy of jet bundle morphisms (even prolonged ones) at the end of this subsection.

To this end, we take a step back and review the construction of the operator basis. This operation is necessary because, being~\eqref{eq.toy-1scalar} ``box-less", EOM redundancies are not manifest. At the same time, we hope to give a useful illustration of the application of derivative field redefinitions in an EFT.

At the 2-derivatives level and with $N$ fields, there are two allowed operators structures:
\begin{align}
O^{(2)}_1 &= (\square \phi)\phi^{N-1} \,,   
&
O^{(2)}_2 &= (\de_\mu \phi\de^\mu\phi)\phi^{N-2}\,,
\end{align}
which are related by the IBP
\begin{equation}
O^{(2)}_1 + (N-2)O^{(2)}_2 = 0  \,.
\end{equation}
To obtain~\eqref{eq.toy-1scalar}, we have used the IBP to remove $O_1^{(1)}$ for each value of $N\geq 2$. This is always the most convenient choice for $N=2$, since in this case $O_2^{(2)}$ is the kinetic term.
However, for $N\geq 3$, we can choose to remove $O_2^{(2)}$ via IBP, and then use the EOM to trade $O_1^{(2)}$ for 0-derivative interactions, effectively obtaining a basis without any 2-derivative operator, besides the kinetic term.
In practice, the EOM removal implies that we start with a Lagrangian
\begin{equation}
\mathcal{L} = \frac{1}{2}\de_\mu \phi\de^\mu\phi - \mathcal{V}(\phi) + \sum_{N=3}^\infty\frac{c_{1,N}^{(2)}}{\Lambda^{N-2}} (\square\phi) \phi^{N-1}+ \dots\,,
\end{equation}
where the scalar potential is as in Eq~\eqref{eq:potential} and the dots stand for other operators that are present in general.
Then we apply the derivative redefinition
\begin{equation}
\label{eq.2deom_redefinition}
 \phi \mapsto  \phi - \sum_{k=0}^\infty \frac{\epsilon_k}{\Lambda^{k+2}}(\square\phi)\phi^k\,,
\end{equation}
and truncate again at 2-derivatives:
\begin{align}
\mathcal{L} &\to  \frac{1}{2}\de_\mu \phi\de^\mu\phi - \mathcal{V}(\phi) 
-  \frac{d\mathcal{V}(\phi)}{d\phi} \sum_k \,
\frac{\epsilon_k}{\Lambda^{k+2}}(\square\phi) \phi^k
+ \sum_{N=3}^\infty\frac{c_{1,N}^{(2)}}{\Lambda^{N-2}} (\square\phi) \phi^{N-1}+ 
\dots
\\
&=\frac{1}{2}\de_\mu \phi\de^\mu\phi - \mathcal{V}(\phi) 
+ (\square\phi)
\left[-\sum_{n,k} \frac{\epsilon_k nV_n}{\Lambda^{n+k-2}} \phi^{n+k-1} +
\sum_{N=3}^\infty\frac{c_{1,N}^{(2)}}{\Lambda^{N-2}}\phi^{N-1}\right]
+\dots
\\
&=\frac{1}{2}\de_\mu \phi\de^\mu\phi - \mathcal{V}(\phi) 
+ \sum_{N=3}^\infty
\left[c_{1,N}^{(2)}-\sum_{n=1}^N \epsilon_{N-n} n V_n\right]\frac{\phi^{N-1}}{\Lambda^{N-2}}
(\square\phi)
+\dots
\end{align}
Choosing the constants $\epsilon_k$ such that $\sum_{n=1}^N \epsilon_{N-n} n V_n=c_{1,N}^{(2)}$ removes the infinite tower of $O_1^{(2)}$ operators with $N\geq 3$.\footnote{This is an infinite system of equations, each labeled by $N$ and depending on $N$ free parameters $\epsilon_0\dots \epsilon_{N-1}$. It always admits a solution that can be found recursively.}

At the level of 4 derivatives and with $N$ fields, we can write the 7 structures
\begin{align}
 O_1^{(4)}&= (\square\square\phi)\phi^{N-1}\,,
 &
 O_5^{(4)}&= (\square\phi)(\de_\mu\phi\de^\mu\phi)\phi^{N-3}\,,
 \\
 O_2^{(4)}&= (\de_\mu\square\phi)(\de^\mu\phi)\phi^{N-2}\,,
 &
 O_6^{(4)}&= (\de_\mu\de_\nu\phi \,\de^\mu\phi\de^\nu\phi)\phi^{N-3}\,,
 \\
 O_3^{(4)}&= (\square\phi)(\square\phi)\phi^{N-2}\,,
 &
 O_7^{(4)}&= (\de_\mu\phi\de^\mu\phi)^2\phi^{N-4}
 \,,
 \\
  O_4^{(4)}&= (\de_\mu\de_\nu\phi)(\de^\mu\de^\nu\phi)\phi^{N-2}\,,
\end{align}
which are related by the 4 IBPs
\begin{align}
O_1^{(4)} + (N-1) O_2^{(4)} &=0\,,\\
O_2^{(4)} + O_3^{(4)} + (N-2) O_5^{(4)} &=0\,,\\
O_2^{(4)} + O_4^{(4)} + (N-2) O_6^{(4)} &=0\,,\\
2O_6^{(4)} + O_5^{(4)} + (N-3) O_7^{(4)} &=0\,.
\end{align}
To obtain Eq~\eqref{eq.toy-1scalar}, we used the four IBP to remove the four operators with boxes. On the other hand, the more widely used procedure to construct minimal operator bases is to first use IBP to remove as many operators \emph{without} boxes as possible, and then employ EOMs to reduce box operators to lower derivative structures, by using field redefinitions similar to Eq~\eqref{eq.2deom_redefinition}. With $N=2,3$ fields, this procedure reveals that all 4-derivative operators are redundant with lower-derivative ones. However, for $N\geq4$, the operator $O_7^{(4)}$ cannot be removed, neither with IBP nor with EOMs. 

In summary, a IBP+EOM reduced Lagrangian with up to 4 derivatives is
\begin{align}
\label{eq.toy-1scalar_eomreduced}
\mathcal{L} &=  \frac{1}{2}\de_\mu\phi\de^\mu\phi - \mathcal{V}(\phi) + 
\frac{1}{\Lambda^{4}}(\de_\mu\phi\de^\mu\phi)^2 \mathcal{F}_3(\phi)\,,
\end{align}
which is simply a subset of the Green's basis in Eq~\eqref{eq.toy-1scalar}, as expected. This indicates that the EOM-reduced Lagrangian can be obtained by pulling back a correspondingly reduced metric $g$ on the 1-jet bundle. Specifically, Eq.~\eqref{eq.toy-1scalar_eomreduced} can be obtained requiring
\begin{align}
C(\phi) +2 G(\phi) +J(\phi) - 2 E'(\phi) &=1\,, \label{eq:can-norm-Rscalar-4d}
\\
V(\phi)&= \mathcal{V}(\phi)/\Lambda^4
\\
A(\phi)&=0\,,
\\
B(\phi)+F_1(\phi) -2F_2(\phi)&= 0\,,
\\
\frac{1}{2}\left(D(\phi) +K(\phi)\right) +H(\phi) -F_2'(\phi)&=\mathcal{F}_3(\phi)\,.
\end{align}
This still leaves much freedom to identify a suitable metric, that respects the conditions~\eqref{eq.invertible_metric_first}-\eqref{eq.invertible_metric_last} as well. A minimal example is
\begin{align}
g^{\mu\nu}_{uu} &= g^\mu_{uu} =g_{\mu u}=0\,,
&
g_{uu} &=  2\frac{u_\rho u^\rho}{\Lambda^4} \mathcal{F}_3(u)\,,
\\
g_{\mu\nu} &=-\frac{\eta_{\mu\nu}}{2}\frac{\mathcal{V}(u)}{\Lambda^4}\,,
&
g^\nu_{\mu u} &=-\frac{\delta^\nu_\mu}{2} \frac{u}{\Lambda}\,.
\end{align}

\subsubsection*{Can prolonged bundle morphisms capture `EOM field redefinitions'?}

Derivative field redefinitions of the form 
\be
\phi \mapsto \psi \quad \text{such that} \quad u \circ \psi = \psi\left(u \circ \phi, \partial_\mu \partial^\mu (u \circ \phi)\right)\, ,
\ee
namely in which the new section $\psi$ is a function of $\phi$ and $\Box \phi$,
have played an important role in our discussion of this 4d real scalar example.
Before our introduction of jet bundles,
in \S \ref{sec:bundle-morphism} we discussed to what extent a morphism on the field space bundle $(E, \Sigma, \pi)$, which recall can capture any {\em non}-derivative field redefinition, can capture the effects of derivative field redefinitions such as this (which we can understand generically to be changes of section). There are indeed fundamental reasons why such morphisms cannot induce derivative maps on sections (see lemma \ref{lemma}). 

Nonetheless, for the particular example of a 4d real scalar with canonical kinetic term and a $\phi^4$ potential, we saw that a bundle morphism of the form $(x^\mu, u) \mapsto (x^\mu, u + \epsilon \lambda(u))$ can replicate the shift of the Lagrangian under a derivative change of section, $\phi \mapsto \psi = \phi - \epsilon \Box \phi$, at least when the Lagrangian is truncated to remain at 2-derivative order, and keeping only operators with up to 4-fields. One might ask whether similar `accidents' can happen when we pass to 4-derivative Lagrangians using our 1-jet bundle formalism. In other words, can a morphism on the 1-jet bundle replicate, at the Lagrangian level, the effects of derivative field redefinitions that depend on $\Box\phi$? We here suggest that the answer to this question is no. (More precisely, we expect that such an equivalence will only occur for very special examples, truncated in a particular way.)

To see this, let us consider the effects of a prolongated bundle morphism, as defined in \S \ref{sec:symmetries-1jet-bundle}, on the 1-jet bundle metrics constructed for the 4d real scalar in this Subsection.
We specialise to the case of a base-point preserving `perturbative morphism' on the bundle $E$ that is consistent with  Poincar\'e symmetry, considered in \S \ref{sec:phi4}, of the form
\begin{equation}
    f_E:(x^\mu, u) \mapsto (x^\mu, f(u)) = (x^\mu, u + \epsilon \lambda(u))\, .
\end{equation}
Using Eq. (\ref{eq:prolonged-morphism}), the prolongation of this bundle morphism to the 1-jet bundle is
\begin{equation}
    j^1 f_E: (x^\mu, u, u_\mu) \mapsto \left(x^\mu, f(u), u_\mu \frac{\partial f}{\partial u}\right) = \left(x^\mu, u + \epsilon \lambda, u_\mu + \epsilon u_\mu \lambda^\prime\right)\, ,
\end{equation}
where $\lambda^\prime := d\lambda/du$ {\em etc}. The coordinate basis 1-forms transform as
\begin{align}
    j^1 f_E: 
    (dx^\mu, du, du_\mu) \mapsto  \left(dx^\mu, du(1+\epsilon \lambda^\prime), du_\mu (1+\epsilon \lambda^\prime) + \epsilon u_\mu \lambda^{\prime\prime} du \right)\, ,
\end{align}
from which we can compute the transformation law for the 1-jet bundle metric. 

Let us start with the 1-jet metric written in the form of Eq. (\ref{eq:1R-scalar-G-basis})
\begin{nalign}
    g^{(1)} = &-\frac{\eta^{\mu\nu} }{2} \frac{\mathcal{V}(u)}{\Lambda^4} dx^\mu dx^\nu + \left( \mathcal{F}_0(u)+ \frac{2u_\mu u^\mu}{\Lambda^4} \mathcal{F}_3(u)\right) du du \\
    &+ \mathcal{F}_2(u) \frac{u^\mu}{\Lambda^3} du du_\mu +  \frac{2\eta^{\mu\nu}}{\Lambda^2} \mathcal{F}_1(u) du_\mu du_\nu\, ,
\end{nalign}
which pulls back to give the Lagrangian in Eq. (\ref{eq.toy-1scalar}). Doing our prolongated jet bundle morphism and keeping only terms that are linear in the perturbation $\epsilon$, we compute the transformed metric to be
\begin{nalign}
    j^1 f_E^\ast g^{(1)} 
    &=-\frac{1}{2} \eta^{\mu\nu} \mathcal{V}(u+\epsilon \lambda) dx^\mu dx^\nu + \mathcal{F}_0(u+\epsilon \lambda)(1+2\epsilon\lambda^\prime) du du \\
    &+ \frac{u_\mu u^\mu}{\Lambda^4}\left[2\mathcal{F}_3(u+\epsilon \lambda)(1+4\epsilon\lambda^\prime) + \epsilon \mathcal{F}_2(u+\epsilon \lambda) \lambda^{\prime\prime} \right] du du \\
    &+ \frac{u^\mu}{\Lambda^2} \left[\mathcal{F}_2(u+\epsilon \lambda) (1+3\epsilon\lambda^\prime) + 4\epsilon \mathcal{F}_1(u+\epsilon \lambda) \lambda^{\prime\prime} \right] du du_\mu  \\
    &+ 2\eta^{\mu\nu} \mathcal{F}_1(u+\epsilon \lambda) (1+2\epsilon \lambda^\prime) du_\mu du_\nu  + O(\epsilon^2)\, .
\end{nalign}
Expanding also each `structure function' $\mathcal{V}$, $\mathcal{F}_{0,1,2,3}$ about $u$, we can capture the effects of this transformation in terms of the following shifts at leading order in $\epsilon$ (the terms neglected are $O(\epsilon^2)$):
\begin{align}
    \mathcal{V} &\mapsto \mathcal{V} + \epsilon \lambda \mathcal{V}^\prime \,  , \\
    \mathcal{F}_0 &\mapsto \mathcal{F}_0 + \epsilon( 2\lambda^\prime \mathcal{F}_0 + \lambda \mathcal{F}_0^\prime) \,  , \\
    \mathcal{F}_1 &\mapsto \mathcal{F}_1+\epsilon( 2\lambda^\prime \mathcal{F}_1 + \lambda \mathcal{F}_1^\prime) \,  , \\
    \mathcal{F}_2 &\mapsto \mathcal{F}_2 + \epsilon (3\lambda^\prime \mathcal{F}_2 + 4\lambda^{\prime\prime} \mathcal{F}_1 +\lambda \mathcal{F}_2^\prime) \,  , \\
    \mathcal{F}_3 &\mapsto \mathcal{F}_3 +\epsilon (4\lambda^\prime \mathcal{F}_3 + \frac{1}{2}\lambda^{\prime\prime} \mathcal{F}_2 +\lambda \mathcal{F}_3^\prime) \,  .
\end{align}
If we then pullback this metric to $\Sigma$ along the original (prolongated) section $j^1 \phi$, we obtain the transformed Lagrangian
\begin{align}
\cL[\phi, (j^1f_E)^\ast g^{(1)}] 
=&\,\,\frac{1}{2}\left( \mathcal{F}_0 + \epsilon( 2\lambda^\prime \mathcal{F}_0 + \lambda \mathcal{F}_0^\prime) \right) (\partial \phi)^2 - (\mathcal{V} + \epsilon \lambda \mathcal{V}^\prime)  \\
&+\frac{1}{\Lambda^2}\left( \mathcal{F}_1+\epsilon( 2\lambda^\prime \mathcal{F}_1 + \lambda \mathcal{F}_1^\prime)\right)\de_\mu\de_\nu\phi\de^\mu\de^\nu\phi \nonumber \\
&+\frac{1}{\Lambda^3}\left(\mathcal{F}_2 + \epsilon (3\lambda^\prime \mathcal{F}_2 + 4\lambda^{\prime\prime} \mathcal{F}_1 +\lambda \mathcal{F}_2^\prime) \right)\de_\mu\de_\nu\phi\de^\mu\phi\de^\nu\phi  \nonumber \\
&+ \frac{1}{\Lambda^4}\left(\mathcal{F}_3 +\epsilon (4\lambda^\prime \mathcal{F}_3 + \frac{1}{2}\lambda^{\prime\prime} \mathcal{F}_2 +\lambda \mathcal{F}_3^\prime) \right) (\de_\mu\phi\de^\mu\phi)^2 \,\,. \nonumber 
\end{align}
Thus, upon doing the prolongated bundle morphism to transform the metric, we see that the transformation of the Lagrangian has a particularly constrained structure; operators with a certain number $D$ of derivatives (here 0, 2, or 4 due to our symmetry assumptions) mix only into operators with the same value of $D$.

In contrast, if we were to perform a derivative change of section, we would see operators of lower-derivative-order mix into operators of higher-derivative-order, as in the $\phi^4$ example considered in \S \ref{sec:phi4}: for example, it is straightforward to compute that $\mathcal{V}$ `mixes into' the 2-derivative structure function $\mathcal{F}_0$ and all three 4-derivative structure functions $\mathcal{F}_{1,2,3}$; the 2-derivative term $\mathcal{F}_0$ likewise mixes into all of $\mathcal{F}_{1,2,3}$. For the 2-derivative toy example considered in \S \ref{sec:phi4} we exploited the fact that there was only a single such operator mixing, namely of the potential into the 2-derivative term, that could be cancelled by a bundle morphism that shifts the kinetic term (explicitly, this fixed $\lambda$ in terms of $\partial_u V$) {\em provided} the potential were high enough order that we could neglect its own variation. This was a very special situation which enabled the derivative change of section to be matched by a bundle morphism.
In this more general 4-derivative example, we must still fix $\lambda$ by matching the shift of the 2-derivative term (even if that is zero, for example in the case of vanishing potential), but then we cannot hope to simultaneously match the shift in any 4-derivative term that is present.
So one cannot even contrive a theory with 2- and 4-derivative terms for which the $\phi \to \psi(\phi, \Box\phi)$ change of section can be described via an (appropriately prolongated) morphism of the jet bundle. Nonetheless, we reiterate that the derivative field redefinition can be implemented as a prolonged change of section, as discussed in \S \ref{sec:1jet-change-section}.

\subsection{Two real scalars in 4d, without internal symmetries} \label{sec:no_symmetry}

We generalize  the previous example by introducing two scalar fields $\phi^1,\phi^2$, without imposing any internal symmetries relating them. The main goals in considering this example are twofold: (i) to cover a case in which non-trivial field indices appear in the metric, and (ii) to check that the 1-jet bundle gives a complete Green's basis for a theory with multiple fields but without internal symmetries.

The formalism is identical to the previous case, except the 1-jet bundle is now a real 14-manifold  with local fibred coordinates
\begin{equation}
j^I = \{ x^\mu,u^i,u^i_\mu\},\quad i=1,2\,,    
\end{equation}
that, along the section $j^1\phi$, evaluate to 
\begin{align}
x^\mu\circ j^1\phi &= x^\mu\,
\\
u^i\circ j^1\phi &= \phi^i(x^\mu)\,
\\
u^i_\mu\circ j^1\phi &= \de_\mu\phi^i(x^\mu)\,
\end{align}
As in the previous cases, we start by writing the EFT Lagrangian containing a complete and IBP-non-redundant basis with up to 4 derivatives, requesting that no more than 2 derivatives act on each field and that there are no boxes.  This identifies the Green's basis:
\begin{align}
\label{eq.toy-2scalars}
\mathcal{L} &=
\frac{1}{2}(\de_\mu\phi^{i} \de^\mu\phi^{j}) \mathcal{F}^0_{ij}
-\mathcal{V}(\phi_1,\phi_2)
\\
&+
\frac{1}{\Lambda^{2}}(\de_\mu\de_\nu\phi^{i} \de^\mu\de^\nu\phi^{j}) \mathcal{F}^1_{ij}
+
\frac{1}{\Lambda^{3}}(\de_\mu\de_\nu\phi^{i} \de^\mu\phi^{j}\de^\nu\phi^{k}) \mathcal{F}^2_{ijk}
+
\frac{1}{\Lambda^{4}}(\de_\mu\phi^{i} \de^\mu\phi^{j})(\de_\nu\phi^{k}\de^\nu\phi^{l}) \mathcal{F}^3_{ijkl}\,.
\nonumber
\end{align}
where, in general, the scalar potential $\mathcal{V}$ and the $\mathcal{F}$ functions have the form
\begin{align}
\mathcal{F}&= f_0 + f_1\frac{\phi^1}{\Lambda} + f_2 \frac{\phi^2}{\Lambda}+ f_{11}\frac{(\phi^1)^2}{\Lambda^2}+\dots
=\sum_{a,b=0}^{\infty} f_{\footnotesize\underbrace{1\dots 1}_{a}\underbrace{2\dots 2}_{b}} \frac{(\phi^1)^a(\phi^2)^b}{\Lambda^{a+b}}
\end{align}
Each such function contains $(n+1)$ independent index assignments for every number of fields $n=a+b$.
Therefore, accounting for the symmetry in $(ij)$, with $N$ field insertions in the Lagrangian, the 2-derivative operator has $3(n+1) = 3(N-1)$ independent index assignments.
For the 4-derivative operators there are a total of $3(5N-11)$ independent assignments for $N\geq 4$.  This number is reduced to 12 for $N=3$ and to 3 for $N=2$, due to the fact that some operator structures are unavailable in these cases. These numbers, as well as the EOM-reduced counting below, have been cross-checked with {\tt BasisGen}.
Canonical kinetic terms require $f^0_{11}=f^0_{22}=1, f^0_{12}=0$.

We now turn to the jet bundle metric from which we would like to recover~\eqref{eq.toy-2scalars}. The generic metric components, truncated as in the previous examples, are
\begin{align}
 g_{ij}^{\mu\nu} &= \eta^{\mu\nu} A_{ij}(u)\,,
 \\
  g_{ij}^\mu &= \frac{u^{k\mu}}{\Lambda^2} B_{ijk}(u) \,,
 \\
  g_{ij} &= C_{ij}(u) 
 + \frac{u^k_\rho u^{l\rho}}{\Lambda^4} D_{ijkl}(u)\,,
 \\
  g_{\mu j}^\nu &= \delta^\nu_\mu E_j(u) 
  + \frac{u^{k\nu} u^l_\mu}{\Lambda^4} F_{1,jkl}(u)
  +\delta^\nu_\mu\frac{u^k_\rho u^{l\rho}}{\Lambda^4} F_{2,jkl}(u)
 \,,
 \\
  g_{\mu j}&= 
 \frac{u^k_\mu}{\Lambda^2} G_{jk}(u) 
 + \frac{u^k_\mu\, u^l_\rho u^{m\rho}}{\Lambda^6} H_{jklm}(u)\,,
 \\
 g_{\mu\nu} &= -\frac{\eta_{\mu\nu}}{2} V(u) 
+\left(\frac{u^k_\mu u^l_\nu}{\Lambda^4}+\frac{\eta_{\mu\nu}}{4}\frac{u^k_\rho u^{l\rho}}{\Lambda^4}\right)
\frac{J_{kl}(u)}{2}
+ \left(\frac{u^k_\mu u^l_\nu}{\Lambda^4}+\frac{\eta_{\mu\nu}}{4}\frac{u^k_\rho u^{l\rho}}{\Lambda^4}\right) \frac{u^m_\rho u^{n\rho}}{\Lambda^4} \frac{K_{klmn}(u)}{2}
\,,
\end{align}
where $u^{i\mu}=\eta^{\mu\nu} u^i_\nu$ and all the functions $V,A\dots K$ are dimensionless functions of $(u^i/\Lambda)$\,.
We also assume that the metric is fully symmetric: $g^{\mu\nu}_{ij}, g^{\mu\nu},g^\nu_{\mu j}$ are symmetric in $\mu\nu$, $g^{\mu\nu}_{ij},g_{ij}$ are symmetric in $ij$,  and $ g_{\nu i}^{(\mu)} = g_{\mu j}^{(\nu)},  g_{ij}^\mu =g_{ij}^\nu$ upon relabeling the free indices.  
All repeated indices are understood to be summed over. Because we are not assuming internal symmetries, they simply represent labels on the functions.

Pulling back along the section $j^1\phi$ and contracting with the inverse metric 
\begin{align}
\cL[\phi,g] &= \frac{1}{2}\left\langle \eta^{-1},\, (j^1\phi)^*g \right\rangle
&
\text{where}\quad
\eta^{-1}&=\eta^{\rho\sigma}\de_\rho\otimes\de_\sigma\,,
\end{align}
we get
\begin{align}
\mathcal{L}[\phi,g] =&\, \frac{\Lambda^4}{2} \eta^{\mu\nu} g_{\mu\nu} + \Lambda^2 g_{\mu j}\de^\mu\phi^j
+\Lambda g_{\mu j}^\nu\de^\mu\de_\nu\phi^j
+\frac{1}{2}g_{ij}\de^\mu\phi^i\de^\mu\phi^j
\\
&
+\frac12\frac{g_{ij}^\mu}{\Lambda}\de_\mu\de_\nu\phi^i\de^\nu\phi^j
+\frac12\frac{g_{ij}^\nu}{\Lambda}\de^\mu\phi^i\de_\mu\de_\nu\phi^j
+\frac{1}{2}\frac{g_{ij}^{\mu\nu}}{\Lambda^2} \de_\rho\de_\mu\phi^i\de^\rho\de_\nu\phi^j
\nonumber\\
=&\,
- \Lambda^4V
+ \frac{1}{2}\de_\mu\phi^i\de^\mu\phi^j\left(C_{ij}+2G_{ij} +J_{ij}\right)
+\Lambda\square\phi^jE_j 
\\
&+\frac{(\de_\mu\de_\nu\phi^i\de^\mu\de^\nu\phi^j)}{\Lambda^2}\frac{A_{ij} }{2}
+\frac{ \de_\mu\de_\nu\phi^i\de^\mu\phi^j\de^\nu\phi^k}{\Lambda^3}\frac{B_{ijk}+B_{jik}+2F_{1,ijk}}{2}
\nonumber\\
&
+\frac{(\square \phi^i) \de_\mu\phi^j\de^\mu\phi^k}{\Lambda^3}F_{2,ijk}(u)
+
\frac{(\de_\mu\phi^i\de^\mu\phi^j)(\de_\nu\phi^k\de^\nu\phi^l)}{\Lambda^4}\frac{D_{ijkl}+2H_{ijkl}+K_{ijkl}}{2}
\,.
\nonumber
\end{align}
This result is again formally equivalent to those obtained in \S\ref{sec:toy} and \S\ref{sec:4dRscalar}, the only difference being the presence of explicit field indices. The $E_j$ function is required to have a non-singular metric at the 2-derivatives level, but it is redundant via IBP:\footnote{As in the previous example, we find that for any number of fields $N$, there are always as many independent IBPs as operators with boxes. This implies that the set of all box-less operators forms a complete and non-redundant Green's basis.}
\begin{align}
\square \phi^j E_j &= - \frac{\de_\mu\phi^i\de^\mu\phi^j}{\Lambda} \frac{dE_j}{d(\phi^i/\Lambda)}\,.
\end{align}
The term proportional to the function $F_{2,ijk}$ is also redundant and can be recast as
\begin{align}
 (\square\phi^i)\de_\mu\phi^j\de^\mu\phi^k F_{2,ijk}
 &= 
 - (\de_\mu\de_\nu\phi^j)\de^\mu\phi^k\de^\nu\phi^i (F_{2,ijk}+F_{2,ikj})
 -\frac{(\de_\mu\phi^j\de^\mu\phi^k)(\de_\nu\phi^i\de^\nu\phi^l)}{\Lambda}\frac{d F_{2,ijk}}{d(\phi^l/\Lambda)}
\end{align}
such that
\begin{align}
\mathcal{L}[\phi, g] =&\, \text{IBP } \circ \frac{1}{2}\left\langle \eta^{-1},\, (j^1\phi)^*g \right\rangle
\\
=&\,
\label{eq.toy-2scalars_ibp}
- \Lambda^4V
+ \frac{1}{2}\de_\mu\phi^i\de^\mu\phi^j\left(C_{ij}+2G_{ij} +J_{ij} - 2\frac{dE_j}{d(\phi^i/\Lambda)}\right)
\\
&+\frac{(\de_\mu\de_\nu\phi^i\de^\mu\de^\nu\phi^j)}{\Lambda^2}\frac{A_{ij} }{2}
+\frac{\de_\mu\de_\nu\phi^i\de^\mu\phi^j\de^\nu\phi^k}{\Lambda^3} \left(\frac{B_{ijk}+B_{jik}}{2}+F_{1,ijk}
-F_{2,kij}-F_{2,kji}\right)
\nonumber\\
&+
\frac{(\de_\mu\phi^i\de^\mu\phi^j)(\de_\nu\phi^k\de^\nu\phi^l)}{\Lambda^4}\left(\frac{D_{ijkl}+K_{ijkl}}{2}+H_{ijkl}
-\frac{dF_{2,kij}}{d(\phi^l/\Lambda)}
\right)\,.
\nonumber
\end{align}
This Lagrangian manifestly captures the complete Green's basis in Eq.~\eqref{eq.toy-2scalars}.

\subsubsection*{Equation-of-Motion redundancies.} 
We can repeat the same exercise illustrated in \S\ref{sec:4dRscalar} in order to identify a minimal operator basis with up to 4 derivatives, accounting for both IBP and EOM reduction. The procedure is formally equivalent to the one illustrated above, and leads to an analogous result, namely:
\begin{align}
\label{eq.toy-2scalars_eomreduced}
\mathcal{L} &=  \frac{1}{2}\de_\mu\phi^i\de^\mu\phi^i - \mathcal{V}(\phi) 
\\
&+\frac{1}{2}\de_\mu\phi^2\de^\mu\phi^2\phi^1\phi^1\mathcal{F}^0_{2211}(\phi)
+
\frac{1}{\Lambda^{4}}(\de_\mu\phi^i\de^\mu\phi^j) (\de_\nu\phi^k\de^\nu\phi^l)\mathcal{F}^3_{ijkl}(\phi)\,.
\nonumber
\end{align}
For $N=2,3$ field insertions it is possible to remove all terms with up to 4 derivatives (except the two kinetic ones) and trade them for non-derivative interactions in the potential. For $N\geq 4$, the structure $(\de\phi)^4$ is always independent. The main difference with the previous examples is that, in addition, $(N-3)$ index assignments of the 2-derivative operator cannot be removed. Here we chose arbitrarily those starting with $(2211)$.

The Lagrangian in Eq.~\eqref{eq.toy-2scalars_eomreduced} is a subset of~\eqref{eq.toy-2scalars}, and therefore it can be obtained pulling back a 1-jet bundle metric. Comparing to Eq.~\eqref{eq.toy-2scalars_ibp}, we see that such a metric should satisfy:
\begin{align}
A_{ij} &=0\,,
\\
B_{ijk} +B_{jik} + 2F_{1,ijk}-2F_{2,kij}-2 F_{2,kji} &=0\,,
\\
V&= \frac{\mathcal{V}(\phi)}{\Lambda^4}
\\
C_{ij}+2G_{ij} +J_{ij}-2\frac{dE_j}{d(\phi^i/\Lambda)}&=\delta_{ij} + \delta_{i2}\delta_{j2}(\phi^1)^2 \mathcal{F}^0_{2211}(\phi)\,,
\\
\frac{1}{2}\left(D_{ijkl}+K_{ijkl}\right) + H_{ijkl} 
-\frac{dF_{2,kij}}{d(\phi^l/\Lambda)}
&=\mathcal{F}^3_{ijkl}(\phi)\,.
\end{align}

\section{Four scalars in 4d, with ${O(4)}$ internal symmetry} \label{sec:HEFTandSMEFT}

We now cover the case of a multiplet of 4 fields in Minkowski space, with an internal $O(4)$ symmetry. This example is representative of SMEFT/HEFT with exact custodial symmetry and in the absence of gauge and fermion fields. 
The discussion in this section follows the same steps taken in the examples in \S\ref{sec:examples}.

\subsection{The EFT Lagrangian up to 4 derivatives}\label{sec:SMEFT_lagrangian}
We start by constructing a complete, IBP-non-redundant set of  $O(4)$-symmetric operators with up to 4 derivatives and arbitrary number of fields. We require that no more than 2 derivatives act on each field and the absence of boxes. As in the toy examples, this identifies a unique Lagrangian: 
\begin{align}
\label{eq.smeft_L}
\cL =&\, \frac{1}{2}\de_\mu\phi \cdot \de^\mu\phi\mathcal{F}_a   + (\de_\mu\phi\cdot \phi)^2\mathcal{F}_b - \mathcal{V}
\\
&
+\frac{1}{\Lambda^2}(\de_\mu\de_\nu\phi\cdot\de^\mu\de^\nu\phi) \mathcal{F}_1
+\frac{1}{\Lambda^4}(\de_\mu\de_\nu\phi\cdot\phi)(\de^\mu\de^\nu\phi\cdot\phi) \mathcal{F}_2
+\frac{1}{\Lambda^4}(\de_\mu\de_\nu\phi\cdot\de^\mu\phi)(\de^\nu\phi\cdot\phi) \mathcal{F}_3
\nonumber
\\
&
+\frac{1}{\Lambda^4}(\de_\mu\de_\nu\phi\cdot\phi)(\de^\mu\phi\cdot\de^\nu\phi) \mathcal{F}_4
+\frac{1}{\Lambda^4}(\de_\mu\phi\cdot\de^\mu\phi)^2 \mathcal{F}_5
+\frac{1}{\Lambda^4}(\de_\mu\phi\cdot\de_\nu\phi)^2 \mathcal{F}_6
\nonumber
\\
&
+\frac{1}{\Lambda^6}(\de_\mu\de_\nu\phi\cdot\phi)(\de^\mu\phi\cdot\phi)(\de^\nu\phi\cdot\phi) \mathcal{F}_7
+\frac{1}{\Lambda^6}(\de_\mu\phi\cdot\de^\mu\phi)(\de_\nu\phi\cdot\phi)^2\mathcal{F}_8
\nonumber\\
&
+\frac{1}{\Lambda^6}(\de_\mu\phi\cdot\de_\nu\phi)(\de^\mu\phi\cdot\phi)(\de^\nu\phi\cdot\phi) \mathcal{F}_9
+\frac{1}{\Lambda^8}(\de_\mu\phi\cdot\phi)^4 \mathcal{F}_{10}\,.
\nonumber
\end{align}
Here the dot indicates the contraction of two 4-vectors: $\phi\cdot\phi=\phi^i\phi^j\delta_{ij}$ and $\mathcal{V},\mathcal{F}_k$ are generic functions of $(\phi\cdot\phi)/\Lambda^2$.
The requirement of a canonical kinetic term translates into ${\mathcal{F}_a(0)=1}$. As for the examples in \S\ref{sec:examples}, the basis structure with and without EOM redundancies has been cross-checked with {\tt BasisGen}. Upon imposing preservation of the custodial symmetry, the number of operators with 2 derivatives + 4 fields, and 4 derivatives + 2 fields in the Green's basis of Eq.~\eqref{eq.smeft_L} matches the number in~\cite{Gherardi:2020det}. The number of operators with 2 derivatives + 6 fields and 4 derivatives + 4 fields match those in~\cite{Chala:2021pll}.\footnote{There are 3 custodial-preserving terms in~\cite{Gherardi:2020det}: $O_{H\square},O'_{HD},O_{DH}$. There are 8 in~\cite{Chala:2021pll}: $O_{\phi^6}^{(1)}$, $O_{\phi^6}^{(3)}$, $O_{\phi^4}^{(3)}$, $O_{\phi^4}^{(4)}$, $O_{\phi^4}^{(10)}$, ($O_{\phi^4}^{(1)}+O_{\phi^4}^{(2)}$), ($O_{\phi^4}^{(8)}+2O_{\phi^4}^{(11)})$, ($O_{\phi^4}^{(6)}+O_{\phi^4}^{(12)}$). }

While the connection to (jet) bundle geometry is best studied considering the 4 scalars as the components of a real vector, it can be useful to make contact with the more familiar form in terms of the $SU(2)$ Higgs doublet, by exploiting the isomorphism
\be
O(4)\cong \left(\frac{SU(2)_L\times SU(2)_R}{\mathbbm{Z}_2}\right)\rtimes\mathbbm{Z}_2\, , 
\ee
where the quotient by $\Z_2$ identifies the central elements of $SU(2)_L$ and $SU(2)_R$; the part in parentheses is $SO(4)$, which recall is a normal subgroup of $O(4)$.

Explicit expressions for the operators in~\eqref{eq.smeft_L} can be obtained straightforwardly from the following dictionary:
\begin{align}
\de_\mu\phi\cdot\phi &= \de_\mu(H^\dag H)\,,
\\
\de_\mu\phi\cdot\de_\nu\phi &=  (\de_\mu H^\dag)( \de_\nu H) + (\de_\nu H^\dag) (\de_\mu H) \,,
\\
\de_\mu\de_\nu\phi\cdot\phi &=
(\de_\mu\de_\nu H^\dag) H + H^\dag(\de_\mu\de_\nu H)\,,
\\
\de_\mu\de_\nu\phi\cdot\de_\rho\de_\sigma\phi&=
(\de_\mu\de_\nu H^\dag)(\de_\rho\de_\sigma H)
+ 
(\de_\rho\de_\sigma H^\dag)(\de_\mu\de_\nu H)\,,
\\
\de_\mu\de_\nu\phi\cdot\de^\mu\phi&=
(\de_\mu\de_\nu H^\dag) (\de^\mu H) + (\de^\mu H^\dag)(\de_\mu\de_\nu H)\,,
\\
\mathcal{F}(\phi\cdot\phi/\Lambda^2) &= \mathcal{F}(H^\dag H/\Lambda^2)\,.
\end{align}

\subsubsection*{Equation-of-Motion redundancies}
If EOM redundancies are accounted for, the Lagrangian in Eq.~\eqref{eq.smeft_L} can be further reduced. Applying the procedure described in \S\ref{sec:4dRscalar}, we find that a possible IBP+EOM -non-redundant basis is:
\begin{align}
\label{eq.smeft_L_eomreduced}
\cL =&\, \frac{1}{2}\de_\mu\phi \cdot \de^\mu\phi  + (\de_\mu\phi\cdot \phi)^2\mathcal{F}_b - \mathcal{V}
\\
&
+\frac{1}{\Lambda^4}(\de_\mu\phi\cdot\de^\mu\phi)^2 \mathcal{F}_5
+\frac{1}{\Lambda^4}(\de_\mu\phi\cdot\de_\nu\phi)^2 \mathcal{F}_6
\nonumber
\\
&
+\frac{1}{\Lambda^6}(\de_\mu\phi\cdot\de^\mu\phi)(\de_\nu\phi\cdot\phi)^2\mathcal{F}_8
+\frac{1}{\Lambda^6}(\de_\mu\phi\cdot\de_\nu\phi)(\de^\mu\phi\cdot\phi)(\de^\nu\phi\cdot\phi) \mathcal{F}_9
\nonumber\\
&
+\frac{1}{\Lambda^8}(\de_\mu\phi\cdot\phi)^4 \mathcal{F}_{10}\,.
\nonumber
\end{align}
Notice that the function $\mathcal{F}_a$ accompanying the kinetic term has now been removed. This is the case because, with $N\geq 4$ fields it is always possible to map via IBP
\begin{equation}
(\de_\mu\phi\de^\mu\phi)(\phi\cdot\phi)^{N-2}    = (2-N)(\de_\mu\phi\cdot\phi)^2(\phi\cdot\phi)^{N-4} - (\square\phi\cdot\phi)(\phi\cdot\phi)^{N-2}\,,
\end{equation}
and then remove the last term via EOM. 
Eq.~\eqref{eq.smeft_L_eomreduced} is consistent with the SMEFT operator bases in the literature: for instance, the Warsaw basis of dimension-6 operators~\cite{Grzadkowski:2010es} contains 1 custodial-preserving operator with 2 derivatives + 4 fields ($O_{H\square}\sim \mathcal{F}_b$), and none with 4 derivatives + 2 fields. The Murphy basis of dimension-8 operators~\cite{Murphy:2020rsh} also contains 1 operator with 2 derivatives + 6 fields ($Q_{H^6}^{(1)}\sim\mathcal{F}_b$), and 3 with 4 derivatives + 4 fields. However, only 2 combinations of those ($Q_{H^4}^{(6)}\sim \mathcal{F}_5$, $(Q_{H^4}^{(1)}+Q_{H^4}^{(2)})\sim \mathcal{F}_6$) preserve the custodial symmetry.
The interactions in the last two lines of Eq~\eqref{eq.smeft_L_eomreduced} correspond to dimension-10 and higher operators in SMEFT and were cross-checked against the results of Ref. \cite{Harlander:2023psl}.

\subsection{The EFT Lagrangian from 1-jet bundle geometry}
The full 4-derivative Lagrangian in Eq.~\eqref{eq.smeft_L} can be captured by pulling back to spacetime a metric from the 1-jet bundle $J^1E$. This is a real 24-manifold, with local fibred coordinates
\begin{align}
    y^I &=\{x^\mu,u^i,u^i_\mu\}\,.
\end{align}
The metric components must be representations of $O(4)$, $g^{\mu\nu},g^\nu_{\mu j}, g^{\mu\nu}_{ij}$ are symmetric in $\mu\nu$ and $g_{ij},g^{\mu\nu}_{ij}$ are symmetric in $ij$, while this is not enforced for $g_{ij}^\mu$.
The generic form is:
\begin{align}
\label{eq.smeft_metric first}
 g_{ij}^{\mu\nu} &= \eta^{\mu\nu} \delta_{ij} A_0(u)+\eta^{\mu\nu} \frac{u_iu_j}{\Lambda^2} A_1(u)\,,
 \\
  g_{ij}^\mu &= 
  \frac{u^{\mu}_i u_j}{\Lambda^3} B_0(u)
 +\frac{u_iu^{\mu}_j}{\Lambda^3} B_1(u)
 +\delta_{ij}\frac{u\cdot u^{\mu}}{\Lambda^3} B_2(u)
 +\frac{u_iu_j (u\cdot u^{\mu})}{\Lambda^5} B_3(u)\,,
 \\
  g_{ij} &= \delta_{ij} C_0(u) + \frac{u_i u_j}{\Lambda^2} C_1(u) 
  \nonumber\\
  &
 +  \delta_{ij}\frac{u_\rho \cdot u^{\rho}}{\Lambda^4} D_0(u)
 +  \frac{u_{i\rho} u_j^{\rho}}{\Lambda^4} D_1(u)
 +  \left[\delta_{ij}\frac{(u\cdot u^{\rho})^2}{\Lambda^6} +  \frac{u_iu_j(u_\rho\cdot u^{\rho})}{\Lambda^6} \right]\frac{D_2(u)}{2}
 \nonumber
 \\
 & 
 +  \frac{(u_iu_{j\rho}+u_ju_{i\rho})(u\cdot u^{\rho})}{2\Lambda^6} D_3(u)
 +  \frac{u_iu_j (u\cdot u_\rho)^2}{\Lambda^8} D_4(u)\,,
 \\
  g_{\mu j}^\nu &= \delta^\nu_\mu \frac{u_j}{\Lambda}E(u) 
  \nonumber\\
  &
 + \frac{u_j (u^{\nu}\cdot u_\mu)}{\Lambda^5} F_{10}(u)
 + \frac{u_j^{\nu} (u\cdot u_\mu)+(u\cdot u^{\nu}) u_{j\mu}}{2\Lambda^5} F_{11}(u)
 + \frac{u_j (u\cdot u^{\nu}) (u\cdot u_\mu)}{\Lambda^7} F_{12}(u)
  \nonumber\\
  &
  + \delta^\nu_\mu\frac{u_j(u_\rho\cdot u^{\rho})}{\Lambda^5} F_{20}(u)
  + \delta^\nu_\mu\frac{u_{j\rho} (u\cdot u^{\rho})}{\Lambda^5} F_{21}(u)
  + \delta^\nu_\mu\frac{u_j (u\cdot u^{\rho})^2}{\Lambda^7} F_{22}(u)
  \,,
 \\
  g_{\mu j}&= 
 \frac{u_{j\mu}}{\Lambda^2} G_0(u) 
 + \frac{u_j (u\cdot u_\mu)}{\Lambda^4} G_1(u) 
 \nonumber\\
 &
 + \frac{u_{j\mu} (u_\rho\cdot u^{\rho})}{\Lambda^6} H_0(u)
 + \frac{ u_{j\rho} (u_\mu\cdot u^{\rho})}{\Lambda^6} H_1(u)
  + \left[\frac{u_j (u\cdot u_\mu) (u_\rho\cdot u^{\rho})}{\Lambda^8} 
 +\frac{u_{j\mu}(u\cdot u_\rho)^2}{\Lambda^8}
 \right]\frac{H_2(u)}{2}
 \nonumber\\
 &
 + \left[\frac{u_j (u_\mu\cdot u_\rho)(u\cdot u^\rho)}{\Lambda^8} 
 +\frac{u_{j\rho}(u\cdot u_\mu)(u\cdot u^\rho)}{\Lambda^8}
 \right]\frac{H_3(u)}{2}
 + \frac{u_j (u\cdot u_\mu)(u\cdot u_\rho)^2}{\Lambda^{10}} H_4(u)\,,
 \\
 g_{\mu\nu} &= -\frac{1}{2}\eta_{\mu\nu} V(u) 
 \nonumber\\
 &
 +\left[ \frac{(u_\mu \cdot u_\nu)}{\Lambda^4} +\frac{\eta_{\mu\nu}}{4}\frac{(u_\rho\cdot u^{\rho})}{\Lambda^4}\right]\frac{J_{0}(u)}{2}
 + \left[\frac{(u\cdot u_\mu)(u\cdot u_\nu)}{\Lambda^6}+\frac{\eta_{\mu\nu}}{4}\frac{(u\cdot u_\rho)^2}{\Lambda^6}\right]\frac{J_{1}(u)}{2}
 \nonumber\\
 &
 + \left[\frac{(u_\mu \cdot u_\nu)}{\Lambda^4} +\frac{\eta_{\mu\nu}}{4}\frac{(u_\rho \cdot u^{\rho})}{\Lambda^4}\right]\frac{ u_\sigma \cdot u^{\sigma}}{\Lambda^4}\frac{K_{0}(u)}{2}
 + \left[\frac{(u_\sigma\cdot u_\mu)(u^\sigma\cdot u_\nu)}{\Lambda^8} 
 +\frac{\eta_{\mu\nu}}{4}\frac{(u_\sigma\cdot u_\rho)^2}{\Lambda^8}\right]\frac{K_{1}(u)}{2}
 \nonumber\\
 &
 + \left[ \frac{(u\cdot u_\mu)(u\cdot u_\nu)(u_\sigma \cdot u^{\sigma})}{\Lambda^{10}} 
 +\frac{\eta_{\mu\nu}}{4}\frac{(u\cdot u_\rho)^2(u_\sigma\cdot u^{\sigma})}{\Lambda^{10}}
 + \frac{(u_\mu\cdot u_\nu)(u\cdot u_\sigma)^2}{\Lambda^{10}} 
 \right]\frac{K_{2}(u)}{3}
 \nonumber\\
 &
 + \left[\frac{(u\cdot u_\mu) (u^\sigma\cdot u_\nu)+(u^\sigma\cdot u_\mu)(u\cdot u_\nu)}{2\Lambda^{7}} 
  +\frac{\eta_{\mu\nu}}{4}\frac{(u\cdot u_\rho) (u^\sigma \cdot u^{\rho})}{\Lambda^{7}}\right]
  \frac{(u\cdot u_\sigma)}{\Lambda^3}
 \frac{K_{3}(u)}{2}
 \nonumber\\
 &
 +\left[ \frac{(u\cdot u_\mu)(u\cdot u_\nu) }{\Lambda^{6}} +\frac{\eta_{\mu\nu}}{4}\frac{(u\cdot u_\rho)^2}{\Lambda^{6}}
 \right]\frac{(u\cdot u_\sigma)^2}{\Lambda^{6}} \frac{K_{4}(u)}{2}
 \,,  
 \label{eq.smeft_metric last}
\end{align}
where $V, A_n\dots K_n$ are functions of $(u\cdot u)/\Lambda^2$, we implicitly defined $u_i=\delta_{ik} u^k$ and the dot denotes a contraction of internal indices.
We can also take $g_{\nu i} = g_{\mu j}, g_{\nu i}^\mu=g_{\mu j}^\nu, g_{ij}^\nu=g_{ij}^\mu$ upon index relabeling and we grouped together structures that yield identical results upon mapping to the Lagrangian.

Pulling back along the section $j^1\phi$ and contracting with the inverse metric 
\begin{align}
\mathcal{L}[\phi,g] &= \frac{1}{2}\left\langle \eta^{-1},\, (j^1\phi)^*g \right\rangle
&
\text{where}\quad
\eta^{-1}&=\eta^{\rho\sigma}\de_\rho\otimes\de_\sigma\,,
\end{align}
we get
\begin{align}
\mathcal{L}[\phi,g] =&\, \frac{\Lambda^4}{2} \eta^{\mu\nu} g_{\mu\nu} + \Lambda^2 g_{\mu j}\de^\mu\phi^j
+\Lambda g_{\mu j}^\nu\de^\mu\de_\nu\phi^j
+\frac{1}{2}g_{ij}\de^\mu\phi^i\de^\mu\phi^j
\\
&
+\frac12\frac{g_{ij}^\mu}{\Lambda}\de_\mu\de_\nu\phi^i \de^\nu\phi^j
+\frac12\frac{g_{ij}^\nu}{\Lambda}\de^\mu\phi^i\de_\mu\de_\nu\phi^j
+\frac{1}{2}\frac{g_{ij}^{\mu\nu}}{\Lambda^2} \de_\rho\de_\mu\phi^i\de^\rho\de_\nu\phi^j
\nonumber
\\[2mm]
=&\,
 (\de_\mu\phi\cdot\de^\mu\phi)\frac{C_0+2G_0+ J_{0}}{2}
+\frac{(\de_\mu\phi\cdot\phi)^2}{\Lambda^2} \frac{C_1+2G_1+J_1}{2}
+(\square\phi\cdot \phi) E
-\Lambda^4 V
\nonumber\\
&
+\frac{(\de_\mu\de_\nu\phi\cdot\de^\mu\de^\nu\phi)}{\Lambda^2}\frac{A_0}{2}
+\frac{(\de_\mu\de_\nu\phi\cdot\phi)(\de^\mu\de^\nu\phi\cdot\phi)}{\Lambda^4}\frac{A_1}{2}
\nonumber\\
&
+\frac{(\de_\mu\de_\nu\phi\cdot\de^\mu\phi)(\de^\nu\phi\cdot\phi)}{\Lambda^4}\frac{B_0+B_1+2B_2+2F_{11}}{2}
+
\frac{(\de_\mu\de_\nu\phi\cdot \phi) (\de^\mu\phi\cdot\de^\nu\phi)}{\Lambda^4}\frac{B_0+B_1+ 2F_{10}}{2}
\nonumber\\
&
+\frac{(\square\phi\cdot\phi)(\de_\mu\phi\cdot\de^\mu\phi)}{\Lambda^4}F_{20}
+\frac{(\square\phi\cdot\de_\mu\phi)(\de^\mu\phi\cdot\phi)}{\Lambda^4} F_{21}
\nonumber\\
&
+\frac{(\de_\mu\phi\cdot\de^\mu\phi)^2}{\Lambda^4} \frac{D_0+2H_0+K_{0}}{2}
+\frac{(\de_\mu\phi\cdot\de_\nu\phi)^2}{\Lambda^4}\frac{D_1+ 2H_1+K_{1}}{2}
\nonumber\\
&
+\frac{(\de_\mu\de_\nu\phi\cdot\phi)(\de^\mu\phi\cdot\phi)(\de^\nu\phi\cdot\phi)}{\Lambda^6}(B_3+F_{12})
+\frac{(\square\phi\cdot\phi)(\de_\mu\phi\cdot\phi)^2}{\Lambda^6} F_{22}
\nonumber\\
&
+\frac{(\de_\mu\phi\cdot\de^\mu\phi)(\de_\nu\phi\cdot\phi)^2}{\Lambda^6} \frac{D_2+2H_2+K_{2}}{2}
\nonumber\\
&
+\frac{(\de_\mu\phi\cdot\de_\nu\phi)(\de^\mu\phi\cdot\phi)(\de^\nu\phi\cdot\phi)}{\Lambda^6}\frac{D_3+2H_3+K_{3}}{2}
\nonumber\\
&
+\frac{(\de_\mu\phi\cdot\phi)^4}{\Lambda^8}\frac{D_4+2H_4+K_{4}}{2}
\label{eq.smeft_jet}\,.
\end{align}
The Lagrangian obtained from the 1-jet metric manifestly matches the complete Green's basis of operators with up to 4 derivatives and arbitrary number of fields, presented in Eq.~\eqref{eq.smeft_L}.
The only leftover IBP redundancies are in the $E$, $F_{20},F_{21}$ and $F_{22}$ terms, that can be recast as\footnote{As in all the previous examples, we find as many independent IBP as operators with boxes, at each fields and derivatives multiplicity (up to 4). Therefore the set of all box-less operators forms a Green's basis.}
\begin{align}
(\square \phi\cdot\phi) E =& - (\de_\mu\phi\cdot\de^\mu\phi) E - 2\frac{(\de_\mu\phi\cdot\phi)(\de^\mu\phi\cdot\phi)}{\Lambda^2}\frac{dE}{d(\phi\cdot\phi/\Lambda^2)}\,, 
\\
(\square\phi\cdot\phi)(\de_\mu\phi\cdot\de^\mu\phi) F_{20}=&
-(\de_\mu\phi\cdot\de^\mu\phi)^2 F_{20}
-2(\de_\nu\de_\mu\phi\cdot\de^\mu\phi)(\de^\nu\phi\cdot\phi)F_{20}
\nonumber\\
&
-\frac{2(\de_\mu\phi\cdot\de^\mu\phi)(\de_\nu\phi\cdot\phi) (\de^\nu\phi\cdot\phi)}{\Lambda^2}\frac{dF_{20}}{d(\phi\cdot\phi/\Lambda^2)}
\\
(\square\phi\cdot\de_\mu\phi)(\de^\mu\phi\cdot\phi) F_{21} =&
-(\de_\mu\de_\nu\phi\cdot\de^\nu\phi)(\de^\mu\phi\cdot\phi)F_{21}
-(\de_\mu\de_\nu\phi\cdot\phi)(\de^\mu\phi\cdot\de^\nu\phi)F_{21}
\nonumber\\
&
-(\de_\mu\phi\cdot\de_\nu\phi)^2 F_{21} 
- \frac{2(\de_\mu\phi\cdot\de_\nu\phi)(\de^\mu\phi\cdot\phi)(\de^\nu\phi\cdot\phi)}{\Lambda^2}\frac{dF_{21}}{d(\phi\cdot\phi/\Lambda^2)}
\\
(\square\phi\cdot\phi)(\de_\mu\phi\cdot\phi)^2 F_{22} =&
-(\de_\nu\phi\cdot\de^\nu\phi)(\de_\mu\phi\cdot\phi)^2 F_{22}
-2(\de_\mu\de_\nu\phi\cdot\phi)(\de^\nu\phi\cdot\phi)(\de^\mu\phi\cdot\phi)F_{22}
\nonumber\\
&
-2(\de_\mu\phi\cdot\de_\nu\phi)(\de^\mu\phi\cdot\phi)(\de^\nu\phi\cdot\phi) F_{22}
\nonumber\\
&
-\frac{2(\de_\mu\phi\cdot\phi)^2(\de_\nu\phi\cdot\phi)^2}{\Lambda^2}\frac{dF_{22}}{d(\phi\cdot\phi/\Lambda^2)}
\end{align}
such that 
\begin{align}
\mathcal{L}[\phi,g] =&\, \text{IBP } \circ \frac{1}{2}\left\langle \eta^{-1},\, (j^1\phi)^*g \right\rangle
\label{eq.smeft_ibp}
\\
=&\,
(\de_\mu\phi\cdot\de^\mu\phi)\frac{C_0+2G_0+ J_{0}-2E}{2}
+\frac{(\de_\mu\phi\cdot\phi)^2}{\Lambda^2} \frac{C_1+2G_1+J_{1} -4E'}{2}
-\Lambda^4 V
\nonumber\\
&
+\frac{(\de_\mu\de_\nu\phi\cdot\de^\mu\de^\nu\phi)}{\Lambda^2}\frac{A_0}{2}
+\frac{(\de_\mu\de_\nu\phi\cdot\phi)(\de^\mu\de^\nu\phi\cdot\phi)}{\Lambda^4}\frac{A_1}{2}
\nonumber\\
&
+\frac{(\de_\mu\de_\nu\phi\cdot\de^\mu\phi)(\de^\nu\phi\cdot\phi)}{\Lambda^4}\frac{B_0+B_1+2B_2+2F_{11}-4F_{20}-2F_{21}}{2}
\nonumber\\
&+
\frac{(\de_\mu\de_\nu\phi\cdot \phi) (\de^\mu\phi\cdot\de^\nu\phi)}{\Lambda^4}\frac{B_0+B_1+ 2F_{10}-2F_{21}}{2}
\nonumber\\
&
+\frac{(\de_\mu\phi\cdot\de^\mu\phi)^2}{\Lambda^4} \frac{D_0+2H_0+K_{10}-2F_{20}}{2}
+\frac{(\de_\mu\phi\cdot\de_\nu\phi)^2}{\Lambda^4}\frac{D_1+ 2H_1+K_{1}-2F_{21}}{2}
\nonumber\\
&
+\frac{(\de_\mu\de_\nu\phi\cdot\phi)(\de^\mu\phi\cdot\phi)(\de^\nu\phi\cdot\phi)}{\Lambda^6}(B_3+F_{12}-2F_{22})
\nonumber\\
&
+\frac{(\de_\mu\phi\cdot\de^\mu\phi)(\de_\nu\phi\cdot\phi)^2}{\Lambda^6} \frac{D_2+2H_2+K_{2}-4F_{20}'-2F_{22}}{2}
\nonumber\\
&
+\frac{(\de_\mu\phi\cdot\de_\nu\phi)(\de^\mu\phi\cdot\phi)(\de^\nu\phi\cdot\phi)}{\Lambda^6}\frac{D_3+2H_3+K_{3}-4F_{21}'-4F_{22}}{2}
\nonumber\\
&
+\frac{(\de_\mu\phi\cdot\phi)^4}{\Lambda^8}\frac{D_4+2H_4+K_{4}-4F_{22}'}{2}\,,
\end{align}
where $E'=dE/d(\phi\cdot\phi/\Lambda^2)$ and analogously for the other functions.

\subsection{Beyond the custodial limit} \label{sec:beyond-custard}
In actual SMEFT/HEFT, the symmetry $O(4)$ is not exact: 
the electroweak vacuum preserves only a $O(3)\sim SU(2)_{L+R}/\mathbbm{Z}_2$ subgroup, namely the custodial symmetry. This is however broken explicitly by the gauging of the hypercharge $U(1)$, which (in the absence of fermions) coincides with the group spanned by the 3rd generator of $SU(2)_R$. While invariance under the gauged $SU(2)_L\times U(1)$ subgroup must clearly be respected, it is generally possible to include operators that are not invariant under the groups spanned by the first two generators of $SU(2)_R$, and therefore break the custodial $O(3)$.

We can ask whether custodial-violating effects can be described in the jet bundle formalism. Indeed, custodial-violating terms can be added to the 1-jet metric without posing any problem. More in general, any internal symmetry structure of the Lagrangian can always be accommodated in the 1-jet bundle by choosing suitable metric entries. Even situations with no symmetry at all can be reproduced, as illustrated in the two scalars example in \S\ref{sec:no_symmetry}.

In the specific case of SMEFT/HEFT custodial violation, they can be introduced allowing for contractions with the 3rd generator of $SU(2)_R$: given two $O(4)$ vectors $u^i, v^j$,
\begin{nalign}
    & u\cdot v = u^i v^j \delta_{ij}\quad &&\text{is custodial preserving}, \\
    &u\, t^{3R} v = u^i v^j t^{3R}_{ij} \quad &&\text{is custodial violating} . \\
\end{nalign}
If $O(4)$ vectors $\phi^i,i=1\dots4$ are defined such that the mapping to $SU(2)$ doublets is
\begin{equation}
    H = \frac{1}{\sqrt2}\binom{\phi^2+i\phi^1}{\phi^4-i\phi^3}\,,
\end{equation}
then an explicit form for $t^{3R}_{ij}$ is 
\begin{align}
    t^{3R} = \frac{1}{2}\begin{pmatrix}
        & 1&&\\-1&&&\\&&&-1\\&&1&
    \end{pmatrix}\,.
\end{align}
While in this work we do not aim at constructing a complete and non-redundant basis including this class of operators, a generalization  would be straightforward.
As a practical example, the metric terms
\begin{align}
g_{ij} &= t^{3R}_{ik} t^{3R}_{jl}\,\frac{u^k u^l}{\Lambda^2}\,, 
&
g_{\mu\nu} &= t^{3R}_{ik} t^{3R}_{jl}\,\frac{u^i_\nu u^{j\nu} u^k_\rho u^{l\rho}}{\Lambda^8} \,,
\end{align}
pull back respectively to the custodial-violating dimension-6 and dimension-8 operators
\begin{align}
  (\de_\mu\phi \, t^{3R} \,\phi)^2 &= (\de_\mu H^\dag H)(H^\dag \de^\mu H)\,,
  \\
  (\de_\mu\phi \cdot t^{3R} \cdot \de_\nu\phi)^2 &= (\de_\mu H^\dag \de_\nu H - \de_\nu H^\dag \de_\mu H)^2 \,,
\end{align}
after using our usual prescription for constructing the Lagrangian.

\subsection{Topological terms} \label{sec:top-terms}

In \S \ref{sec:other_terms} we mentioned the possibility of topological terms in scalar EFTs. The Higgs EFTs we consider in this Section admit such topological terms (with a similar story playing out for an $O(N)$-invariant theory of $N$ real scalars in $d=N$ spacetime dimensions). We briefly digress in this Subsection to sketch the construction of these topological terms, for the 4d $O(4)$-invariant theory of 4 real scalars considered in this Section. Given such terms are not the focus of our geometric jet bundle formalism -- the topological terms are, after all, those that explicitly do {\em not} require a geometry -- our discussion here is neither rigorous nor general. But we believe it is sufficient for the particular example we consider, up to the omission of possible torsion effects that cannot na\"ively be captured by differential forms (but can be captured, for example, by differential cohomology~\cite{Davighi:2020vcm}).

To understand these topological terms, it is enough to stick with the field space bundle $(E,\Sigma, \pi)$ introduced in \S \ref{sec:fib-bundles}, {\em i.e.} we do not need to pass to the 1-jet bundle $J^1 E$.
Locally, we adopt our usual fibred coordinate system $(x^\mu, u^i)$ and choose a section $\phi \in \Gamma_x(\pi)$ with which to pull back objects. The topological term is proportional to the following operator in the Lagrangian:\footnote{Note that $\delta_{ij}$ and $\epsilon_{ijkl}$ are the only $O(4)$ invariant tensors we can use to contract fundamental $O(N)$ indices to form singlets. While the Kronecker-delta is used ubiquitously in our constructions of invariant metric terms and EFT operators, the epsilon tensor appears only in the topological terms discussed in this Subsection, and so it does not appear when formulating our EFTs perturbatively.} 
\begin{equation}
    \mathcal{O}_{\rm{top}} \propto \frac{1}{\Lambda^4}\epsilon_{\mu\nu\rho\sigma}\epsilon_{ijkl}\de^\mu\phi^i\de^\nu\phi^j\de^\rho\phi^k\de^\sigma\phi^l\, .
\end{equation}
The corresponding term in the action is `topological' because it is in fact the integral of an $O(4)$-invariant differential form. This means it is independent of any metric -- even that on spacetime which appears in all other Lagrangian terms captured by our construction. Here the differential form in question is a 4-form $A \in \Omega^4(E) $ that is closed ($dA=0$) but need not be exact (its integral over a closed submanifold can be non-zero), which can be written in our local coordinate chart as
\begin{equation}
    A|_{\mathcal{U}_\Sigma \times \mathcal{F}_M} = du^1 \wedge du^2\wedge du^3\wedge du^4  \, .
\end{equation}
The corresponding action is obtained by pulling back along the section $\phi$ and integrating,
\begin{equation} \label{eq:top-term-example}
    S_{\rm{top}} = \frac{\theta}{2\pi V}\int_{\Sigma} \phi^\ast A, \qquad \theta \in \R/2\pi \Z\, ,
\end{equation}
which will give the operator structure $\mathcal{O}_{\rm{top}}$.
Here $V \propto \Lambda^4$ is an appropriate normalisation factor whose precise value is subtle, but which can be roughly thought of as the (dimensionful) `volume' of a typical fibre $M$, such that the action is dimensionless and the coupling constant $\theta$ is $2\pi$ periodic, 
analogous to a `theta-angle' in gauge theories such as QCD. 

Because the 4-form $A$ that we pull back and integrate is closed, it is locally (but not necessarily global) exact by the Poincar\'e lemma. This means that, on our local coordinate patch, the Lagrangian term $\mathcal{O}_{\rm{top}}$ is a total derivative. This term in the action therefore gives no contribution to the classical equations of motion, nor does it contribute to any Feynman diagrams. We therefore choose to neglect such terms in our main discussion -- but we caution that this does not preclude their having physical effects {\em non}-perturbatively.\footnote{Indeed, depending on the topology of the field space fibre $M$, and because $A$ is not exact, the topological term can seemingly be activated by instantons, corresponding to maps in non-trivial homotopy classes in $\pi_4(M)$. For example, if $M=S^4$ as in the Minimal Composite Higgs model~\cite{Agashe:2004rs}, we have $\pi_4(S^4) = \Z$ characterised by maps of different integer winding number. A scaling argument~\cite[Appendix A]{Davighi:2018xwn} suggests these instanton effects are expected to be important close to the UV cut-off scale (in contrast to the more familiar story for 4d gauge theory, where instantons give important effects in the IR).}

The same is true for a (seemingly) more general class of terms that we can try to build using the $\epsilon$ tensor. For example, consider a more general operator structure of the form
\begin{align}
\mathcal{O}_{\rm top}^{(1)} =\frac{1}{\Lambda^{4}} \, f\left(\frac{\phi\cdot \phi}{\Lambda^2}\right)\, \epsilon_{\mu\nu\rho\sigma}\epsilon_{ijkl}\,\de^\mu\phi^i\de^\nu\phi^j\de^\rho\phi^k\de^\sigma\phi^l \,,
\end{align}
for any suitably smooth function with compact support $f(x)$. This is again topological, obtained by pulling back a differential form $A^{(1)}$ along the section $\phi$, where $A^{(1)}$ has the (local) form
\begin{equation}
    A^{(1)}|_{\mathcal{U}_\Sigma \times \mathcal{F}_M} = f\left(\frac{u\cdot u}{\Lambda^2}\right)\, du^1 \wedge du^2\wedge du^3\wedge du^4  \, .
\end{equation}
Again, it is easy to verify that the differential form $A^{(1)}$ is closed, {\em viz.} $dA^{(1)} \propto f^\prime u^i du^i \wedge A = 0$, where $f^\prime = df/dx$, and where the last equality holds because $du^i \wedge du^i$ (no sum) vanishes for any $i \in \{1,...,4\}$.
Therefore, again by the Poincar\'e lemma, any operator like $\mathcal{O}_{\rm top}^{(1)}$ is also a total derivative.\footnote{One might ask if there are any other contractions we can play with using the $\epsilon$ tensor that give terms which are {\em not} total derivatives, and the answer is no. For example, one might consider an operator structure $\mathcal{O}_{\rm top}^{(2)} = \epsilon_{\mu\nu\rho\sigma}\epsilon_{ijkl}\,  \phi^i \de^\nu\phi^j\de^\rho\phi^k\de^\sigma\phi^l\de^\mu\phi^m \,\phi^m$. But $\mathcal{O}_{\rm top}^{(2)} d^4 x \propto \phi^\ast (\epsilon_{ijkl} u^i du^j \wedge du^k \wedge du^l) \wedge du^m u^m = (u^1 du^2 \wedge du^3 \wedge du^4) \wedge du^1 u^1 + \text{perms} \propto (u \cdot u) \, du^1 \wedge du^2 \wedge du^3 \wedge du^4$, and so this is the same form as $A^{(1)}$. Finally, we remark that a differential form like $A^{(1)}$, for non-constant function $f$, will likely be not just closed but also exact, in which case it has no effects even in the presence of instanton configurations, since it evaluates identically to zero on any closed manifold. }

\section{From Jets to Amplitudes} \label{sec:amplitudes}

To conclude the paper, in this Section we seek to connect the higher-derivative EFT Lagrangians that we have constructed using geometry on 1-jet bundles with physical observables -- namely, some basic amplitudes in these scalar theories.
Our study here is far from comprehensive, but we nonetheless already see some advantages of going to the jet bundle geometry. 

\subsection{Normal coordinates on the jet bundle (and why they fail)}

In Section \ref{sec:amplitudes-part1} we reviewed the use of normal coordinates in the geometric formulation of EFTs, showcasing their ability to relate amplitudes to geometric invariants by offering an efficient coordinate system in which to expand the metric around a particular point. 

In this paper, we have re-formulated scalar EFTs first using bundles $(E,\Sigma,\pi)$, which (besides other general advantages) allows one to incorporate the 0-derivative potential terms into the metric, alongside the 2-derivative terms. Additionally, the bundle construction provides a starting point for passing to higher jet bundles which allow one to further incorporate higher-derivative terms into geometry. In particular, we have seen in detail that geometry on the 1-jet bundle $J^1 E$ allows one to capture the full 4-derivative EFT. A natural question to ask is whether normal coordinates can be used to expand our jet bundle metric, in a similar fashion to the 2-derivative case considered in \S \ref{sec:amplitudes-part1}, in an effort to obtain generalised formulae relating $n$-point correlators to components of the jet bundle Riemann tensor. 

It turns out that such an attempt would be doomed to fail, in general. This is because, even though normal coordinates still exist (since the total spaces of the various bundles we consider are themselves pseudo-Riemannian manifolds), the map to normal coordinates is not guaranteed to respect the bundle structure of $(E,\Sigma,\pi)$,\footnote{
A simple but technical explanation lies in the fact that the map to normal coordinates around a point $m \in E$ is defined by the exponential from some neighbourhood $V \subset T_m E$ to $U \subset E$, as described above in footnote~\ref{foot:NC}. But for the transformation to preserve the fibres, we require $V \subset \pi^* (T_{\pi(m)}\Sigma)$, which is not necessarily true.} likewise of the 1-jet bundle.
The additional structure of a bundle $(E,\Sigma,\pi)$
is crucial to our formulation of the EFT, and this bundle structure imposes some restrictions. In particular, the fields are defined to be sections of the bundle, $\phi \in \Gamma(\pi)$. 
{\em Any} physically meaningful transformation that we can do must therefore map the bundle to another bundle; if not, then we can no longer even define fields (because we will have lost the notion of sections).
For example, we have already discussed how field redefinitions, including derivative field redefinitions, may be defined as suitable changes of section, $\phi \to \psi$ with $\phi,\psi \in \Gamma(\pi)$; the non-derivative field redefinitions can moreover be identified with bundle morphisms $(E,\Sigma,\pi) \to (F,\Omega,\rho)$, which induce maps $\Gamma(\pi) \to \Gamma(\rho)$ at the level of sections. 
In either case, it is straightforward to see that 
the condition $x^\mu \circ \phi = x^\mu$ on our local fibred coordinate system $(x^\mu, u^i)$, and the simple requirement that any section is locally an inverse of the projection $\pi$, restricts us to transformations that act on coordinates as
\begin{nalign} \label{eq:bundle-preserving}
    x^\mu \mapsto x'^\mu(x^\nu)\, ,\\
    u^i \mapsto u'^i(x^\nu,u^j)\, .\\
\end{nalign}
This is manifestly true in the case of non-derivative redefinitions, as can be seen from the commutative diagram in (\ref{eq:CD-bundle}), but should hold more generally for any physical `transformation' of the bundle $(E,\Sigma,\pi)$ that we use to define our EFT.

The map to normal coordinates can violate the basic requirement (\ref{eq:bundle-preserving}) for preserving the (essential) bundle structure.
To illustrate this, we consider two counter-examples in which one cannot satisfy the conditions for normal coordinates while preserving the bundle structure, both in the case of a single real scalar in 4d (the subject of \S \ref{sec:4dRscalar}). 

\paragraph{Counter-example: condition 1 for normal coordinates.}

Recall from \S \ref{sec:amplitudes-part1} that the first condition (\ref{eq:NCs_1}) for normal coordinates is that the metric is diagonal (with entries further normalised to $\pm 1$, determining the signature). Given an initial set of coordinates $x^I = (x^\mu,u,u_\mu)$ on the 1-jet bundle, there are choices of invariant metrics written in those coordinates that are not diagonal (with off-diagonal elements indeed being necessary to match onto certain EFTs via our construction of the Lagrangian). This means that to pass to coordinates satisfying condition 1, we need to first do a rotation, and this rotation can mix $x^\mu$ with the other coordinates, violating (\ref{eq:bundle-preserving}).

For an explicit example,
using the general expressions (\ref{eq.1scalar_metric_first}--\ref{eq.1scalar_metric_last}) for the invariant metric in the case of one real scalar in 4d, 
and evaluating at a point $p \in J^1E$ such that $u(p)= 0 =u_\mu(p) $, we see that the metric tensor $g_{IJ}(p)$ takes the following block form
\begin{equation} \label{eq:eg_NC1}
    g_{IJ} = \begin{pmatrix}
        \eta_{\mu \nu} V(0) & 0 & \delta^\mu_\nu E(0)\\
        0 & C(0) & 0\\
        \delta^\mu_\nu E(0) & 0 & \eta^{\mu \nu} A(0)
    \end{pmatrix}
\end{equation}
Assuming that none of the component functions vanish, it is clear that diagonalizing the metric requires a rotation of the form $(x^\mu,u_\mu) \mapsto O(x^\mu, u_\mu)$ which mixes $x^\mu$ with $u_\mu$, thus violating the bundle condition (\ref{eq:bundle-preserving}). 

Now, the attentive reader might retort that, for the EFT described by this metric (with all functions non-zero), the function $E(0)$ appearing on the off-diagonal is actually redundant, in that it pulls back to the $\box \phi$ operator that is equivalent to the kinetic term after using IBPs; so if we take $E(0)$ to zero, the metric is already (block) diagonal without the need to mix $x^\mu$ with other coordinates. However, there are EFTs where we {\em need} to keep the $E(0)$ function; in particular, if $A(0)=0$, in which case the EFT is purely 2-derivative and the $E(0)$ term is required for invertibility of the metric. In that case, diagonalising the metric clearly requires a non-trivial rotation $(x^\mu,u_\mu) \mapsto O(x^\mu, u_\mu)$, violating the bundle condition.

\paragraph{Counter-example: condition 3 for normal coordinates.}

We now suppose that one has a metric for which condition 1 is already satisfied, for example by taking the metric (\ref{eq:eg_NC1}) and setting $E(0)=0$ and $A(0) \neq 0$, and try to thence find coordinates that further satisfy condition 3 (\ref{eq:NCs_3}).
The determination of the explicit coordinate transformation that takes us to normal coordinates on the total space can be found order-by-order via an iterative procedure, but for our demonstration it will suffice to stop at the first order, at which point the obstruction shall already arise.

We start with a local fibred coordinate system $x^I = (x^\mu,u,u_\mu)$ on a local patch of the 1-jet manifold. To try to construct normal coordinates in a perturbative fashion, consider a new coordinate system $y^I = (y^\mu,v,v_\mu)$, defined by $x^I = y^I + a^I_{JK} y^J y^K$, where $a^I_{JK}$ is a tensor of constants satisfying $a^I_{JK} = a^I_{KJ}$. When evaluating the metric components $\bar{g}_{IJ}$ in the $y_I$ coordinate system, at the origin in these coordinates, we have
\begin{nalign}
    \frac{\partial}{\partial y^K} \bar{g}_{IJ}(0) &= \frac{\partial}{\partial y^K} \bigg( \frac{\partial x^R}{\partial y^I} \frac{\partial x^S}{\partial y^J} g_{RS} \bigg) \bigg|_{y=0}\\
    &= a^R_{IK} g_{RJ}(0) + a^S_{JK} g_{IS} (0) + \frac{\partial}{\partial x^K} g_{IJ}(0)
\end{nalign}
We see that the number of indices and the symmetries of $a^I_{JK}$ and $ \frac{\partial}{\partial x^K} g_{IJ}(0)$ match. This implies it is possible to choose the components $a^I_{JK}$ such that $\frac{\partial}{\partial y^K} \bar{g}_{IJ}(0) = 0$, thus satisfying a necessary condition for normal coordinates (see \S \ref{sec:amplitudes-part1}), by requiring
\begin{equation}
\label{equation for a}
    \frac{\partial}{\partial x^K} g_{IJ}(0) = - \bigg(a^R_{IK} g_{RJ}(0) + a^R_{JK} g_{RI} (0) \bigg)\, .
\end{equation}
From (\ref{equation for a}) we see that $a^{I}_{JK} = -\Gamma^I_{JK}(0)$, where the symmetries dictate that there are 18 different structures. The Christoffel symbols (with respect to the original $x^I$ coordinates) are as usual given by
\begin{equation}
    \Gamma^I_{JK} = \frac{1}{2}g^{IM} (g_{MJ,K} + g_{MK,J} - g_{JK,M} )\, .
\end{equation}
For our particular example, using the general expressions in Eqs. (\ref{eq.1scalar_metric_first}--\ref{eq.1scalar_metric_last}) for the metric components and lowering all indices of the Christoffel symbols, we find
\begin{nalign}
    &\Gamma_{\rho \mu \nu}(0) = 0, \quad \Gamma_{\rho \mu u}(0) = -\frac{1}{2} \partial_u V(0), \quad \Gamma_{\rho uu} (0) = 0,\\
    &\Gamma_{\rho \mu u}^{\;\;\;\; \nu}(0) = 0, \quad \Gamma_{\rho u u}^{\; \; \; \; \mu}(0) = \frac{1}{2} \delta^\mu_\rho \bigg(\frac{1}{\Lambda^2}G(0) + \partial_u E(0) \bigg), \quad \Gamma_{\rho u u}^{\; \; \mu \nu} (0)=0,\\
    &\Gamma_{uuu}(0) = \frac{1}{2} \partial_\mu C(0),\quad \Gamma_{uu \mu} (0) = 0,\quad \Gamma_{u \mu \nu} = \frac{1}{2} \partial_u V(0),\\
    &\Gamma_{uuu}^{\;\;\;\; \mu} (0) = 0, \quad \Gamma_{u\mu u}^{\;\;\;\; \nu} (0) = \frac{1}{2} \delta^\nu_\mu  \bigg(\frac{1}{\Lambda^2}G(0) - \partial_u E(0) \bigg), \quad \Gamma_{uuu}^{\;\; \mu \nu} (0) = \eta^{\mu \nu} \bigg( \frac{1}{\Lambda^2}B(0) - \frac{1}{2} \partial_u A(0) \bigg),\\
    &\Gamma_{uuu}^{\rho \mu \nu} (0) = 0, \quad \Gamma_{uuu}^{\mu \nu} = \frac{1}{2} \eta^{\mu \nu} \partial_u A(0), \quad  \Gamma_{uu \rho}^{\mu \nu}(0)= 0\\
    & \Gamma_{uuu}^{\mu} (0) = 0, \quad \Gamma_{u \nu \rho}^{\mu}(0) = 0, \quad \Gamma_{uu \nu}^{\mu}(0) = \frac{1}{2} \delta^\mu_\nu \bigg(\frac{1}{\Lambda^2} B(0) + \partial_u E(0) - \frac{1}{\Lambda^2} G(0) \bigg)\, .
\end{nalign}
Now let us consider a section $j^1 \psi$, obtained by prolongation of a section $\psi \in \Gamma(\pi)$, with
\begin{nalign}
    y^\mu \circ j^1 \psi &= y^\mu\\
    v \circ j^1 \psi &= \psi\\
    v_\mu \circ j^1 \psi &= \frac{\partial \psi}{\partial y^\mu}
\end{nalign}
But when evaluated on the coordinate map $x^\mu$, this section yields
\begin{nalign}
\label{normal coordinates section x}
    x^\mu \circ j^1 \psi &= (y^\mu - \frac{1}{2}g^{\mu L}(0)\Gamma_{LJK}(0) y^J y^K) \circ j^1 \psi\\
    &= (y^\mu - \frac{1}{2}(g^{\mu \nu}(0) \Gamma_{\nu JK}(0) + g^{\mu u}_{\; \; \nu} (0) \Gamma_{u JK}^{\nu} (0))y^J y^K) \circ j^1 \psi\\
    &= (y^\mu  - (g^{\mu \nu}(0) \Gamma_{\nu \rho u}(0) y^\rho v + g^{\mu \nu}(0) \Gamma_{\nu u u}^{\;\;\rho}(0)v_\rho v\\  &\,\,\,\,\,+ g^{\mu u}_{\; \; \nu} (0) \Gamma_{u uu}^{\nu \rho} (0)) v v_\rho  + g^{\mu u}_{\; \; \nu} (0) \Gamma_{u u \rho}^{\nu} (0))y^\rho v) \circ j^1 \psi \\
    &= y^\mu - (g^{\rho \nu}(0) \Gamma_{\nu \mu u}(0) y^\mu \psi + g^{\rho \nu}(0) \Gamma_{\nu u u}^{\;\;\mu}(0) \psi \partial_\mu \psi \\  &\,\,\,\,\,+ g^{\rho u}_{\; \; \nu} (0) \Gamma_{u uu}^{\nu \mu} (0)) \psi \partial_\mu \psi + g^{\rho u}_{\; \; \nu} (0) \Gamma_{u u \mu}^{\nu} (0))y^\mu \psi)\, , \\
\end{nalign}
thus failing to satisfy the condition (\ref{eq:bundle-preserving}). 

This means that the change of coordinates we wish to implement does not preserve the bundle structure that is essential in formulating the EFT.\footnote{Lest there is any confusion here, thanks to the coordinate independent nature of the metric tensor it is possible to use normal coordinates to obtain the same result as in any other coordinate chart, since the transformation $g_{IJ}(x) \to \bar{g}_{IJ}(y)$ is compensated by $dx^I \otimes dx^J \to dy^I \otimes dy^J$, such that the combination preserves the fibres, although there is little value in using normal coordinates in this way. The map to normal coordinates not being fibre preserving only causes an issue if we try to use it as a smooth map, since then there is no compensating factor.} 
Of course, one can always find example geometries for which going to normal coordinates does happen to be consistent with the bundle structure; from Eq.~\ref{normal coordinates section x} it is easy to see that if the Christoffel symbols or metric components appearing were set to zero, then the map to normal coordinates would indeed satisfy (\ref{eq:bundle-preserving}). But this situation is far from generic, meaning that normal coordinates are not especially useful in our formalism.

\bigskip

Another important reason why normal coordinates are not generally useful in our formalism is that passing to normal coordinates would be inconsistent with the inclusion of any non-constant potential $V$. This can be seen by supposing that we have gone to normal coordinates, and
expanding the metric up to second order, giving
\begin{equation}
    g = g_{IJ}(u) dx^I \otimes dx^J = \bigg(\bar{g}_{IJ} + \frac{1}{3} R_{IJKL}(m) x^K x^L + \dots \bigg) dx^I \otimes dx^J\, .
\end{equation}
If we wish to impose Poincar\'e invariance, then a term such as $R_{i \mu \nu j}x^\mu x^\nu du^i \otimes du^j$ must be set to zero. However, the symmetries of the Riemann tensor state that $R_{i \mu \nu j} = R_{\mu i j \nu}$ and $R_{\mu i j \nu} u^i u^j dx^\mu \otimes dx^\nu$ corresponds to the mass term in the Lagrangian. Therefore, the map to normal coordinates in conjunction with Poincar\'e invariance requires the mass term to vanish and it is easy to see that all terms in the potential are forced to vanish beyond the leading order in the expansion.

\bigskip

Before moving on, we wish to make one last comment about normal coordinates in relation to our jet bundle formalism, which is that the {\em motivation} for passing to normal coordinates is also weaker than it was in the geometric approach that we reviewed in \S \ref{sec:preliminaries}.
An important motivation for passing to normal coordinates comes from the inclusion of the potential $V$ contributions to the geometric formulae for amplitudes: the problem is that the second derivative of the potential, which enters for example in four point amplitudes, does not in general transform like a tensor on field space manifold $M$ due to the appearance of Christoffel symbols in the expression. Normal coordinates circumvent this non-tensorial behaviour because the Christoffel symbols vanish. In contrast, in our bundle-based reformulation of EFTs the potential is itself coming from components of the metric tensor, rather than being added in by hand, so we never face this issue.

\subsection{Expansion of the jet bundle metric}

We now show how the metric on the jet bundle may be expanded, explicitly {\em not} in normal coordinates, but in a way that {\em is} consistent with the iterated fibre bundle structure. As usual, we work with the local fibred coordinate system $\{x^\mu, u^i, u^i_\mu\}$ on a general jet manifold $J^1 E$.\footnote{To avoid any confusion in this Section, we remind the reader that the coordinates $\{x^\mu,u^i,u^i_\mu \}$ are all independent on the jet manifold; in particular $\frac{\partial}{\partial x^\mu} u^i = 0$ (with no relation to $u_\mu^i$).} 
The choice of origin point in the 1-jet manifold around which we expand the metric will be determined by the section $j^1 \phi$ according to
\begin{nalign}
\label{origin for section definition}
    u^i (j^1\phi(p)) = 0\, ,\\
    u^i_\mu (j^1\phi(p)) = 0\, ,\\
\end{nalign}
{\em i.e.} $p \in \Sigma$ is a point in spacetime where the scalar field values $\phi^i$, evaluated in our choice of local fibred coordinates, vanish.
We can consider a completely general scalar theory, with arbitrary spacetime dimensions $d$ and an arbitrary number of fields. While we do not impose any internal symmetries, we will impose Poincar\'e invariance, which simplifies the expansion substantially since:
\begin{itemize}
    \item None of the metric tensor components may depend explicitly on $x^\mu$, {\em i.e.} $\frac{\partial}{\partial x^\mu} g_{IJ}=0$ $\forall I,J$.
    \item All spacetime indices must be contracted into singlets, including those labelling derivative coordinates thanks to (\ref{eq:prolonged-Poincare-formula}), thus any terms in the expansion with an odd number of spacetime indices must vanish.
\end{itemize}
We adopt the following notational conventions for the expansion
\begin{equation}
    \bar{g}_{IJ} := g_{IJ}(p) 
    \quad \textrm{and} \quad 
    g_{IJ,K} := \frac{\partial}{\partial x^K} g_{IJ}\, .
\end{equation}
By Taylor expanding the metric and keeping all contributions to operators with up to four derivatives, we find 
\begin{nalign}
\label{metric expansion jet manifold}
    g_{IJ}(x^K) &dx^I \otimes dx^J =  \sum_{n = 0}^\infty \frac{1}{n!}\bigg( \bar{g}_{\mu \nu,k_1 \dots k_n} u^{k_1} \dots u^{k_n} + \bar{g}_{\mu\nu,qr k_1 \dots k_{n-2}}^{\;\;\;\;\:  \rho \sigma} u^q_\rho u^r_\sigma u^{k_1} \dots u^{k_{n-2}} \\&+ \bar{g}_{\mu\nu,qrst k_1 \dots k_{n-4}}^{\;\;\;\;\:  \rho \sigma \alpha \beta} u^q_\rho u^r_\sigma u^s_\alpha u^t_\beta u^{k_1} \dots u^{k_{n-4}} \bigg) dx^\mu \otimes dx^\nu\\
    &+\sum_{n = 0}^\infty \frac{1}{n!}\bigg(\bar{g}_{ij,k_1 \dots k_n} u^{k_1} \dots u^{k_n} + \bar{g}_{ij,qr k_1 \dots k_{n-2}}^{\;\;\;\; \rho \sigma} u^q_\rho u^r_\sigma u^{k_1} \dots u^{k_{n-2}} \bigg) du^i \otimes du^j\\
    & +\sum_{n = 0}^\infty \frac{1}{n!} \bar{g}_{ij,k_1 \dots k_n}^{\mu \nu} u^{k_1} \dots u^{k_n} du^i_\mu \otimes du^j_\nu \\
    & +2\sum_{n = 0}^\infty \frac{1}{n!} \bigg(\bar{g}_{i \mu,qk_1 \dots k_{n-1}}^{\;\;\;\;\:\rho} u^q_\rho u^{k_1} \dots u^{k_{n-1}} + \bar{g}_{i \mu,qrs k_1 \dots k_{n-3}}^{\;\;\;\;\:\rho \sigma \alpha} u^q_\rho u^r_\sigma u^s_\alpha u^{k_1} \dots u^{k_{n-3}} \bigg) du^i \otimes dx^\mu\\
    & +2\sum_{n = 0}^\infty \frac{1}{n!} \bigg(\bar{g}_{i \mu,k_1 \dots k_n}^{\nu} u^{k_1} \dots u^{k_n} + \bar{g}_{i \mu,qr k_1 \dots k_{n-2}}^{\nu \;\;\:\rho \sigma} u^q_\rho u^r_\sigma u^{k_1} \dots u^{k_{n-2}} \bigg) du^i_\nu \otimes dx^\mu\\
    & + 2\sum_{n = 0}^\infty \frac{1}{n!}  \bar{g}_{i j,qk_1 \dots k_{n-1}}^{\mu \;\;\rho} u^q_\rho u^{k_1} \dots u^{k_{n-1}} du^i_\mu \otimes du^j\\
\end{nalign}
where the factors of 2 appearing in front of off-diagonal blocks match our previous conventions.

We re-iterate that, in the field space approach, the potential poses a problem in such an expansion because it does not transform like a tensor. But within the jet bundle approach, the potential is included in a component of the metric tensor, and thus is packaged into an object that transforms covariantly. The derivatives of the metric that appear in the expansion may be re-expressed in terms of Christoffel symbols and curvature invariants if we so desire, but the translation would be tedious and quickly becomes unwieldy as we pass to higher jets.

\subsection{Geometrizing amplitudes on the jet bundle}
\label{sec:geometrizing amplitudes example}

We are now almost ready to extract the lowest degree $n$-point amplitudes from the expansion (\ref{metric expansion jet manifold}) by recalling our definition of the Lagrangian 
\begin{equation}
    \mathcal{L} = \frac{1}{2} \langle \eta^{-1}, (j^1 \phi)^* g \rangle
\end{equation}
where composing the section $j^1 \phi$ with the coordinate charts yields \begin{nalign}
    x^\mu \circ j^1 \phi = x^\mu \, ,\\
    u^i \circ j^1 \phi = \phi^i\, ,\\
    u^i_\mu \circ j^1 \phi = \partial_\mu \phi^i\, .
\end{nalign}

For the propagator we find 
\begin{nalign}
\label{propagator amplitude}
    \begin{tikzpicture}[baseline = (a.base)]
  \begin{feynman} [inline = (a.base)]
      \vertex (a) {$i$};
      \vertex [right= 2cm of a] (b){$j$};

      \diagram*{
      (a) -- [scalar] (b)};
  \end{feynman}
\end{tikzpicture} \,&= \frac{i}{
\bar{g}_{ij}p^2
+\frac{\eta^{\mu \nu}}{2} \bar{g}_{\mu \nu,i j}^{\;\;\;\;\: \rho \sigma} p_\rho p_\sigma 
+ 2 \left(\bar{g}_{i \mu,j}^{\;\;\;\; \nu}-  \bar g_{i \mu,j}^{\nu}\right) p_\nu p^\mu 
+ \frac{\eta^{\mu \nu}}{2} \bar{g}_{\mu\nu,ij} 
+ \bar{g}_{ij}^{\mu \nu}p^2 p_\mu p_\nu}
\end{nalign}
Requiring a canonical 2-derivative term enforces the relation 
\begin{equation}
    \bar{g}_{ij}+\frac{1}{2d}\eta^{\mu \nu} \eta_{\rho \sigma}\bar{g}_{\mu \nu,i j}^{\;\;\;\;\: \rho \sigma} + \frac{2}{d}\delta^\mu_\nu\left( \bar{g}_{i \mu,j}^{\;\;\;\; \nu}- \bar g_{i \mu,j}^{\nu}\right) = \delta_{ij}\, ,
\end{equation}
reproducing {\em e.g.} (\ref{eq:can-norm-Rscalar-4d}),
which can be thought of as a physics constraint on the jet bundle geometry. More precisely, for a metric that describes a physical EFT, we must be able to find a section $\phi$ such that the kinetic term is canonically normalised in this way, and it is implicit that we have already performed this change of section.

For $\bar g_{ij}^{\mu\nu}\neq0$, avoiding tachyonic instabilities in the propagator~\cite{Brivio:2014} additionally requires that
\begin{equation}
    \eta_{\mu \nu} \bar{g}_{ij}^{\mu \nu} = \begin{pmatrix}
        \lambda_1 & 0 & \dots & 0\\
        0 & \lambda_2 & \dots & 0\\
        \dots & \dots & \dots & \dots\\
        0 & 0 & \dots & \lambda_n
    \end{pmatrix}, 
    \quad
    \lambda_i < 0 \,\,\, \forall i \in \{1,\dots,n=\text{dim}(M)\} \, ,
\end{equation}
Again, it is implicit that we have done a derivative field redefinition, or change of section, to further bring $\eta_{\mu\nu}\bar{g}_{ij}^{\mu\nu}$ into diagonal form, which is always possible up to order $p^6$ corrections. This inequality partly fixes the {\em signature} of the jet bundle metric (in the derivative-coordinate directions), and is the manifestation of {\em positivity} in our geometric formulation. For example, in the case of a single real scalar in 4d (\S \ref{sec:4dRscalar}), and using the parametrization (\ref{eq:1R-scalar-G-basis}), the above conditions become simply
\begin{align}
    \mathcal{F}_0(0) &= 1 \quad \text{(2-derivative term)}, \\
    \mathcal{F}_1(0) &< 0 \quad \text{(4-derivative term)}\, , \label{eq:positivity-F_1}
\end{align}
with the latter being a known positivity bound~\cite{Brivio:2014}.

Continuing, the $3$-point amplitude is\footnote{
In these expressions for the amplitudes, a `dot product' denotes a contraction in spacetime indices, {\em i.e.} $p_q \cdot p_r := \eta^{\mu\nu} (p_q)_\mu (p_r)_\nu$. Furthermore,
Lorentz symmetry, together with the symmetry properties of the metric, fixes the space-time structure of metric components.
For example, $\bar{g}_{\mu \nu,i j k}^{  \;\;\;\;\;\alpha \beta} \eta^{\mu \nu} \sim \eta^{\alpha \beta}$ and $\bar{g}_{i \mu, j k}^{ \;\;\;\;\alpha} \sim \delta_\mu^\alpha$.}
\begin{nalign}
\label{3 point amplitude}
\begin{tikzpicture}[baseline = (b.base)]
  \begin{feynman} [inline = (b.base)]
      \vertex (a) {1};
      \vertex [below= 1cm of a] (f);
      \vertex [right=1 cm of f] (b);
      \vertex [right= 1cm of b] (c) {2} ;
      \vertex [below = 2cm of a] (d) {3} ;

      \diagram*{
      (a) -- [scalar] (b) -- [scalar] (c),
      (d) -- [scalar] (b)
      };
  \end{feynman}
\end{tikzpicture} =
\begin{aligned}[t] & i\bigg(\frac{1}{12}\bar{g}_{\mu \nu, i j k} \eta^{\mu \nu} - \frac{1}{12}\bar{g}_{\mu \nu,i j k}^{  \;\;\;\;\;\alpha \beta} \eta^{\mu \nu} \sum_{q,r} (p_q)_\alpha (p_r)_\beta -\frac{1}{2} \bar{g}_{i j,k} \sum_{q,r} p_q \cdot p_r\\ &- \frac{1}{2}\bar{g}_{i \mu, j k}^{ \;\;\;\;\alpha} \sum_{q,r} (p_q)_\alpha (p_r)^\mu - \frac{1}{2}\bar{g}_{i \mu , j k}^{\nu}  \sum_{q} (p_q)_\nu (p_q)^\mu\\
&+ \frac{1}{2} \bar{g}_{i j,k}^{\mu \nu}  \sum_{q,r} (p_q \cdot p_r) (p_q)_\mu (p_r)_\nu + \bar{g}_{i j,k}^{\mu \;\;  \nu}  \sum_{q,r,s} (p_q \cdot p_r) (p_q)_\mu (p_s)_\nu \\& + \frac{1}{2}\bar{g}_{i \mu,jk}^{\nu  \;\;\; \alpha  \beta} 
    \sum_{q,r,s} (p_{q})_\alpha  (p_{r})_\beta (p_s)_\nu (p_s)^\mu \bigg)\, .
\end{aligned}
\end{nalign}
The $4$-point amplitude is
\begin{nalign}
\label{4 point amplitude}
\begin{tikzpicture}[baseline = (b.base)]
  \begin{feynman} [inline = (b.base)]
      \vertex (a) {1};
      \vertex [below= 1cm of a] (f);
      \vertex [right=1 cm of f] (b);
      \vertex [right= 1cm of b] (c)  ;
      \vertex [below = 2cm of a] (d) {2} ;
      \vertex [right = 2cm of a] (h) {3} ;
      \vertex [below = 2cm of h] (g) {4} ;

      \diagram*{
      (a) -- [scalar] (b) -- [scalar] (g),
      (d) -- [scalar] (b) -- [scalar] (h),
      };
  \end{feynman}
\end{tikzpicture} = \begin{aligned}[t] & i \bigg(\frac{1}{48}\bar{g}_{\mu \nu, i j l k} \eta^{\mu \nu} - \frac{1}{48}\bar{g}_{\mu \nu, i j k l}^{\;\;\;\;\; \alpha  \beta} \eta^{\mu \nu} \sum_{q,r} (p_q)_\alpha (p_r)_\beta - \frac{1}{4}\bar{g}_{i j, k l} \sum_{q,r} p_q \cdot p_r \\& - \frac{1}{6}\bar{g}_{i \mu, j k l}^{\;\;\;\;\,\nu} \sum_{q,r} (p_q)^\mu (p_r)_\nu - \frac{1}{6} \bar{g}_{i \mu,j k l}^{\nu} \sum_{q} (p_q)^\mu (p_q)_\nu \\ &+ \frac{1}{4} \bar{g}_{i j,k l}^{\mu \nu} \sum_{q,r} (p_q \cdot p_r) (p_q)_\mu (p_r)_\nu  + \frac{1}{48}\bar{g}_{\mu \nu, i j k l}^{\;\;\;\; \gamma  \delta  \alpha  \beta}\eta^{ \mu \nu} \sum_{q,r,s,t} (p_q)_\gamma (p_r)_\delta (p_s)_\alpha (p_t)_\beta  \\ &+ \frac{1}{4} \bar{g}_{i j,k l}^{\;\;\;   \alpha  \beta} \sum_{q,r,s,t} (p_q \cdot  p_r) (p_s)_\alpha (p_t)_\beta + \frac{1}{6} \bar{g}_{ i \mu, j k l}^{\;\;\;\;\, \nu  \alpha  \beta} \sum_{q,r,s,t} (p_q)^\mu (p_r)_\nu (p_s)_\alpha (p_t)_\beta  \\ &+ \frac{1}{6}\bar{g}_{i \mu,j k l}^{\nu \;\;\, \alpha  \beta} \sum_{q,r,s}  (p_q)^\mu (p_q)_\nu (p_r)_\alpha (p_s)_\beta + \frac{1}{2}\bar{g}_{i j, k l}^{\mu \;\;  \nu}  \sum_{q,r,s} (p_q \cdot  p_r)  (p_q)_\mu (p_s)_\nu  \bigg)\, .
\end{aligned}
\end{nalign}
To explain a little our notation here: sums over two indices run over values $1 \leq q < r \leq n$, where $n$ is the number of `points' in the amplitude; a sum over $n$ indices runs over all possible permutations, {\em e.g.} $(q,r,s) \in \sigma(123)$ when appearing in $3$-point amplitudes, and $(q,r,s,t) \in \sigma(1234)$ in $4$-point amplitudes; lastly, the sums over three indices in Eq. (\ref{4 point amplitude}) for the  $4$-point amplitude run over all possible combinations of three unique values out of four objects (a total of 12 combinations).

\subsection{Comment on the role of geometric invariants}

We conclude with a brief discussion of the role of curvature invariants on the jet manifold, and their possible relevance to scattering amplitudes.
For any (pseudo)-Riemannian metric that we define on the jet manifold we can calculate the curvature tensors associated to its Levi-Civita connection. The Riemann tensor, Ricci tensor, and Ricci scalar of a given metric are all objects we can compute on the jet manifold. All of them can only be related to physical fields after being pulled back along a section $j^1 \phi$ of the jet bundle, to form a Lagrangian functional of $\phi$.

Since a smooth map moves points around on the manifold, it should be clear that none of these object are invariant under one. We have seen many examples of this throughout the paper. For the Riemann and Ricci tensors, the story is the same one we saw back in \S \ref{sec:preliminaries}, whereby a smooth map changes the metric tensor $g$ (see {\em e.g.} Eq. (\ref{Transformation of metric under diffeomorphism})). The story for the Ricci scalar (which has no `components') is perhaps more subtle and thus deserving of a closer look. The isometry invariance of the Ricci scalar (as is relevant, for example, in showing that it is invariant under coordinate transformations) follows simply from the manipulations
\begin{equation}
    f^* R((f^*)^{-1}g) = R(f^* (f^*)^{-1}g) = R(g)\, .
\end{equation}
In contrast, under a smooth map $f$ (that actually moves points, rather than just relabelling them) we do not have the compensating factor of $(f^*)^{-1}$, thus giving
\begin{equation}
\label{ricci scalar under map f}
    f^* R(g) = R(f^* g) \neq R(g)\, ,
\end{equation}
unless, of course, the smooth map $f$ is itself an isometry with respect to the metric $g$ (in this case it would be a `non-trivial isometry', like the internal $O(4)$ symmetry we have discussed in the context of the Higgs EFTs).
If we further take the smooth map $f$ to be a diffeomorphism, which we have seen in preceding Sections can replicate a {\em non-derivative field redefinition}, then $f$ will generally change the values of the curvature invariants -- but the requirement that $f$ and $f^{-1}$ be well defined and differentiable ensures that a singularity cannot be introduced by $f$. This helps to make sense of the criterion suggested in Ref.~\cite{Cohen:2020xca} for determining whether SMEFT is enough.

More generally, a field redefinition can be viewed as a change of section, which is more general than a diffeomorphism (indeed, it cannot in general be implemented as a bundle morphism, as discussed quite generally in \S \ref{sec:lemma}). Such a derivative change of section {\em can} introduce a singularity into any of the curvature invariants, which is precisely the case in the example discussed in Appendix E of ~\cite{Cohen:2020xca}. Finally, we observe that a change of basis at the level of the Lagrangian is typically done through a field redefinition involving derivatives (such as the `$\Box \phi$'-dependent redefinitions we frequently encountered in this paper), which can completely change the values of curvature invariants, and potentially even introduce singularities. This makes the use of curvature invariants a somewhat unreliable ({\em i.e.} fundamentally basis-dependent) way of comparing theories at the level of the Lagrangian.\footnote{We nonetheless offer some related comments concerning curvature invariants, and their role in detecting a certain subset of redundancies in our metric to Lagrangian maps, in Appendix~\ref{app}.}

\section{Conclusion}\label{sec:conclusions}

In this paper we used geometry on jet bundles, which are a sequence of manifolds that one can construct given a set of scalar fields $\phi^i$ for which spacetime derivatives $\partial_{\mu_1 \dots \mu_r} \phi^i$ are treated as independent coordinates, as a source of Lagrangians defining scalar field theories. In particular, we showed that Lagrangians containing operators with up to 4 spacetime derivatives can be obtained by pulling back a metric on the appropriate 1-jet bundle given the field content of the theory. The jet bundle geometry is taken to be the most general geometry that is invariant under the symmetries of the EFT, which we saw are naturally extended (or `prolongated') to the jet bundle. 

In all the examples we have explored (\S \ref{sec:examples} and \S \ref{sec:HEFTandSMEFT}), the Lagrangian obtained by pulling back the 1-jet bundle metric contains a complete basis of operators up to 4 derivatives, where all IBP redundancies have been removed. The IBP redundancies are accounted for automatically in the formalism, meaning that the jet bundle metric naturally pulls back to a Lagrangian without $\square \phi$ terms -- one could say that the jet bundle metric naturally takes us to the most general possible EFT Lagrangian (with up to 4-derivatives) expressed in a Green's basis.
On the other hand, equation-of-motion redundancies, which stem from invariance of the $S$-matrix under certain derivative field redefinitions, can be identified and implemented, but in a more {\em ad hoc} fashion.
An important thing to stress, in our view, is that the map from 1-jet metrics to 4-derivative EFT Lagrangians surjects, in the sense that one can find metrics that map to every possible EFT.
We suspect that this statement can be generalized to  invariant Lagrangians with up to $2(r+1)$ derivatives, in correspondence with invariant metrics on the $r$-jet bundle -- but we leave a proof to future work.

Along the way, we hope to have clarified various issues related to field redefinitions, which we understand to be smooth changes of (possibly prolongated) section of the (jet) bundle. This applies to both non-derivative and derivative field redefinitions. In the former case, one is afforded an equivalent description in terms of bundle morphisms, which means the field redefinition can be implemented by doing a diffeomorphism on the jet bundle metric. We find no such correspondence between derivative changes of section and bundle morphisms.

Finally, we expand our metric in the vicinity of a point in the 1-jet bundle. Translating this to the Lagrangian, we organise an expansion of $\cL$ in terms higher- and higher-point interactions, inspired by the ideas of~\cite{Cohen:2021ucp}, keeping all interactions with up to 4 derivatives. This facilitates a connection between components of the 1-jet bundle metric (and its derivatives) and propagators/amplitudes of the EFT. We do not elucidate this correspondence comprehensively in this paper, but we already highlight some nice features -- for example, we obtain the lowest-point correlation functions explicitly, and infer {\em e.g.} how positivity (unitarity) constrains the signature of the jet bundle metric. 

In the future, we aim to investigate more systematically the structure of amplitudes in this jet-bundle formalism, especially in the important case of the Higgs EFTs. Other elements that we plan to incorporate into our jet bundle formalism, to describe higher-derivative EFTs geometrically, include the incorporation of fermions, and the (partial) gauging of symmetries -- both of which are necessary to connect with the rich phenomenology of the EFTs that describe elementary particles at the energy frontier.

\acknowledgments 

JD is grateful to Ben Gripaios and Joseph Tooby-Smith for passing on their knowledge of jet bundles a number of years ago. JD further thanks J. Tooby-Smith for comments on the first version of this manuscript, and Nakarin Lohitsiri for discussions. IB thanks J.C. Criado, R.V. Harlander and M. Schaaf for their support with the validation of the dimension-10 and 12 SMEFT bases. We thank Javier Lizana, Ken Mimasu, Dave Sutherland, Tevong You, and other participants of the HEFT 2023 workshop, for helpful discussions. We are also grateful to Nathaniel Craig and Yu-Tse Lee for their patience in waiting for us to complete a draft of our manuscript, and to coordinate with their release of~\cite{Craig:2023hhp}. 
The work of JD is funded by the European Research Council (ERC) under the European Union’s Horizon 2020 research and innovation programme under grant agreement 833280 (FLAY), and by the Swiss National Science Foundation (SNF) under contract 200020-204428. 
IB and MA acknowledge funding
from SNF through the PRIMA grant no. 201508.

\appendix

\section{Diagnosing redundancies in the metric using invariants} \label{app}

Although the geometric invariants are unlikely to be useful when it comes to distinguishing Lagrangians, they do have an application when it comes to redundancies at the level of the metric.

Our expression for the Lagrangian is coordinate free, which has the benefit of allowing us to express the metric in our preferred coordinate system, but introduces ambiguities as it is difficult at a first glance to determine whether two metrics $g$ and $g'$ will yield the same Lagrangian.
We saw earlier that a coordinate transformation leaves the Lagrangian unchanged, thus if we can show that $g$ and $g'$ are related by a coordinate transformation, then we know that they will generate the same Lagrangian.
Writing down the most general coordinate transformation is quite challenging and quickly becomes unreasonable as the dimension of our jet manifold increases. Comparing scalar geometric invariants gives an approach to see if two metrics are dissimilar~\cite{Mergulhao:2020awr}.

Scalar geometric invariants are coordinate independent, but they are typically expressed in coordinates making them difficult to compare directly. However, an alternative approach exists. One can start by calculating several scalar invariants, such as:
\begin{itemize}
    \item Ricci scalar: $\mathcal{R}_1 = g^{\mu \nu} g^{\rho \sigma} R_{\mu \nu \rho \sigma} $
    \item Kretschmann scalar: $\mathcal{R}_2= R^{\mu \nu \rho \sigma} R_{\mu \nu \rho \sigma}$
    \item Cubic curvature invariant: $\mathcal{R}_3 = R^{\mu \nu}_{\alpha \beta} R^{\alpha \beta}_{\rho \sigma} R^{\rho \sigma}_{\mu \nu}$
\end{itemize}

Then one can establish functional relations among these invariants. If at any point the functional relations of the two metrics disagree, then we conclude that they are not related by a coordinate transformation. However, agreement in all computed functional relations is not proof that the metrics are related by a transformation, since there is an infinite number of scalar invariants.

In Eq. (\ref{general metric toy model}) we saw that several components of the metric pullback to the same term resulting in redundancies in the final expression. To identify the terms that could be removed by a coordinate transformation it is not enough to compare scalar invariants, rather we need a way to prove that two metrics generate the same geometry, which can be done using the Cartan-Karlhede algorithm~\cite{cartan1951leccons,brans1965invariant,karlhede1980review,Mergulhao:2020awr,McNutt:2017zxm}.

The first step in applying the formalism is to choose a symmetric metric with constant signature ({\em e.g.} $\delta_{ij}, \eta_{ij}, \dots$) and introduce a frame of vector fields $\{e_a = e^i_a \partial_i\}$ such that
\begin{equation}
    g_{ij} e^i_a e^j_b = \delta_{ab}
\end{equation}
the choice of frame is not unique since
\begin{equation}
    e^a_i \to e'^a_i = \Lambda^a_b e^b_i
\end{equation}
with
\begin{equation}
    \delta_{ab} \Lambda^a_c \Lambda^b_d = \delta_{cd}
\end{equation}
is also a valid frame of vector fields.

The matrices $\Lambda$ form a Lie group with dimension $\frac{n(n-1)}{2}$. Now we compute the Riemann tensor and project it along these frames looking for the subgroup $H_0$ of matrices that preserve the form of the Riemann tensor
\begin{equation}
    R_{a b c d} = R'_{\alpha \beta \gamma \delta}
\end{equation}
and determine the number $\tau_0$ of independent parameters upon which the components of the Riemann tensor depend.

Afterwards we repeat the procedure for the covariant derivatives of the Riemann tensor and determine subgroups $H_1, H_2, \dots$ as well as numbers of independent variables $\tau_1, \tau_2, \dots$ stopping once we have reached an order $q$ at which $H_q = H_{q-1}$ and $\tau_q = \tau_{q-1}$. If a frame exists at this order such that all the derivatives of Riemann tensor for each of the two metrics $g$ and $g'$ agree then we conclude that they are equivalent and produce the same geometry and thus yield the same Lagrangian upon pullback.

The redundancies captured by this algorithm are only the ones that could be removed by a coordinate transformation. Redundancies that need to be removed via integration by parts relations and field redefinitions are only seen after pullback and thus are not related to coordinate transformations.

\section{Higher jet bundles for higher derivatives: a completeness proof} \label{app:proof}

One of the main points of this paper is that scalar EFTs with up to 4 derivatives admit a geometric interpretation on a 1-jet bundle.
More precisely, we showed in a number of explicit examples that the Lagrangian obtained pulling back to spacetime a 1-jet bundle metric always contains a redundant set of effective operators with 0, 2 and 4 derivatives, that can be reduced down to a minimal basis by applying a small number of IBP (and EOM, if desired) relations. 

In this Appendix we provide a general proof that the  Lagrangian obtained by pulling back to spacetime the most general $r$-jet bundle metric always contains a redundant set of effective operators with up to $2(r+1)$ derivatives and arbitrary number of field insertions. Moreover, within this set, it is always possible to identify a complete and non-redundant Green's basis that does not contain operators with boxes.

This result proves that the empirical observations in \S\S\ref{sec:examples}, \ref{sec:HEFTandSMEFT} generalize to any scalar theory, and that operators with 6 or more derivatives also admit a geometric interpretation in terms of higher-jet bundle metrics. 
The proof formally holds for a Lorentz-invariant scalar EFT in any number of spacetime dimensions (with Lorentz indices being everywhere contracted via the Minkowski metric, as justified generally in \S \ref{sec:prolonging-symmetres-1jet-bundle}), although the interpretation of the EFT expansion is of course different in different dimensions.\footnote{For example, in the special case of 2d EFTs, the scalar field is dimensionless and so the counting of derivatives coincides precisely with the mass dimension counting of EFT operators.}

\bigskip

Our argument consists of two points, that we prove individually below:
\begin{enumerate}
\item Consider pure scalar interactions among $N$ fields and with a total of $2(r+1)$ derivatives. It is always possible to construct a complete and IBP-non-redundant basis of operators at this order, that does not contain operators with boxes, nor operators with more than $(r+1)$ derivatives acting on a single field. 

\item Any operator containing an arbitrary number of scalar fields $N$ and up to $2(r+1)$ derivatives, such that no more than $(r+1)$ derivatives act on one field and that boxes are absent, can be obtained by pulling back to spacetime the most general metric of a $r$-jet bundle. 
\end{enumerate}
Put together, these statements imply that a non-redundant Green's basis can always be found within the set of box-less operators with $N$ fields and up to $2(r+1)$ derivatives produced by the $r$-jet bundle metric. Vice versa, any arbitrary scalar EFT Lagrangian always admits a geometric interpretation on the appropriate jet bundle.

In practice, the metric will always produce a redundant Lagrangian, that contains all possible box-less structures with up to $2(r+1)$ derivatives, plus some (but not all!) of those with boxes. 
In general, there will be several options to reduce the resulting operator set down to a minimal basis. Here we choose, as an algorithmic rule, to remove all operators with boxes. This choice is convenient for a number of reasons: (i)  statement 1. above guarantees that boxes can always be safely removed in any scalar theory at any order;
(ii) statement 2. above ensures that 
a box-less basis will be obtained directly from the jet bundle metric, without requiring manipulations at the Lagrangian level;
(iii) empirically we find that this choice removes most ambiguities in the basis construction. In fact, in all the examples considered, there turned out to be a unique basis satisfying the box-less requirement; 
(iv) any alternative choice aimed at retaining operators with boxes would need to distinguish between box operators that can and cannot be obtained from the metric, making the construction more cumbersome. 

In the statements above we are also requesting the absence of operators with more than half of the derivatives $(r+1)$ acting on a single field. This condition is strictly necessary in order to obtain the Lagrangian from a $r$-jet bundle. In principle one could have evaded this rule by choosing a higher-jet bundle, which however would have also produced terms with more than $2(r+1)$ derivatives. It is interesting that the minimal (and more natural) choice of a $r$-jet bundle suffices to produce a complete basis.

In the next two subsections we provide proofs of points 1. and 2. above.
The results in this Appendix hold independently of the number of scalar flavours present and on whether an internal symmetry is imposed or not, as they are insensitive to the presence of  flavour indices on the scalar fields. Arbitrary contractions of internal indices can always be inserted appropriately in the metric functions.

\subsection{Proof of point 1.}
Operators with more than $(r+1)$ derivatives on a single field can always be removed by trivially moving derivatives across the operator via IBP, until there is a maximum of $(r+1)$ derivatives on each field.\footnote{Of course, this requires at least 2 fields to be inserted in one operator. On the other hand, a term with 1 field only would automatically be a total derivative.} Therefore they are always redundant.

Operators with boxes acting on one or more fields can also be removed similarly, by using IBP to move one of the two derivatives contracted in a box to other fields. This operation requires to apply, in sequence, an IBP for each box, \emph{e.g.}
\begin{align}
  (\de_\mu\square\phi^{i_1})(\square\phi^{i_2}) (\de^\mu\phi^{i_3}) A(\phi) 
  &= -(\de_\mu\de_\rho\phi^{i_1})(\square \phi^{i_2})(\de^\mu\phi^{i_3})(\de^\rho A(\phi)) + \dots
  \\
  &= (\de_\sigma\de_\mu\de_\rho\phi^{i_1})(\de^\sigma \phi^{i_2})(\de^\mu\phi^{i_3})(\de^\rho A(\phi)) + \dots
  \nonumber
\end{align}
where  $A$ is some analytic function of the fields, $i_k$ is a  potential flavour index carried by the $k$-th field, and the dots stand for other terms stemming from the IBPs.
A key point is that this operation can never reintroduce boxes because, at each step, it moves a derivative with a different Lorentz index. Therefore operators with boxes can always be fully removed from an operator basis.

To conclude our proof, we need to check that the two redundancy classes can be treated independently, \emph{i.e.} that the IBPs used to remove boxes do not re-introduce operators with more than $(r+1)$ derivatives on a single field. Let us consider an operator with $N$ fields and $2(r+1)$ derivatives, such that $d_k$ derivatives act on the $k$-th field, and they are contracted to form $b_k$ boxes, with $2b_k\leq d_k$. 

The boxes can be removed with a sequence of IBPs as shown above. They send
\begin{align}
\label{eq.counting_d_after_boxIBP}
d_k &\to d_k-b_k+s\,,
&\text{with}\quad
0&\leq s \leq \sum_{j\neq k} b_j\,.
\end{align}
The term $-b_k$ accounts for the fact that, for every box on the $k$-th field, one derivative is removed and placed elsewhere. In addition,  the $k$-th field receives $s$ derivatives coming from other boxes in the product. IBPs replace the initial operator with a sum over several terms, in which $s$ takes all possible values in a range from 0 to the total number of boxes originally acting on fields other than the $k$-th.
The latter number is bounded by
\begin{align}
\sum_{j\neq k} b_j&\leq \frac{2(r+1)-d_k-(d_k-2 b_k)}{2}
= r-d_k+b_k+1\,.
\label{eq.max_boxes}
\end{align}
The numerator on the righthand side counts how many Lorentz indices on $\phi_{j\neq k}$ can be contracted into boxes: there are at most $2(r+1)-d_k$, from which we need to subtract $(d_k-2b_k)$ that must be contracted with open derivatives on $\phi_k$.
Putting together~\eqref{eq.counting_d_after_boxIBP} and~\eqref{eq.max_boxes} we see that
\begin{align}
   d_k-b_k+s \leq d_k-b_k+\sum_{j\neq k} b_j \leq  r+1\,.
\end{align}
This proves that the IBP required to remove boxes can never reintroduce terms with more than $r+1$ derivatives on one field.
Therefore both classes of operators can always be fully removed using IBPs, \emph{i.e.} it is always possible to construct a complete and non-redundant Green's basis of operators that does not contain either.

\subsection{Proof of point 2.}
To prove point 2, we write down the most general $r$-jet bundle metric, extending Eq.~\eqref{eq:general_g1}:
\begin{align}
g^{(r)}&= \begin{pmatrix}
  dx^\mu & du^i & du^i_{\mu_1} & du^i_{\mu_1\mu_2} &\cdots &du^i_{\mu_1\dots \mu_{r}}
 \end{pmatrix}
 [g^{(r)}]
  \begin{pmatrix}
  dx^\nu \\ du^j \\ du^j_{\nu_1} \\ du^j_{\nu_1\nu_2} \\\vdots\\ du^j_{\nu_1\dots \nu_{r}}
 \end{pmatrix}\,,
 \\[2mm]
  [g^{(r)}] &=\begin{pmatrix}
  g_{\mu\nu} & g_{\mu j} & g_{\mu j}^{\nu_1} & g_{\mu j}^{\nu_1\nu_2} & \cdots & g_{\mu j}^{\nu_1\dots \nu_{r}}
  \\[2mm]
  g_{\nu i} & g_{ij} & g_{ij}^{\nu_1} & g_{ij}^{\nu_1\nu_2} & \cdots & g_{ij}^{\nu_1\dots \nu_{r}}
  \\[2mm]
  g_{\nu i}^{\mu_1} & g_{ij}^{\mu_1} & g_{ij}^{\mu_1\nu_1}& g_{ij}^{\mu_1\nu_1\nu_2}& \cdots & g_{ij}^{\mu_1\nu_1\dots \nu_{r}}\\[2mm]
  g_{\nu i}^{\mu_1\mu_2} & g_{ij}^{\mu_1\mu_2} & g_{ij}^{\mu_1\mu_2\nu_1}& g_{ij}^{\mu_1\mu_2\nu_1\nu_2}& \cdots & g_{ij}^{\mu_1\mu_2\nu_1\dots \nu_{r}}\\[2mm]
  \vdots& \vdots& \vdots& \vdots& \ddots& \vdots\\[2mm]
  g_{\nu i}^{\mu_1\dots \mu_{r}} & g_{ij}^{\mu_1\dots\mu_{r}} & g_{ij}^{\mu_1\dots\mu_{r}\nu_1}& g_{ij}^{\mu_1\dots\mu_{r}\nu_1\nu_2}& \cdots & g_{ij}^{\mu_1\dots\mu_{r}\nu_2\dots \nu_{r}}\\[2mm]
  \end{pmatrix}\,,
  \label{eq.gr_metric}
\end{align}
where implicitly all the metric entries are functions of $u^i,u_{\mu_1}^i, \dots u^i_{\mu_1\dots \mu_{r}}$ and we can assume
\begin{align}
 g_{\nu i} &= g_{\mu j}\,,
 &
 g_{\mu j}^{\nu_1\dots \nu_s} &= g_{\nu i}^{\mu_1\dots \mu_s} \,,
\end{align}
upon relabeling indices, by symmetry of the metric tensor.

Pulling this metric back to spacetime along $j^1\phi$ and contracting with $\eta^{-1}$ gives 
\begin{align}
\label{eq.general_L}
 \mathcal{L}[\phi,g^{(r)}]&= 
 \frac{1}{2}\eta^{\mu\nu} g_{\mu\nu}
+\frac{1}{2}g_{ij} (\de_\mu \phi^i)(\de^\mu\phi^j)
+ g_{\mu j} (\de^\mu\phi^j)
\\
&+\sum_{s=1}^{r}
g_{\mu j}^{\nu_1\dots \nu_s} (\de^\mu\de_{\nu_1}\cdots\de_{\nu_s}\phi^j)
+\frac{1}{2}g_{ij}^{\mu_1\dots \mu_s} (\de_\rho\de_{\mu_1}\cdots\de_{\mu_s}\phi^i)(\de^\rho\phi^j)
+\frac{1}{2}g_{ij}^{\nu_1\dots \nu_s} (\de_\rho\phi^i)(\de^\rho\de_{\nu_1}\cdots\de_{\nu_s}\phi^j)
\nonumber\\
&
+\frac{1}{2}\sum_{m=1}^{r}\sum_{s=1}^{r} 
g_{ij}^{\mu_1\dots \mu_m\nu_1\dots\nu_s} (\de_\rho\de_{\mu_1}\cdots\de_{\mu_m}\phi^i)(\de^\rho\de_{\nu_1}\cdots\de_{\nu_s}\phi^j)\,.
\nonumber
\end{align}
At this point it easy to show that any effective operator with exactly $2(r+1)$ derivatives and an arbitrary number of fields $N$ can be matched to at least one of the terms in Eq.~\eqref{eq.general_L}.

To do so, it is convenient to classify the operators according to the number of derivatives acting on each field, namely a string
\begin{align}
   &d_1d_2\dots d_N\,, 
   &\text{such that}\quad \sum_k d_k &= 2(r+1)\,,
\end{align}
that we take to be sorted from largest to smallest $d_k$.
In practice, the operator classes are labeled by the integer partitions of $2(r+1)$ into at most $N$ terms, and that do not contain numbers higher than $(r+1)$. For instance, with 8 derivatives and 4 fields, there are 8 classes:
\begin{align}
&4400\,,&
&4310\,,&
&4220\,,&
&4211\,,&
\\
&3320\,,&
&3311\,,&
&3221\,,&
&2222\,.
\end{align}
Each class can be easily put in correspondence with a metric element. 
For instance, operators of class $(r+1)(r+1)0\dots$ can be obtained by pulling back $g_{ij}^{\mu_1\dots\mu_r\nu_1\dots\nu_r}$. 
In general, operators of each class can be obtained by pulling back multiple metric elements, that give identical expressions. To prove our hypothesis, it is enough to find one example for each, which is particularly simple in this notation:
\begin{center}
\renewcommand{\arraystretch}{1.4}
\begin{tabular}{>{$}l<{$}@{$\quad\to\quad$}>{$}l<{$}>{$}l<{$}}
  g_{ij}^{\mu_1\dots\mu_{m-1}\nu_1\dots\nu_{s-1}} &
  ms\dots  & 2\leq m,s\leq r+1 \,,
  \\
  g_{ij}^{\mu_1\dots\mu_{s-1}} &
  s 1\dots  & 2\leq s\leq r +1 \,,
  \\
  g_{ij} & 111\dots 
\end{tabular}
\end{center}
where the dots stand for factors in the integer decomposition of $2(r+1)-m-s$. For instance, considering the example above with $r=3, N=4$:
\begin{align}
g_{ij}^{\mu_1\mu_2\mu_3\nu_1\nu_2\nu_3}&\to 4400\,,
&
g_{ij}^{\mu_1\mu_2\mu_3\nu_1\nu_2}&\to 4310\,,
&
g_{ij}^{\mu_1\mu_2\mu_3\nu_1}&\to 4220,4221\,,\nonumber
\\
g_{ij}^{\mu_1\mu_2\nu_1\nu_2}&\to 3320,3311\,,
&
g_{ij}^{\mu_1\mu_2\nu_1}&\to 3221\,,
&
g_{ij}^{\mu_1\nu_1}&\to 2222\,.
\label{eq.metric_to_class_example}
\end{align}
This correspondence assumes that the metric entries are general functions of $u^i,u^i_\mu,\dots u^i_{\mu_1\dots\mu_r}$, such that insertions of fields and derivatives in the $(N-2)$ rightmost terms in the string can be obtained by extracting the appropriate dependence of the metric on these variables. 
Let us translate this into concrete examples for the cases in~\eqref{eq.metric_to_class_example}. For simplicity, we omit the dependence on $\Lambda$ and take the case of a single real scalar:
\begin{align}
  g_{uu}^{\mu_1\mu_2\mu_3\nu_1\nu_2\nu_3}  &= \eta^{\mu_1\nu_1}\eta^{\mu_2\nu_2}\eta^{\mu_3\nu_3}u^2+\dots 
  &&\to& 
  &(\de_\mu\de_\nu\de_\rho\de_\sigma \phi) (\de^\mu\de^\nu\de^\rho\de^\sigma \phi)\phi^2\,,
  \\
  g_{uu}^{\mu_1\mu_2\mu_3\nu_1\nu_2}  &= \eta^{\mu_1\nu_1}\eta^{\mu_2\nu_2}u^{\mu_3}u+\dots 
  &&\to& 
  &(\de_\mu\de_\nu\de_\rho\de_\sigma \phi) (\de^\mu\de^\nu\de^\rho \phi)(\de^\sigma\phi)\phi\,,
  \label{eq.8d-example_2}
  \\
  g_{uu}^{\mu_1\mu_2\mu_3\nu_1}  &= \eta^{\mu_1\nu_1}u^{\mu_2\mu_3}u+\dots 
  &&\to& 
  &(\de_\mu\de_\nu\de_\rho\de_\sigma \phi) (\de^\mu\de^\nu \phi)(\de^\rho\de^\sigma\phi)\phi\,,
  \\
  g_{uu}^{\mu_1\mu_2\mu_3\nu_1}  &= \eta^{\mu_1\nu_1}u^{\mu_2}u^{\mu_3}u+\dots 
  &&\to& 
  &(\de_\mu\de_\nu\de_\rho\de_\sigma \phi) (\de^\mu\de^\nu \phi)(\de^\rho\phi)(\de^\sigma\phi)\phi\,,
  \\
  g_{uu}^{\mu_1\mu_2\nu_1}  &= \eta^{\mu_1\nu_1}u^{\mu_2\sigma}u_\sigma+\dots 
  &&\to& 
  &(\de_\mu\de_\nu\de_\rho \phi) (\de^\mu\de^\nu \phi)(\de^\rho\de^\sigma\phi)(\de_\sigma\phi)\,,
  \\
  g_{uu}^{\mu_1\nu_1}  &= u^{\mu_1\sigma}u^{\nu_1}_\sigma+\dots 
  &&\to& 
  &(\de_\mu\de_\nu\phi)(\de^\mu\de^\rho\phi) (\de_\rho\de_\sigma \phi)(\de^\nu\de^\sigma\phi)\,.
  \end{align}
  With a larger number of field insertions, more classes are present and metric terms with fewer Lorentz indices become relevant as well, \emph{e.g.}:
  \begin{align}
  g_{uu}^{\mu_1\mu_2\mu_3}  &= u^{\mu_1}u^{\mu_2}u^{\mu_3}+\dots 
  &&\to& 
  &(\de_\mu\de_\nu\de_\rho\de_\sigma\phi)(\de^\mu\phi)(\de^\nu\phi)(\de^\rho\phi)(\de^\sigma\phi)\,,
  \\
   g_{uu}  &= (u_\rho)^6+\dots 
  &&\to& 
  &(\de_\mu\phi)^8\,.
\end{align}
Within each operator class, it is always possible to choose an explicit form of the metric function that gives any arbitrary Lorentz structure in the Lagrangian, provided that at least one pair of derivatives acting on two different fields are contracted with each other (the $\rho$ indices in~\eqref{eq.general_L}). If we stick to box-less operators, then this condition is always verified. Therefore \emph{any} box-less operator with $2(r+1)$ derivatives and maximum $(r+1)$ derivatives on each field can be obtained from a $r$-jet bundle metric. It is not hard to see that several operators \emph{with} boxes, that satisfy that minimal condition, can be generated as well. For instance, we could modify~\eqref{eq.8d-example_2} into
\begin{align}
   g_{uu}^{\mu_1\mu_2\mu_3\nu_1\nu_2}  &= \eta^{\mu_1\mu_2}\eta^{\nu_1\nu_2}u^{\mu_3}+\dots 
  &&\to& 
  &(\de_\mu\de_\sigma\square \phi) (\de^\mu\square \phi)(\de^\sigma\phi)\,.
\end{align}
However, it is not possible to obtain structures such as $(\square\phi)^4$.

In the presence of multiple scalar flavours, arbitrary index assignments or contractions can be achieved. For instance, taking $r=2$, we can have $O(n)$ invariant contractions:
\begin{align}
  g_{ij}^{\mu_1\mu_2\nu_1\nu_2}  &= \delta_{ij}\eta^{\mu_1\nu_1}\eta^{\mu_2\nu_2}\eta^{\mu_3\nu_3}+\dots 
  &&\to& 
  &(\de_\mu\de_\nu\de_\rho \phi \cdot \de^\mu\de^\nu\de^\rho\phi)\,,
  \\
  g_{ij}^{\mu_1\nu_1}  &= \delta_{ij}\eta^{\mu_1\nu_1}(u\cdot  u_\sigma)^2+\dots 
  &&\to& 
  &(\de_\mu\de_\nu\phi\cdot\de^\mu\de^\nu\phi)(\phi\cdot\de_\sigma\phi)^2\,,
  \\
   g_{ij}  &= u_i u_j (u_\rho \cdot u^{\rho}) (u_\sigma \cdot u^{\sigma})+\dots 
  &&\to& 
  &(\phi\cdot\de_\mu\phi)^2(\de_\rho\phi\cdot\de^\rho\phi)^2\,,
\end{align}
or arbitrary flavour assignments $i_1i_2\dots i_N$ without any symmetry imposed (in this case the indices are just labels, they do not need to be contracted)
\begin{align}
  g_{ij}^{\mu_1\mu_2\nu_1\nu_2}  &= \delta_{ii_1}\delta_{ji_2}\eta^{\mu_1\nu_1}\eta^{\mu_2\nu_2}\eta^{\mu_3\nu_3}+\dots 
  &&\to& 
  &(\de_\mu\de_\nu\de_\rho \phi^{i_1})( \de^\mu\de^\nu\de^\rho\phi^{i_2})\,,
  \\
  g_{ij}^{\mu_1\nu_1}  &= \delta_{ii_1}\delta_{ji_2}\eta^{\mu_1\nu_1}u^{i_3}_\sigma u^{i_4\sigma}+\dots 
  &&\to& 
  &(\de_\mu\de_\nu\phi^{i_1})(\de^\mu\de^\nu\phi^{i_2})(\de_\sigma\phi^{i_3})(\de^\sigma\phi^{i_4})\,,
  \\
   g_{ij}  &= \delta_{ii_1}\delta_{ji_2}u^{i_3}_\rho u^{i_4\rho} u^{i_5}_\sigma u^{i_6\sigma}+\dots 
  &&\to& 
  &(\de_\mu\phi^{i_1})(\de^\mu\phi^{i_2})(\de_\rho\phi^{i_3})(\de^\rho\phi^{i_4})(\de_\sigma\phi^{i_5})(\de^\sigma\phi^{i_6})\,.
\end{align}
This concludes the proof for scalar operators with exactly $2(r+1)$ derivatives. The last step is extending it to lower derivatives. This is trivial: by the argument just concluded, operators with $2n<2(r+1)$ derivatives can be obtained from a $(n-1)$-jet bundle metric. Since $(n-1)<r$, the latter is contained as a sub-block in the $r$-jet bundle metric (see Eq.~\eqref{eq.gr_metric}). Finally, operators without derivatives can always be obtained from $g_{\mu\nu}$, in any $r$-jet bundle metric.

\bibliographystyle{JHEP}
\bibliography{refs}
\end{document}